\documentclass[AMA,STIX2COL]{MRM}

\newcommand{\Papertitle}{Rapid quantitative magnetization transfer imaging: utilizing the hybrid state and the generalized Bloch model}
\articletype{Pre-Print -- published in Magnetic Resonance in Medicine}%

\received{15 June 2023}
\revised{30 October 2023}
\accepted{14 November 2023}
\newif\ifOL
\OLfalse

\usepackage{filecontents}

\ifOL
\else
\usepackage{animate}
\usepackage{tikz}
\usepackage{pgfplots}

\usetikzlibrary{spy,backgrounds,arrows,decorations.pathmorphing,backgrounds,positioning,fit,matrix,calc,shapes.multipart,shapes.misc,pgfplots.statistics, pgfplots.colorbrewer}

\pgfplotsset{compat=1.18,
	boxplot/estimator=R1,
	colormap={parula}{
			rgb255=(53,42,135)
			rgb255=(15,92,221)
			rgb255=(18,125,216)
			rgb255=(7,156,207)
			rgb255=(21,177,180)
			rgb255=(89,189,140)
			rgb255=(165,190,107)
			rgb255=(225,185,82)
			rgb255=(252,206,46)
			rgb255=(249,251,14)},
	colormap={mybluered}{
			rgb255(0cm)=(0,0,180)
			rgb255(1cm)=(0,180,180)
			rgb255(2cm)=(70,180,0)
			rgb255(3cm)=(180,180,0)
			rgb255(4cm)=(255,0,0)
			rgb255(5cm)=(128,0,0)},
	colormap={mybluewhitered_m10_2}{
			rgb255(0cm)=(0,0,255)
			rgb255(10cm)=(255,255,255)
			rgb255(12cm)=(255,0,0)},
	colormap={mybluewhitered_m1_2}{
			rgb255(0cm)=(0,0,255)
			rgb255(1cm)=(255,255,255)
			rgb255(3cm)=(255,0,0)},
	colormap={mybluewhitered_0_1}{
			rgb255(0cm)=(255,255,255)
			rgb255(1cm)=(255,0,0)},
	colormap={gist_earth}{
			rgb255=(0.0,0.0,0.0)
			rgb255=(1.7085000000000001,0.0,62.424)
			rgb255=(3.3914999999999997,0.0,89.8365)
			rgb255=(5.1000000000000005,1.4789999999999999,113.985)
			rgb255=(6.8085,7.343999999999999,116.4075)
			rgb255=(8.4915,13.209,116.8665)
			rgb255=(10.200000000000001,19.074,117.3255)
			rgb255=(11.9085,24.939,117.7845)
			rgb255=(13.5915,30.804000000000002,118.2435)
			rgb255=(15.299999999999999,36.669000000000004,118.7025)
			rgb255=(16.983,42.534,119.1615)
			rgb255=(18.6915,48.3735,119.6205)
			rgb255=(20.400000000000002,53.677499999999995,120.10499999999999)
			rgb255=(22.083,58.981500000000004,120.564)
			rgb255=(23.7915,64.2855,121.02300000000001)
			rgb255=(25.5,69.58949999999999,121.482)
			rgb255=(27.183,74.8935,121.941)
			rgb255=(28.8915,79.917,122.39999999999999)
			rgb255=(30.599999999999998,84.6855,122.859)
			rgb255=(32.282999999999994,89.454,123.3435)
			rgb255=(33.9915,94.2225,123.8025)
			rgb255=(35.7,98.838,124.2615)
			rgb255=(37.383,102.86699999999999,124.7205)
			rgb255=(39.091499999999996,106.896,125.1795)
			rgb255=(40.774499999999996,110.925,125.63850000000001)
			rgb255=(42.483,114.954,126.0975)
			rgb255=(44.191500000000005,118.983,126.5565)
			rgb255=(45.8745,123.012,127.041)
			rgb255=(47.583,127.041,127.5)
			rgb255=(48.909,129.13199999999998,125.23049999999999)
			rgb255=(50.1075,130.5855,122.1195)
			rgb255=(51.306,132.0645,118.983)
			rgb255=(52.5045,133.518,115.872)
			rgb255=(53.7285,134.99699999999999,112.7355)
			rgb255=(54.927,136.476,109.6245)
			rgb255=(56.125499999999995,137.92950000000002,106.51350000000001)
			rgb255=(57.324,139.4085,103.377)
			rgb255=(58.5225,140.862,100.266)
			rgb255=(59.721,142.341,97.12950000000001)
			rgb255=(60.945,143.7945,94.0185)
			rgb255=(62.1435,145.27349999999998,90.882)
			rgb255=(63.342000000000006,146.7525,87.771)
			rgb255=(64.5405,148.20600000000002,84.66000000000001)
			rgb255=(65.73899999999999,149.685,81.5235)
			rgb255=(66.9375,151.1385,78.4125)
			rgb255=(68.136,152.6175,75.27600000000001)
			rgb255=(69.8955,154.071,72.16499999999999)
			rgb255=(75.5565,155.54999999999998,70.5585)
			rgb255=(81.2175,157.029,72.063)
			rgb255=(86.904,158.48250000000002,73.542)
			rgb255=(92.565,159.9615,75.021)
			rgb255=(98.226,161.415,76.5)
			rgb255=(103.887,162.894,78.00450000000001)
			rgb255=(109.57350000000001,164.06699999999998,79.48349999999999)
			rgb255=(115.23450000000001,165.18900000000002,80.9625)
			rgb255=(120.74249999999999,166.311,82.11)
			rgb255=(124.95,167.4075,82.926)
			rgb255=(129.1575,168.5295,83.742)
			rgb255=(133.365,169.6515,84.5325)
			rgb255=(137.5725,170.77349999999998,85.3485)
			rgb255=(141.78,171.8955,86.16449999999999)
			rgb255=(145.9875,173.01749999999998,86.95500000000001)
			rgb255=(150.195,174.1395,87.771)
			rgb255=(154.4025,175.2615,88.5615)
			rgb255=(158.60999999999999,176.3835,89.3775)
			rgb255=(162.792,177.48,90.1935)
			rgb255=(166.9995,178.602,90.984)
			rgb255=(171.207,179.724,91.8)
			rgb255=(175.41449999999998,180.846,92.59049999999999)
			rgb255=(179.622,181.968,93.40650000000001)
			rgb255=(183.192,182.42700000000002,94.2225)
			rgb255=(184.263,180.2085,95.01299999999999)
			rgb255=(185.334,178.0155,95.82900000000001)
			rgb255=(186.405,175.8225,96.6195)
			rgb255=(187.476,173.60399999999998,97.4355)
			rgb255=(188.5215,171.411,98.2515)
			rgb255=(189.5925,169.218,99.042)
			rgb255=(190.6635,166.9995,99.858)
			rgb255=(191.7345,164.8065,100.6485)
			rgb255=(193.1115,163.2255,104.5755)
			rgb255=(195.9675,164.3985,110.823)
			rgb255=(198.8235,165.75,117.0705)
			rgb255=(201.67950000000002,167.94299999999998,123.318)
			rgb255=(204.5355,170.1615,129.5655)
			rgb255=(207.366,172.38000000000002,135.813)
			rgb255=(210.222,174.573,142.06050000000002)
			rgb255=(213.078,176.7915,148.308)
			rgb255=(215.934,179.469,154.5555)
			rgb255=(218.79,183.141,160.80300000000003)
			rgb255=(221.646,186.66,167.0505)
			rgb255=(224.4765,190.4595,174.1395)
			rgb255=(227.33249999999998,195.45749999999998,181.815)
			rgb255=(230.18849999999998,200.43,189.516)
			rgb255=(233.0445,205.428,197.1915)
			rgb255=(235.90050000000002,210.42600000000002,204.89249999999998)
			rgb255=(238.75650000000002,216.1635,212.568)
			rgb255=(241.6125,222.0285,220.2435)
			rgb255=(244.443,228.7605,227.9445)
			rgb255=(247.299,236.2065,235.62)
			rgb255=(250.155,243.6525,243.32100000000003)
			rgb255=(253.011,250.9965,250.9965)
		},
	colormap={berlin}{
			rgb255=(158.37591,175.99641,254.87428500000001)
			rgb255=(156.100035,175.75313999999997,253.82037)
			rgb255=(153.81651,175.50375,252.765945)
			rgb255=(151.521,175.250535,251.70846)
			rgb255=(149.21707500000002,174.99324,250.64511)
			rgb255=(146.90244,174.73161,249.57691499999999)
			rgb255=(144.57505500000002,174.46233,248.50361999999998)
			rgb255=(142.236705,174.18999,247.42344)
			rgb255=(139.89045000000002,173.90796,246.33408)
			rgb255=(137.52838500000001,173.619045,245.23452)
			rgb255=(135.15867,173.32120500000002,244.12425000000002)
			rgb255=(132.775185,173.01342,243.00020999999998)
			rgb255=(130.38022500000002,172.69365,241.85883)
			rgb255=(127.97506499999999,172.35654,240.70036499999998)
			rgb255=(125.55384,172.00413,239.520735)
			rgb255=(123.12216,171.63412499999998,238.31637)
			rgb255=(120.675945,171.24015,237.086505)
			rgb255=(118.22055,170.82399,235.82553000000001)
			rgb255=(115.752405,170.37876,234.532425)
			rgb255=(113.274315,169.90012499999997,233.20209000000003)
			rgb255=(110.7822,169.38910499999997,231.82968000000002)
			rgb255=(108.284475,168.84060000000002,230.41443)
			rgb255=(105.77859000000001,168.246705,228.950475)
			rgb255=(103.268625,167.608695,227.43577499999998)
			rgb255=(100.759935,166.917645,225.86625)
			rgb255=(98.25048000000001,166.17686999999998,224.23884)
			rgb255=(95.750715,165.37668,222.553035)
			rgb255=(93.26421,164.51682,220.805265)
			rgb255=(90.795045,163.59576,218.994255)
			rgb255=(88.345515,162.61146000000002,217.120005)
			rgb255=(85.93041,161.56136999999998,215.18302500000001)
			rgb255=(83.54871,160.443195,213.184335)
			rgb255=(81.214185,159.260505,211.12393500000002)
			rgb255=(78.932445,158.012535,209.00514)
			rgb255=(76.69992,156.696735,206.83254000000002)
			rgb255=(74.538795,155.324325,204.606645)
			rgb255=(72.44499,153.888675,202.33485)
			rgb255=(70.43227499999999,152.39667,200.01843)
			rgb255=(68.49172499999999,150.8529,197.661465)
			rgb255=(66.63354,149.26042500000003,195.2739)
			rgb255=(64.86384,147.62153999999998,192.85548)
			rgb255=(63.177015,145.94364,190.413345)
			rgb255=(61.57332,144.22698,187.95183)
			rgb255=(60.056325,142.48074,185.474505)
			rgb255=(58.60971,140.70951,182.988)
			rgb255=(57.248265,138.91125,180.490275)
			rgb255=(55.968675000000005,137.09514,177.98949)
			rgb255=(54.74697,135.26041500000002,175.48845)
			rgb255=(53.59386,133.414215,172.986135)
			rgb255=(52.501695,131.55373500000002,170.48739)
			rgb255=(51.451605,129.69249,167.990685)
			rgb255=(50.45889,127.82078999999999,165.50265)
			rgb255=(49.51386,125.94526499999999,163.01818500000002)
			rgb255=(48.59178,124.067955,160.54086)
			rgb255=(47.71356,122.19115500000001,158.07398999999998)
			rgb255=(46.85676,120.31563000000001,155.611455)
			rgb255=(46.027499999999996,118.44087,153.15861)
			rgb255=(45.228075,116.564835,150.714435)
			rgb255=(44.43732,114.69594000000001,148.278165)
			rgb255=(43.662119999999994,112.83086999999999,145.85133)
			rgb255=(42.90171,110.96886,143.43393)
			rgb255=(42.14946,109.11041999999999,141.020355)
			rgb255=(41.421945,107.25504,138.618765)
			rgb255=(40.683975,105.40935,136.22355)
			rgb255=(39.968444999999996,103.56748499999999,133.83828)
			rgb255=(39.245774999999995,101.72766,131.45862)
			rgb255=(38.53611,99.89803500000001,129.08865)
			rgb255=(37.828230000000005,98.070705,126.72786)
			rgb255=(37.138455,96.25281000000001,124.376505)
			rgb255=(36.434145,94.43746499999999,122.02872)
			rgb255=(35.73519,92.630535,119.694195)
			rgb255=(35.05383,90.83202,117.360945)
			rgb255=(34.368135,89.03630999999999,115.042485)
			rgb255=(33.680145,87.24825,112.731675)
			rgb255=(32.997254999999996,85.469115,110.42571)
			rgb255=(32.317425,83.68972500000001,108.12918)
			rgb255=(31.642950000000003,81.921045,105.84310500000001)
			rgb255=(30.97128,80.158485,103.56672)
			rgb255=(30.319245000000002,78.4023,101.29467)
			rgb255=(29.66058,76.65504,99.03588)
			rgb255=(29.001405,74.91415500000001,96.78423)
			rgb255=(28.352684999999997,73.1799,94.54074)
			rgb255=(27.696315000000002,71.455335,92.31102)
			rgb255=(27.070545,69.74173499999999,90.0864)
			rgb255=(26.43585,68.02788000000001,87.87147)
			rgb255=(25.801665,66.32754,85.66775999999999)
			rgb255=(25.18788,64.634085,83.47221)
			rgb255=(24.568485,62.94675,81.289665)
			rgb255=(23.985045,61.26732,79.11808500000001)
			rgb255=(23.405939999999998,59.600384999999996,76.94829)
			rgb255=(22.82403,57.947475,74.79609)
			rgb255=(22.271955000000002,56.304,72.65307)
			rgb255=(21.711209999999998,54.6618,70.52687999999999)
			rgb255=(21.182595000000003,53.035155,68.403495)
			rgb255=(20.67999,51.415905,66.29796)
			rgb255=(20.178150000000002,49.817055,64.204155)
			rgb255=(19.707929999999998,48.22968,62.115195)
			rgb255=(19.270605,46.650465,60.05301)
			rgb255=(18.868215,45.092925,57.99567)
			rgb255=(18.46455,43.550174999999996,55.955414999999995)
			rgb255=(18.116474999999998,42.022725,53.9325)
			rgb255=(17.790585,40.519754999999996,51.92514)
			rgb255=(17.49759,39.025200000000005,49.933589999999995)
			rgb255=(17.2278,37.566345,47.97162)
			rgb255=(16.999575,36.126104999999995,46.01526)
			rgb255=(16.810364999999997,34.69938,44.093835)
			rgb255=(16.661444999999997,33.327225,42.19179)
			rgb255=(16.552305,31.963994999999997,40.333095)
			rgb255=(16.48218,30.63366,38.49123)
			rgb255=(16.451835000000003,29.342850000000002,36.691694999999996)
			rgb255=(16.46127,28.10661,34.924034999999996)
			rgb255=(16.510995,26.883885000000003,33.208650000000006)
			rgb255=(16.6005,25.716495,31.510095)
			rgb255=(16.672665,24.599595,29.878349999999998)
			rgb255=(16.721369999999997,23.54619,28.30704)
			rgb255=(16.80246,22.491255,26.770410000000002)
			rgb255=(16.92894,21.45417,25.31844)
			rgb255=(17.11254,20.413005,23.926395)
			rgb255=(17.389215,19.405245,22.559849999999997)
			rgb255=(17.7786,18.432165,21.171375)
			rgb255=(18.267944999999997,17.506770000000003,19.77372)
			rgb255=(18.86439,16.58979,18.38805)
			rgb255=(19.53198,15.722534999999999,16.996005)
			rgb255=(20.307434999999998,14.93025,15.588915)
			rgb255=(21.155565,14.19483,14.214975)
			rgb255=(22.066935,13.514235000000001,12.83568)
			rgb255=(23.030325,12.928245,11.485199999999999)
			rgb255=(24.0363,12.432015,10.142115)
			rgb255=(25.071345,11.995455,8.844165)
			rgb255=(26.12679,11.63412,7.66887)
			rgb255=(27.216659999999997,11.399775,6.63306)
			rgb255=(28.30143,11.21286,5.706645)
			rgb255=(29.387475,11.11698,4.88325)
			rgb255=(30.484485,11.109585000000001,4.156245)
			rgb255=(31.57206,11.184555,3.5182350000000002)
			rgb255=(32.666775,11.337045,2.9549399999999997)
			rgb255=(33.740325,11.533394999999999,2.430405)
			rgb255=(34.795004999999996,11.77182,2.013225)
			rgb255=(35.861925,12.08037,1.65801)
			rgb255=(36.945420000000006,12.40167,1.358385)
			rgb255=(38.05365,12.708179999999999,1.10823)
			rgb255=(39.189674999999994,13.004235,0.901935)
			rgb255=(40.368795,13.29315,0.73491)
			rgb255=(41.56857,13.570590000000001,0.602565)
			rgb255=(42.791805,13.831199999999999,0.5005649999999999)
			rgb255=(44.04768,14.068859999999999,0.425595)
			rgb255=(45.339254999999994,14.28459,0.374595)
			rgb255=(46.630065,14.4891,0.3417)
			rgb255=(47.95479,14.68137,0.32181000000000004)
			rgb255=(49.274415,14.921069999999999,0.31263)
			rgb255=(50.608065,15.18525,0.312885)
			rgb255=(51.96339,15.427755,0.32130000000000003)
			rgb255=(53.31846,15.67893,0.33711)
			rgb255=(54.68985,15.991050000000001,0.36006)
			rgb255=(56.073735,16.274865000000002,0.389895)
			rgb255=(57.462975,16.581885,0.42712500000000003)
			rgb255=(58.86828,16.905735,0.472515)
			rgb255=(60.28761,17.249475,0.52734)
			rgb255=(61.71408,17.61846,0.5928749999999999)
			rgb255=(63.158655,17.966790000000003,0.67116)
			rgb255=(64.61445,18.35643,0.76449)
			rgb255=(66.08988000000001,18.7782,0.8759250000000001)
			rgb255=(67.574235,19.185435,1.0085250000000001)
			rgb255=(69.08817,19.626075,1.165605)
			rgb255=(70.61664,20.099610000000002,1.351755)
			rgb255=(72.169335,20.608845,1.5710549999999999)
			rgb255=(73.74498,21.134145,1.828605)
			rgb255=(75.34383,21.694125,2.128995)
			rgb255=(76.97379000000001,22.3023,2.48013)
			rgb255=(78.63868500000001,22.92756,2.9210249999999998)
			rgb255=(80.33112000000001,23.59515,3.3976200000000003)
			rgb255=(82.06053,24.32496,3.930315)
			rgb255=(83.82818999999999,25.09098,4.5339)
			rgb255=(85.63027500000001,25.902900000000002,5.214494999999999)
			rgb255=(87.47418,26.769135,5.9772)
			rgb255=(89.35531499999999,27.7032,6.826605)
			rgb255=(91.27648500000001,28.70382,7.76628)
			rgb255=(93.235395,29.74779,8.815605)
			rgb255=(95.23485000000001,30.847604999999998,9.974324999999999)
			rgb255=(97.273065,32.02953,11.141715000000001)
			rgb255=(99.343665,33.266535,12.360105)
			rgb255=(101.450475,34.54587,13.54968)
			rgb255=(103.5861,35.902725000000004,14.75124)
			rgb255=(105.74595000000001,37.29987,15.992325000000001)
			rgb255=(107.923395,38.754645000000004,17.259674999999998)
			rgb255=(110.118435,40.26603,18.62622)
			rgb255=(112.31322,41.82714,20.048099999999998)
			rgb255=(114.516675,43.418595,21.58422)
			rgb255=(116.71452,45.04983,23.171595)
			rgb255=(118.91007,46.719314999999995,24.820425)
			rgb255=(121.09949999999999,48.42144,26.53632)
			rgb255=(123.27210000000001,50.152635,28.314944999999998)
			rgb255=(125.43705,51.89658,30.138450000000002)
			rgb255=(127.58211000000001,53.660415,32.002755)
			rgb255=(129.71595,55.443375,33.910664999999995)
			rgb255=(131.829135,57.23016,35.858865)
			rgb255=(133.92523500000001,59.043465,37.828995)
			rgb255=(136.003995,60.84912,39.846555)
			rgb255=(138.0672,62.667525,41.879414999999995)
			rgb255=(140.117655,64.495365,43.927575)
			rgb255=(142.15281000000002,66.323205,46.002765000000004)
			rgb255=(144.17139,68.150025,48.1032)
			rgb255=(146.18436,69.987555,50.215619999999994)
			rgb255=(148.18356,71.826615,52.335435000000004)
			rgb255=(150.1746,73.66797,54.474375)
			rgb255=(152.16258000000002,75.50907,56.62377)
			rgb255=(154.14342,77.352975,58.784895000000006)
			rgb255=(156.11814,79.207335,60.958259999999996)
			rgb255=(158.09387999999998,81.056085,63.14259)
			rgb255=(160.06554,82.90866,65.328195)
			rgb255=(162.03669,84.772965,67.527825)
			rgb255=(164.009115,86.63497500000001,69.73995000000001)
			rgb255=(165.983835,88.50131999999999,71.955645)
			rgb255=(167.95855500000002,90.37072500000001,74.176185)
			rgb255=(169.936845,92.246505,76.40820000000001)
			rgb255=(171.91947,94.12585499999999,78.645825)
			rgb255=(173.903625,96.01209,80.890845)
			rgb255=(175.894665,97.89959999999999,83.14096500000001)
			rgb255=(177.88698,99.79526999999999,85.40689499999999)
			rgb255=(179.88567,101.69247,87.66798)
			rgb255=(181.88844,103.599105,89.94359999999999)
			rgb255=(183.89529,105.50803499999999,92.22381)
			rgb255=(185.90877,107.42104499999999,94.50759000000001)
			rgb255=(187.92684,109.34298,96.80208)
			rgb255=(189.9495,111.26899499999999,99.102945)
			rgb255=(191.97700500000002,113.200365,101.4084)
			rgb255=(194.011905,115.13556000000001,103.72048500000001)
			rgb255=(196.049355,117.07661999999999,106.03869)
			rgb255=(198.09521999999998,119.02278,108.36531000000001)
			rgb255=(200.144145,120.973785,110.69345999999999)
			rgb255=(202.198425,122.9304,113.033595)
			rgb255=(204.257295,124.889565,115.378575)
			rgb255=(206.32305,126.85893,117.72636000000001)
			rgb255=(208.39161000000001,128.82778499999998,120.08205)
			rgb255=(210.46629000000001,130.80531000000002,122.44335)
			rgb255=(212.546835,132.78665999999998,124.808475)
			rgb255=(214.63146,134.774385,127.184565)
			rgb255=(216.720675,136.765425,129.56448)
			rgb255=(218.81346,138.761055,131.94924)
			rgb255=(220.91262,140.763315,134.340375)
			rgb255=(223.01484,142.769145,136.73559)
			rgb255=(225.121395,144.779055,139.13896499999998)
			rgb255=(227.23305,146.79585,141.54591)
			rgb255=(229.348785,148.814175,143.96025)
			rgb255=(231.46758,150.84015,146.37969)
			rgb255=(233.587905,152.87046,148.80576)
			rgb255=(235.71384,154.905615,151.239225)
			rgb255=(237.84207,156.9423,153.67549499999998)
			rgb255=(239.97438,158.988165,156.118395)
			rgb255=(242.10898500000002,161.03556,158.56716)
			rgb255=(244.24716,163.088565,161.024085)
			rgb255=(246.38584500000002,165.14514,163.48381500000002)
			rgb255=(248.527845,167.20809,165.95196)
			rgb255=(250.67264999999998,169.274865,168.42418500000002)
			rgb255=(252.81924,171.34444499999998,170.90508)
			rgb255=(254.967615,173.41836,173.38725000000002)
		}
}

\pgfdeclarelayer{background}
\pgfdeclarelayer{foreground}
\pgfsetlayers{background,main,foreground}

\usepgfplotslibrary{external,fillbetween}
\RequirePackage{shellesc}
\fi

\usepackage{xcolor}
\definecolor{UKLred} {RGB}{207, 25,  59}
\definecolor{UKLblue}{RGB}{ 47, 63, 157}
\definecolor{NYUpurple} {RGB}{88,15,139}
\definecolor{Pastrami} {RGB}{229,85,79}
\definecolor{TheLake}{RGB}{72,159,223}
\definecolor{EastRiver}{RGB}{0,115,152}
\definecolor{SpicyMustard}{RGB}{203,160,82}
\definecolor{CentralPark}{RGB}{0,108,91}
\definecolor{ProspectPark}{RGB}{64,192,172}
\definecolor{turquois}{rgb}{0,0.75,0.75}%

\usepackage{colortbl}

\usepackage{amssymb}
\usepackage{upgreek}

\usepackage{subcaption}

\usepackage{xr}

\usepackage{nameref}

\makeatletter
\newcommand*{\addFileDependency}[1]{
\typeout{(#1)}
\@addtofilelist{#1}
\IfFileExists{#1}{}{\typeout{No file #1.}}
}\makeatother

\newcommand*{\myexternaldocument}[1]{%
\externaldocument{#1}%
\addFileDependency{#1.tex}%
\addFileDependency{#1.aux}%
}

\myexternaldocument{./SupportingInformation}

\begin{document}

\title{\Papertitle}

\author[1,2]{Jakob Assl\"ander*}{\orcid{0000-0003-2288-038X}}
\author[3]{Cem Gultekin}{\orcid{0000-0002-2562-371X}}
\author[1,2,4]{Andrew Mao}{\orcid{0000-0002-1398-0699}}
\author[1,2]{Xiaoxia Zhang}{\orcid{0000-0003-3620-6994}}
\author[5]{Quentin Duchemin}{\orcid{0000-0003-3636-3770}}
\author[6]{Kangning Liu}{\orcid{0000-0002-0187-4602}}
\author[7]{Robert W Charlson}{}
\author[1]{Timothy M Shepherd}{\orcid{0000-0003-0232-1636}}
\author[3,6]{Carlos Fernandez-Granda}{\orcid{0000-0001-7039-8606}}
\author[1,2]{Sebastian Flassbeck}{\orcid{0000-0003-0865-9021}}

\authormark{Jakob Assl\"ander \textsc{et al}}

\address[1]{\orgdiv{Center for Biomedical Imaging, Dept. of Radiology}, \orgname{NYU School of Medicine}, \orgaddress{\state{NY}, \country{USA}}}
\address[2]{\orgdiv{Center for Advanced Imaging Innovation and Research (CAI\textsuperscript{2}R), Dept. of Radiology}, \orgname{NYU School of Medicine}, \orgaddress{\state{NY}, \country{USA}}}
\address[3]{\orgdiv{Courant Institute of Mathematical Sciences}, \orgname{New York University}, \orgaddress{\state{NY}, \country{USA}}}
\address[4]{\orgdiv{Vilcek Institute of Grad. Biomed. Sciences}, \orgname{NYU School of Medicine}, \orgaddress{\state{NY}, \country{USA}}}
\address[5]{\orgdiv{Laboratoire d’analyse et de mathématiques appliquées}, \orgname{Université Gustave Eiffel}, \orgaddress{\country{Fr.}}}
\address[6]{\orgdiv{Center for Data Science}, \orgname{New York University}, \orgaddress{\state{NY}, \country{USA}}}
\address[7]{\orgdiv{Department of Neurology}, \orgname{NYU School of Medicine}, \orgaddress{\state{NY}, \country{USA}}}

\corres{*Jakob Assl\"ander, Center for Biomedical Imaging, NYU School of Medicine, 227 E 30 Street, New York, NY 10016, USA.\\ \email{jakob.asslaender@nyumc.org}}

\finfo{This work was supported by the \fundingAgency{NIH/NIBIB} grant \fundingNumber{R21 EB027241} and \fundingNumber{P41 EB017183}. AM acknowledges support from the \fundingAgency{NIH/NIA} \fundingNumber{T32 GM136573} and \fundingNumber{F30 AG077794}.}

\abstract[Abstract]{
	\textbf{Purpose:}
	To explore efficient encoding schemes for quantitative magnetization transfer (qMT) imaging with few constraints on model parameters.

	\textbf{Theory and Methods:}
	We combine two recently proposed models in a Bloch-McConnell equation: the dynamics of the \textit{free spin pool} are confined to the hybrid state, and the dynamics of the \textit{semi-solid spin pool} are described by the generalized Bloch model. We numerically optimize the flip angles and durations of a train of radio frequency pulses to enhance the encoding of three qMT parameters while accounting for all 8 parameters of the 2-pool model. We sparsely sample each time frame along this spin dynamics with a 3D radial koosh-ball trajectory, reconstruct the data with subspace modeling, and fit the qMT model with a neural network for computational efficiency.

	\textbf{Results:}
	We extracted qMT parameter maps of the whole brain with an effective resolution of 1.24mm from a 12.6-minute scan. In lesions of multiple sclerosis subjects, we observe a decreased size of the semi-solid spin pool and longer relaxation times, consistent with previous reports.

	\textbf{Conclusion:}
	The encoding power of the hybrid state, combined with regularized image reconstruction, and the accuracy of the generalized Bloch model provide an excellent basis for efficient quantitative magnetization transfer imaging with few constraints on model parameters.
}

\keywords{quantitative MRI, qMRI, parameter mapping, relaxometry, MRF}


\maketitle
\footnotetext{Word Count: 7541}

\begin{figure*}[b]
	\centering
	\ifOL
		\includegraphics[]{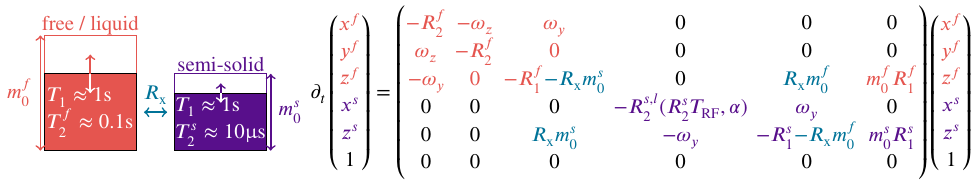}
	\else
		\begin{tikzpicture}[scale = 0.65, every text node part/.style={align=left}]
			\node [color=Pastrami] at (1.2,3.25) {free / liquid};
			\draw [color=Pastrami] (0,2) rectangle (2.4,3);
			\draw [fill=Pastrami] (0,0) rectangle (2.4,2) node[pos=.5, text = white] {$T_1 \approx 1$s \\ $T_2^f \approx 0.1$s};
			\draw [->, color = Pastrami, thick] (1.2, 2) -- (1.2, 2.5);
			\draw [->, color = white    , thick] (1.2, 2) -- (1.2, 1.5);

			\draw [<->, color = Pastrami, thick] (-0.1, 0) -- (-0.1, 3) node[left, midway] {$m_0^f$};

			\draw [<->, color = EastRiver, thick] (2.6, 1) -- (3.2, 1) node[above, midway, text = EastRiver] {$R_{\text{x}}$};

			\node [color=NYUpurple] at (4.6,2.25) {semi-solid};
			\draw [color=NYUpurple] (3.4,1.5) rectangle (5.8,2);
			\draw [fill=NYUpurple]  (3.4,0)   rectangle (5.8,1.5) node[pos=.5, text = white] {$T_1 \approx 1$s \\ $T_2^s \approx 10\upmu$s};
			\draw [->, color = NYUpurple, thick] (4.6, 1.5) -- (4.6, 1.75);
			\draw [->, color = white    , thick] (4.6, 1.5) -- (4.6, 1.25);

			\draw [<->, color = NYUpurple, thick] (5.9, 0) -- (5.9, 2) node[right, midway] {$m_0^s$};

			\node [anchor = west] at (6.75, 1.5) {
				$
					\partial_t \begin{pmatrix} \color{Pastrami} x^f \\ \color{Pastrami} y^f \\ \color{Pastrami} z^f \\ \color{NYUpurple} x^s \\ \color{NYUpurple} z^s \\ 1 \end{pmatrix} = \begin{pmatrix}
						\color{Pastrami} -R_2^f    & \color{Pastrami} -\omega_z & \color{Pastrami} \omega_y                                         & 0                                                      & 0                                                                  & 0                             \\
						\color{Pastrami} \omega_z  & \color{Pastrami} -R_2^f    & \color{Pastrami} 0                                                & 0                                                      & 0                                                                  & 0                             \\
						\color{Pastrami} -\omega_y & \color{Pastrami} 0         & {\color{Pastrami} -R_1^f} {\color{EastRiver}- R_{\text{x}} m_0^s} & 0                                                      & \color{EastRiver} R_{\text{x}} m_0^f                               & \color{Pastrami} m_0^f R_1^f  \\
						0                          & 0                          & 0                                                                 & \color{NYUpurple} -R_2^{s,l}(R_2^s T_\text{RF},\alpha) & \color{NYUpurple} \omega_y                                         & 0                             \\
						0                          & 0                          & \color{EastRiver} R_{\text{x}} m_0^s                              & \color{NYUpurple} -\omega_y                            & {\color{NYUpurple}-R_1^s} {\color{EastRiver} - R_{\text{x}} m_0^f} & \color{NYUpurple} m_0^s R_1^s \\
						0                          & 0                          & 0                                                                 & 0                                                      & 0                                                                  & 0
					\end{pmatrix} \begin{pmatrix} \color{Pastrami} x^f \\ \color{Pastrami} y^f \\ \color{Pastrami} z^f \\ \color{NYUpurple} x^s \\ \color{NYUpurple} z^s \\ 1 \end{pmatrix}
				$};
		\end{tikzpicture}
	\fi
	\caption{Two-pool magnetization transfer model.\cite{Henkelman1993} This model jointly describes all magnetization arising from protons bound in liquids by the \textit{free} spin pool of size $m_0^f$, and all magnetization arising from protons bound in macromolecules by the \textit{semi-solid} spin pool of size $m_0^s$. The latter has a transversal relaxation time that is several orders of magnitude shorter compared to the free pool. We normalize the thermal equilibrium magnetization to $m_0^f + m_0^s = 1$ and describe the magnetization transfer between the pools by the rate $R_{\text{x}}$. The equation replicates Eq.~\eqref{eq:Bloch_McConnell} with color-coding.}
	\label{fig:Model}
\end{figure*}

\section{Introduction}
Magnetization transfer\cite{Edzes1977, Wolff1989} (MT) between protons bound in water and protons bound in large molecules, such as proteins and lipids, is a major source of contrast in many clinical MRI sequences.\cite{Melki1992}
This effect can be emphasized with saturation pulses to add diagnostic value for many pathologies throughout the body.\cite{Wolff1994}
Quantification of MT parameters may improve specificity compared to standard clinical protocols\cite{Henkelman.2001, Stanisz.2004}
by disentangling aqueous and non-aqueous spins with a two-pool model, as originally proposed by Henkelman et al.\cite{Henkelman1993}
In this model, one describes all aqueous protons by the so-called free pool with a characteristic transversal relaxation time $T_2^f$ in the order of $50$ms or longer. Such comparably slow relaxation results from motional narrowing, hence the name \textit{free} pool and the superscript $f$.
The so-called semi-solid spin pool (superscript $s$), in contrast, captures non-aqueous spins with restricted motion that, in general, have a non-exponential decay\cite{Henkelman1993, Morrison1995a} with a much shorter characteristic $T_2^s \approx 10\upmu\text{s}$.

Despite the increased specificity, quantitative MT (qMT) is seldom used in routine clinical imaging due to its long scan time and limited resolution. For example, a recent approach performs a full-brain qMT scan with a voxel volume of 14.5mm\textsuperscript{3} in 18 minutes with good image quality, or in 7 minutes with reduced image quality.\cite{Cronin2020} While this is a substantial improvement over previous approaches that were often limited to single slices,\cite{Sled2001, Gelderen2016, Gochberg.2003, Gochberg2007, Gloor2008, Dortch.2011} it still does not meet the clinical standard of 1mm\textsuperscript{3} isotropic.
In his recent work, Yarnykh constrained the exchange rate $R_x$, $T_2^f/T_1$, and $T_2^s$ to fixed values, which enabled further reductions of the voxel volume to 2mm\textsuperscript{3} (nominal resolution of 1.25mm isotropic) and whole brain coverage in 20 minutes in adults\cite{Yarnykh.2012, Yarnykh.2016} and to 11 minutes in infants.\cite{Corrigan.2022}
While Yarnykh's approach allows for improvements in scan efficiency, it bears the risk of biases when any of the fixed parameters change in pathology.
Further, it reduces specificity as different pathologies can result in similar changes in the few free parameters.

In this paper, we aim to explore a new avenue for efficient qMT imaging without constraints. For this purpose, we tailor the MT model, pulse sequence, and image reconstruction to one another.
Similar to balanced steady-state free precession\cite{Carr1958} (bSSFP) pulse sequences, we use balanced gradient moments, but we drive the free spin pool into a \textit{hybrid state}.\cite{Asslander2019a} The hybrid state is characterized by efficient encoding and disentanglement of biophysical parameters,\cite{Asslander2020} combined with tractable spin dynamics that allow for a description of the inhomogeneously broadened lineshape in each voxel by a single spin isochromat.
We encode the semi-solid spin pool's characteristics with a combination of the selective inversion-recovery technique\cite{Gochberg.2003, Gochberg2007} and the saturation of its magnetization with the on-resonant radio-frequency (RF) pulses that excite the free pool.\cite{Gloor2008}
We describe the resulting dynamics of the semi-solid pool's magnetization with the \textit{generalized Bloch model}, which we recently proposed for a more accurate description of the non-exponential signal decay in  semi-solids.\cite{Asslander.2022u6f}
We numerically optimize the flip angles and durations for a train of RF pulses to improve the estimated parameters' signal-to-noise ratio (SNR).\cite{Jones1996, Jones1997, Zhao2019, Asslander2019b}
Similar to MR-Fingerprinting,\cite{Ma2013} we highly undersample each time frame.
In line with more recent MR-Fingerprinting approaches,\cite{Zhao2015b,Tamir2017,Asslander2018,Zhao2018} we use an iterative image reconstruction for reduced image artifacts and combine it with a neural network to fit the biophysical model with more computational efficiency.\cite{Cohen2018, Zhang.2022} Combining these measures enables us to quantify MT parameters without constraints on $R_x$, $T_2^f/T_1$, or $T_2^s$ and with an effective resolution of 1.24mm and whole brain coverage in 12.6 minutes.
We tested the approach with phantom and in vivo scans of participants with multiple sclerosis (MS) and healthy controls.

\section{Theory}
\subsection{Henkelman's two-pool model} \label{sec:HenkelmanModel}
The proposed MT model builds on Henkelman's two-pool spin model.\cite{Henkelman1993} As sketched in Fig.~\ref{fig:Model}, one pool describes all \textit{free} protons, i.e., protons bound in liquids where motional narrowing leads to a relaxation time of the transversal magnetization of 50ms and longer. The second pool, the so-called \textit{semi-solid} pool, describes the magnetization of protons bound in large molecules such as lipids. The motion of such molecules is restricted, resulting in a much faster and non-exponential relaxation with a characteristic time constant $T_2^s \approx 10\upmu$s.
Throughout this report, we assume a super-Lorentzian lineshape.\cite{Morrison1995a}
Standard MRI approaches cannot directly detect any signal from the semi-solid pool as it decays below the noise level before we can measure it. However, the longitudinal magnetization of the semi-solid pool exchanges with the free pool, providing indirect detection.

Similar to Henkelman,\cite{Henkelman1993} we use a Bloch-McConnell equation,\cite{McConnell1958} which we denote here in matrix form:

\begin{widetext}
	\begin{equation}
		\partial_t \begin{pmatrix} x^f \\ y^f \\ z^f \\ x^s \\ z^s \\ 1 \end{pmatrix} = \begin{pmatrix}
			-R_2^f    & -\omega_z & \omega_y                    & 0                                    & 0                           & 0           \\
			\omega_z  & -R_2^f    & 0                           & 0                                    & 0                           & 0           \\
			-\omega_y & 0         & -R_1^f - R_{\text{x}} m_0^s & 0                                    & R_{\text{x}} m_0^f          & m_0^f R_1^f \\
			0         & 0         & 0                           & -R_2^{s,l}(R_2^s,\alpha,T_\text{RF}) & \omega_y                    & 0           \\
			0         & 0         & R_{\text{x}} m_0^s          & -\omega_y                            & -R_1^s - R_{\text{x}} m_0^f & m_0^s R_1^s \\
			0         & 0         & 0                           & 0                                    & 0                           & 0
		\end{pmatrix} \begin{pmatrix} x^f \\ y^f \\ z^f \\ x^s \\ z^s \\ 1 \end{pmatrix}.
		\label{eq:Bloch_McConnell}
	\end{equation}
\end{widetext}
Fig.~\ref{fig:Model} replicates the equation with color-coding: the red color highlights the elements corresponding to the Bloch equation of the free pool, whose magnetization is described by the Cartesian coordinates $x^f$, $y^f$, $z^f$. The Larmor frequency is described by $\omega_z$ and the Rabi frequency of the RF pulses by $\omega_y$. For readability, we here use relaxation rates ($R_{1,2}^{f,s} = 1 / T_{1,2}^{f,s}$).
Purple highlights the semi-solid spin pool's magnetization components $x^s,z^s$. As detailed below, we describe its spin dynamics with the \textit{generalized Bloch model},\cite{Asslander.2022u6f} which is here captured by the \textit{linearized} relaxation rate $R_2^{s,l}(R_2^s,\alpha,T_\text{RF})$ that depends, in addition to the biophysical parameter $R_2^s$, on the flip angle $\alpha$ and the duration $T_\text{RF}$ of respective RF pulse (cf. Section \ref{sec:TheorygBloch}). We neglect the $y^s$ component, assuming (without loss of generality) $\omega_x=0$ and given that $R_2^{s,l} \gg \omega_z$.
Exchange processes between the pools, highlighted in turquoise, are captured by the exchange rate $R_{\text{x}}$.
We added the sixth dimension to allow for a compact notation of the longitudinal relaxation to a non-zero thermal equilibrium.

This two-pool model entails overall 9 parameters: The four relaxation rates, $R_{1,2}^{f,s}$, the size of the semi-solid spin pool $m_0^s$, the magnetic fields $B_{0,1}^{(+)}$, which are here represented by the frequencies $\omega_{z,y}$, and the real and imaginary parts of the complex-valued scaling factor $M_0$. The size of the free spin pool is not a free parameter due to the constraint $m_0^f + m_0^s = 1$. As described in Section~\ref{sec:R1a}, we further constrain the longitudinal relaxation rate of the semi-solid spin pool to $R_1^s = R_1^f$, which reduces the number of free parameters to 8 for this two-pool model.

\subsection{Hybrid state of the free pool}
As detailed in Ref. \citen{Asslander2019a}, the hybrid state combines the transient state's sensitivity to biophysical parameters with the steady-state's tractable off-resonance characteristics, in particular, the refocusing of intra-voxel dephasing.\cite{Carr1958, Scheffler2003}
It is possible to obtain these two properties simultaneously because the sensitivities to biophysical parameters and magnetic field variations are dominated by the spin dynamics along orthogonal dimensions: in spherical coordinates, relaxation and exchange primarily act on the magnetization's magnitude, while magnetic field variations act on its direction.\cite{Hargreaves2001,Ganter2004,Asslander2019a}

One can show that the magnetization's direction follows the steady state adiabatically if the condition
\begin{equation}
	\max \{ |\Delta \alpha|, |\Delta \phi| \} \ll \sin^2 \frac{\alpha}{2} + \sin^2 \frac{\phi}{2}.
	\label{eq:hybrid_state_condition}
\end{equation}
(Eq.~(4) in Ref. \citen{Asslander2019a}) is fulfilled. Here, $\alpha$ denotes the flip angle, $\phi = \omega_z T_{\text{R}}$ the phase accumulation in one repetition time $T_{\text{R}}$, and $\Delta$ their change in consecutive repetitions.
Changing a flip angle slower than this limit ensures a steady state of the magnetization's direction. Consequently, hybrid-state magnetization is only slightly sensitive to magnetic field inhomogeneities, similar to steady-state magnetization.
In other words, the hybrid state preserves the spin echo nature of bSSFP sequences,\cite{Scheffler2003} which allows us to neglect intra-voxel dephasing and to describe the inhomogeneously broadened lineshape of a voxel by a single isochromat.
Furthermore, it is possible to simultaneously induce a transient state of the magnetization's magnitude, which enables efficient encoding of relaxation and MT parameters.\cite{Asslander2019a, Asslander2019b, Asslander2020}
The Supporting Information describes the spin dynamics in the hybrid state and the exchange processes and how the flip angle controls them.

\subsection{Description of the semi-solid pool with the generalized Bloch model} \label{sec:TheorygBloch}
Henkelman's original MT model assumes a steady state of magnetization in the presence of a continuous RF wave. In clinical MRI, however, one commonly uses RF pulses, i.e., waves of finite duration. Established models for MT during RF pulses\cite{Graham1997, Sled2000} build on Henkelman's theory and model a saturation of the z-magnetization that depends on the amplitude, duration, and shape of the pulse. These models describe experimental observations well for off-resonant saturation pulses of several milliseconds. However, these models fail to accurately describe the spin dynamics for short RF pulses, which becomes apparent when approaching the extreme case of a hard, i.e., instantaneous pulse. Manning et al.\cite{Manning2021} demonstrated experimentally that the semi-solid pool is inverted by a short inversion pulse, which contrasts existing MT models that predict a saturation of the semi-solid pool.

The recently proposed generalized Bloch model overcomes this limitation of existing MT models\cite{Asslander.2022u6f}
and we use it here to describe the dynamics of the semi-solid spin pool. Like a quantum-mechanical description of spin dynamics and the original Bloch equations, but unlike established MT models, the generalized Bloch model is based on the algebra of angular momentum in the sense that it explicitly models the rotations induced by RF pulses.
It is a generalization of the Bloch model to non-exponential decays or, equivalently, to non-Lorentzian spectral lineshapes.

The key to the generalized Bloch model is to denote the original Bloch equation for $x$ in integral form:
\begin{equation}
	x^s(t) = \int_0^t G(t,\tau) \omega_y(\tau) z^s(\tau) d\tau
	\label{eq:gBloch_x}
\end{equation}
The original Bloch model is solved by the Green's function
\begin{equation}
	G(t,\tau) = \exp(- R_2^s (t-\tau)) \;\; \forall \;\; t\geq\tau .
\end{equation}
This reformulation exposes the exponential free induction decay (FID) that is inherent to the Bloch model and equal to the Fourier transform of the Lorentzian lineshape. For non-Lorentzian lineshapes, such as the super-Lorentzian lineshape,\cite{Wennerstrom1973} which describes the semi-solid pool in brain white matter well,\cite{Morrison1995a} we make the ansatz that we can replace the Green's function with the Fourier transform of respective lineshape. For a super-Lorentzian lineshape, the Green's function is
\begin{equation}
	G(t,\tau) = \int_0^{1} \exp \left( - R_2^s (t-\tau)^2 \cdot \frac{(3\zeta^2-1)^2}{8} \right) d\zeta.
\end{equation}
Inserting this Green's function in Eq.~\eqref{eq:gBloch_x} facilitates the description of the spin dynamics in the semi-solid pool during RF pulses.

As Eq.~\eqref{eq:gBloch_x} is numerically challenging to solve, we pre-compute the magnetization at the end of RF pulses with different flip angles and durations, assuming different $R_2^s$ rates.
By interpolating between the pre-computed grid points, we approximate the generalized Bloch model with exponential decays, whose decay rate $R_2^{s,l}(R_2^s,\alpha,T_\text{RF})$ is a function of the flip angle $\alpha$ and pulse duration $T_\text{RF}$, which allows for incorporating this linear approximation of the generalized Bloch model into Eq.~\eqref{eq:Bloch_McConnell}. Refer to Ref. \citen{Asslander.2022u6f} for more details.

For comparison, we also fitted Graham's spectral model\cite{Graham1997} to the in vivo data. This model assumes an exponential saturation of the semi-solid pool's longitudinal magnetization.
The rate of this saturation is defined by Eq.~(4) in Ref. \citen{Graham1997} and takes the integral over the entire lineshape, multiplied by the pulse's power spectral density.\cite{Stremler1982} Since the lineshape has, per definition, a finite or, more precisely, a unit area under the curve, this integral is well-defined and finite despite the super-Lorentzian's divergence on resonance.
We note that we do not use the more widespread single frequency approximation proposed by Graham (Eq.~(8) in Ref. \citen{Graham1997}) in this paper, as it is not well-defined for a super-Lorentzian lineshape in combination with on-resonant RF pulses.\cite{Gloor2008}
In order to highlight differences between the generalized Bloch and Graham's spectral model, we also show the Bloch-McConnell equation for the latter in Appendix \ref{sec:AppendixGraham}.

\subsection{Apparent $R_1$-relaxation rate} \label{sec:R1a}
There is currently an active discussion in the field about the size of $R_1^f$ and $R_1^s$. While most MT literature assumes that both are in the range of 1/s and either fix $R_1^s := 1/$s\cite{Henkelman1993, Morrison1995a}
or $R_1^s := R_1^f$,\cite{Dortch.2011,Yarnykh.2012} more recent studies have suggested that $R_1^s$ is substantially larger.\cite{Helms2009,Gelderen2016,Manning2021,Samsonov2021}
While we are actively investigating this question ourselves,\cite{Flassbeck.2022,Asslander.2023} we follow the more established literature in the present publication and assume an apparent $R_1 := R_1^s = R_1^f$.

\section{Methods}
\subsection{Pulse sequence design and optimization} \label{sec:Methods_Optimization}
As outlined above, we build our approach on the balanced SSFP sequence.
We jointly use two established mechanisms to encode magnetization transfer: first, we invert the free pool with a rectangular RF pulse,
which saturates the semi-solid pool only slightly if the pulse duration is much larger than $T_2^s$.\cite{Gochberg.2003, Asslander.2022u6f} As described by Gochberg et al.,\cite{Gochberg.2003} this induces a bi-exponential inversion recovery curve composed of $T_1$-relaxation and magnetization transfer between the pools, and the bi-exponential nature is most pronounced when the magnetization of the two pools is least equal, i.e., when the free pool is inverted and the semi-solid pool remains unchanged.\cite{Reynolds.2023}
After that, we apply on-resonant rectangular RF pulses with a $\pi$-phase increment in consecutive RF pulses, as is typical for bSSFP sequences.
After 1142 RF pulses with varying flip angles and pulse durations spaced $T_\text{R} = 3.5\text{ms}$ apart, i.e., after 4 seconds, we invert the remaining magnetization with a rectangular $\pi$-pulse, flanked by crusher gradients, and repeat the same pulse train.
As described in the Theory section, relaxation, and exchange processes are predominantly controlled by the flip angle, while the saturation of the semi-solid spin pool is additionally controlled by the pulse duration.
This encoding mechanism resembles Gloor's bSSFP-based qMT approach.\cite{Gloor2008}

We optimized the flip angle and the duration of each of the 1142 RF pulses for a minimal Cram\'er-Rao bound\cite{Rao1945, Cramer1946} (CRB) of selected model parameters.
The CRB of the $i^\text{th}$ parameter $\theta_i$ is given by
\begin{equation}
	\text{CRB}(\theta_i) = \sigma^2 (\mathbf{F}^{-1})_{i,i}
	\label{eq:CRB}
\end{equation}
and provides a lower bound for the noise variance in the estimated parameter $\hat{\theta}_i$ for a given input noise variance $\sigma^2$, assuming the model-fitting routine is unbiased. $\mathbf{F}$ denotes the Fisher information matrix with the elements $\mathbf{F}_{i,j}= (\partial \mathbf{s} / \partial \theta_i)' \cdot (\partial \mathbf{s} / \partial \theta_j)$ that are given by the inner products of the signal's derivatives with respect to each model parameter.

The here-used simulation and optimization framework, whose source code is publicly available (cf. Appendix \ref{sec:data_availability}), differs from our previous implementations.\cite{Asslander2019a, Asslander2019b}
Formerly, we described the hybrid-state magnetization in spherical coordinates (cf. Eq.~\eqref{eq:Hybrid_State_Model_gBloch}).
CRB-based optimizations in spherical coordinates without constraints robustly converge to smooth flip-angle trains with few violations of the hybrid-state conditions (cf. Supporting Fig.~\ref{fig:Spin_dynamics_v3p2}, which we used for the experiments).\cite{Asslander2019a}
For the optimizations depicted in Fig.~\ref{fig:MT_Spin_Dynamics} and Supporting Figs.~\ref{fig:Spin_dynamics_Inv_constFA_constTRF}--\ref{fig:Spin_dynamics_noInv_varyFA_varyTRF}, the analysis in Tab.~\ref{tab:CRB_seq}, and all fits in this paper, were, instead, performed with the standard Cartesian Bloch-McConnell formulation (Eq.~\eqref{eq:Bloch_McConnell}) as it allows for a more straightforward description of magnetization transfer and the spin dynamics during finite RF pulses.
Its disadvantage, however, is that optimizations often converge to local minima with comparably poor (high) CRBs and strong flip angle fluctuations that destroy the hybrid state.
This is overcome by an additive penalty of second-order flip angle changes $(2\alpha_{i} - \alpha_{i+1} - \alpha_{i-1})^2$ that is scaled by a heuristically chosen regularization parameter.
This penalty improved the final CRB and provided a more robust regularization than the more commonly used total variation penalty.
Additionally, we added a total variation penalty of the pulse durations that had a negligible impact on the final CRB but improved the robustness to imperfections in the RF hardware, such as amplifier non-linearities.

Several groups have previously used the CRB for numerically optimizing quantitative MRI sequences.\cite{Jones1996,Jones1997, Zhao2019, Asslander2019b}
Our implementation follows the one described in Ref. \citen{Asslander2019b} closely: we normalize the CRB of each parameter with the squared parameter value to resemble the inverse squared SNR of the estimated parameter and sum over the normalized CRB of the three parameters of interest ($m_0^s$, $R_1$, and $R_2^f$).

We initialized all presented optimizations with a sinusoidal flip angle pattern ($\alpha = 0.7 |\sin \eta|$ with $\eta \in [0,2\pi]$) and a constant $T_\text{RF} = 500\upmu$s. While our optimization is non-convex, we observed a robust convergence from different initializations to local minima with similar performance and features in the RF pattern. We refer the interested reader to Ref. \citen{Asslander2019b} for a detailed analysis of these convergence properties.

While this framework is general, we optimize the RF pattern only for good performance in the three parameters $m_0^s$, $R_1$, and $R_2^f$ that we selected based on prior reports of their utility for distinguishing demyelination from inflammation.\cite{Stanisz.2004}
We calculate the CRB for these three parameters at $m_0^s = 0.1$, $R_1 = 0.625/\text{s}$, $R_2^f = 15/\text{s}$, $R_\text{x} = 30/\text{s}$, $T_2^s = 10\upmu\text{s}$, $\omega_z = 0$, and $B_1^+ = 1$.
Each parameter's CRB accounts for all model parameters, i.e., we optimize the acquisition for precision in the estimates of $m_0^s$, $R_1$, and $R_2^f$ during a fit of all model parameters.

Rectangular RF pulses were chosen here due to their efficient rotation of the free pool with minimal saturation of the semi-solid spin pool for a given flip angle and $T_\text{RF}$. The latter property increases the dynamic range of different saturation levels and improves the MT parameters' encoding. We note, however, that these methods work with any pulse shape, such as slice- or slab-selective sinc pulses.
This flexibility also includes the linear approximation of the generalized Bloch model that can easily be translated to approximate the dynamics of the semi-solid spin pool during shaped RF pulses with varying flip angles and pulse durations.

\subsection{Cram\'er-Rao bound analysis} \label{sec:CRBanalysis}
In order to identify the individual contributions of pulse sequence features, we performed the above-described optimizations with four different modifications: with and without inversion pulse, varying or fixed flip angles, and pulse duration, as tabulated in Tab.~\ref{tab:CRB_seq}. We note that the inversion-recovery bSSFP\cite{Schmitt2004} sequence with fixed flip angle and $T_\text{RF}$ resulted in very large CRB values, and we had to constrain several parameters for a reasonable CRB estimation.

In addition, we also calculated the CRB of the selective inversion recovery (SIR)\cite{Gochberg.2003, Gochberg2007} technique, more specifically, of the turbo-spin-echo (TSE) version of the latest implementation as described in Refs. \citen{Dortch.2011,Cronin2020}. This technique assumes that all magnetization is destroyed at the end of a TSE pulse train. In 5 different scans, the magnetization relaxes towards thermal equilibrium for $T_\text{D} \in \{3270, 4489, 1652, 2922, 10\}$ms, after which it is inverted with a 1ms rectangular $\pi$-pulse, followed by an undisturbed recovery for $T_\text{I} \in \{10, 50, 56, 277, 843\}$ms, respectively, and a TSE readout of 22 k-space lines, spaced 5.9ms apart. We calculated the CRB assuming that each k-space line is acquired with the same bandwidth as our radial k-space trajectory.
We simulated the dynamics of the coupled spin pool with our recently-proposed generalized Bloch model\cite{Asslander.2022u6f} and assuming a super-Lorentzian lineshape,\cite{Morrison1995a} which suggested a saturation of the semi-solid spin pool to 0.51 of its original value. Importantly, this value is substantially lower than the 0.83 estimated in Refs. \citen{Gochberg.2003,Gochberg2007,Dortch.2011,Cronin2020} based on Graham's single frequency approximation\cite{Graham1997} (cf. Section \ref{sec:TheorygBloch}) and an assumed Gaussian lineshape.
We note that this saturation depends on $T_2^s$ and $B_1^+$, and we neglected these respective gradients for the calculation of the CRB, which is both in line with Refs. \citen{Gochberg.2003,Gochberg2007,Dortch.2011,Cronin2020} and required to avoid divergent CRB values at $B_1^+ = 1$.

We also calculated the CRB of the (to our knowledge) currently fastest version\cite{Yarnykh.2012, Yarnykh.2016} of an off-resonance saturation qMT approach,\cite{Henkelman1993} which samples only a single saturation frequency (4kHz) and a single saturation flip angle (560$^\circ$).\cite{Yarnykh.2016}
We simulate the spin dynamics with the generalized Bloch model during the 12ms sinc-shaped saturation pulse, followed by a 1ms, 10$^\circ$ rectangular excitation pulse, and free precession through the remainder of the $T_\text{R}=28$ms. Before each RF pulse, we set all transversal magnetization to zero in the assumption of perfect RF spoiling,\cite{Zur1991} and repeated this simulation until the magnetization reached a steady state.
Similarly, we simulated the dynamics of the coupled spin system during variable-flip angle $T_1$ mapping measurements\cite{Deoni.2003} with $\alpha \in \{4^\circ, 25^\circ\}$, $T_\text{R}=21$ms, assuming a 1ms rectangular excitation pulse and perfect RF-spoiling.
We calculated the CRB of $m_0^s$, $R_1$, and the scaling factor $M_0$, assuming heuristically fixed $R_\text{x}$ and $T_2^s$ values, a fixed ratio $R_1/R_2^f$, and assuming that $B_0$ and $B_1^+$ are known from external measurements.
The calculated CRB values account for a two-echo per $T_\text{R}$ readout,\cite{Yarnykh.2016} and we assume the same readout bandwidth as for our experiments. We neglect any uncertainty or bias in $B_0$ or $B_1^+$, and we also neglect the scan time required to estimate these values (5:34min in Ref. \citen{Yarnykh.2016}).

\subsection{Pulse sequence implementation and data acquisition} \label{sec:MethodsAcquisition}
We performed all experiments on a 3T Prisma scanner (Siemens, Erlangen, Germany) with a 64-channel receive coil.
We use the flip angle pattern depicted in Supporting Fig.~\ref{fig:Spin_dynamics_v3p2}, which was optimized with the legacy framework in spherical coordinates.
The pattern was repeated 189 times---amounting to 12.6 minutes of total scan time---and throughout these cycles, we acquire 3D radial k-space spokes with a nominal resolution of 1.0mm isotropic (defined by $|k_{\max}| = \pi/$voxel).
The sampled k-space covers the insphere of the typically acquired 1.0mm k-space cube. By comparing the covered k-space volume, we estimate an effective resolution of 1.24mm, which we report throughout this paper.\cite{Pipe.2011}
All sampled k-space spokes amount approximately to sampling the outmost k-space area twice at the Nyquist rate.
When reconstructing 13 coefficient images (cf. Section~\ref{sec:MethodsImageRecon}), this equates, in a simplified calculation, to a maximum 2D undersampling factor of 6.5, or 2.5 in each of the two angular k-space dimensions, that will be addressed with parallel imaging and compressed sensing.
We note, however, that the complexity of the subspace reconstruction obscures a simple and accurate metric of undersampling and we resort to judging the resulting image quality.
We changed the direction of the k-space spokes with a 2D golden means pattern\cite{Winkelmann2007, Chan.2009} that was reshuffled to improve the k-space coverage for each of the 1142 time points and to minimize eddy current artifacts.\cite{Flassbeck2022}

\subsection{Image reconstruction} \label{sec:MethodsImageRecon}
We used the Berkeley Advanced Reconstruction Toolbox (BART) \cite{Uecker2015} to reconstruct coefficient images in the low-dimensional space spanned by singular vectors of a coarse dictionary of signals (or fingerprints).\cite{McGivney.2014,Tamir2017,Asslander2018}
We reconstructed 13 coefficient images with the FISTA algorithm,\cite{Beck2009} sensitivity encoding,\cite{Sodickson1997, Pruessmann.2001} and a locally low-rank constraint\cite{Lustig2007, Trzasko2011, Zhang2015} to reduce residual undersampling artifacts and noise. We calculated the coil sensitivity maps with ESPIRiT\cite{Uecker.2014} and ran 500 FISTA iterations. More details on the reconstruction can be found in Refs. \citen{Tamir2017,Asslander2018}.

\subsection{Model fitting} \label{sec:MethodsFitting}
For computational efficiency and robustness, we used a neural network (NN) to fit the MT model to the reconstructed coefficient images.\cite{Cohen2018, Nataraj.2018, Duchemin2020,Zhang.2022}
In comparison to non-linear least square fitting, this approach provides substantial computational benefits: once the network is trained, estimating the parameter maps of a whole brain volume takes only a few seconds instead of many hours. Further, the feed-forward estimation of the biomarkers is robust since it does not require choosing an initialization and cannot get stuck in local minima during the parameter estimation. The non-convexity is, instead, shifted to the network training, where lifting to a high dimension reduces the risk of adverse local minima.

Our network, implemented using Julia's \textit{Flux.jl} package, closely follows the design described in Fig.~2 of Ref. \citen{Zhang.2022}, retaining a similar overall architecture that up-samples the input vector to size 1024 before down-sampling again over 10 fully connected layers with skip connections. We jointly input the real and imaginary parts of the 13 coefficients to the NN after normalizing them by the first (complex-valued) coefficient. The NN outputs estimates of all five biophysical MT parameters, where each parameter is constrained with a ReLU function capped at the maximum value expected in vivo. We trained the NN for 2000 epochs using the rectified ADAM optimizer\cite{Liu.2019} with a learning rate of $10^{-3}$ and an inverse time decay rate of $5\cdot 10^{-4}$.

The neural network includes a data-driven $B_0$ and $B_1^+$ correction, achieved simply by varying $B_0$ and $B_1^+$ in the training data. This correction does not require the network to estimate $B_0$ and $B_1^+$, but to validate this concept, we trained a modified second network that does estimate $B_0$ and $B_1^+$. Further, we trained a third network where $B_0=0$ and $B_1^+=1$ are kept constant in the training data.
Lastly, we trained a network with training data simulated using Graham's spectral model (Eq.~\eqref{eq:Bloch_McConnell_Graham}; also with data-driven $B_0$ and $B_1^+$ correction).
For more details on the neural network fitting, refer to Ref.~\citen{Zhang.2022}.

\subsection{Phantom Experiments}
\subsubsection{Comparison to reference methods}
We performed validations in two different phantoms.
The first validation aims to compare our method to established methods and, for this purpose, we utilize a combination of thermally cross-linked bovine serum albumin (BSA) and MnCl\textsubscript{2} doping to create diverse conditions.
We mixed the BSA powder (approximately 10\%, 15\%, and 20\% of the total weight) with distilled water and stirred it at 30$^\circ$C until the BSA was fully dissolved.
We divided the solution in half and added MnCl\textsubscript{2} to one half to create a concentration of approximately 0.1mM.
We filled six cylindrical 50mL tubes with the solutions and thermally cross-linked them in a water bath at approximately 90$^\circ$C for 10 minutes.
We embedded all tubes in a container filled with distilled water and 0.1mM MnCl\textsubscript{2}.

We scanned the phantom at room temperature with the proposed hybrid-state pulse sequence. For comparison, we also scanned the phantom with a SIR gradient recalled echo (SIR-GRE) sequence.\cite{Gochberg.2003} In this sequence, the magnetization is inverted by a 1ms rectangular pulse, and after an inversion time $T_\text{I}$, we excite the magnetization with a slice selective $\pi/2$-pulse followed by a gradient echo readout of a single line in k-space with an echo time of 3.4ms. Thereafter, the magnetization relaxes to the thermal equilibrium during a 10s waiting period before we repeat the inversion pulse.
Overall, 20 exponentially spaced $T_\text{I}$s were measured, spanning from 4.4ms to 5s
($T_\text{I} \in \{3.2, 4.4, 6.6, 9.7, 14, 22, 31, 46, 68, 100, 150, 220, 325, 480,$ $710, 1050, 1550, 2290, 3380, 5000\}$ms). To account for a frequency drift during the approximately 7h of scan time, a 5\textsuperscript{th} order polynomial was fitted pixel-wise to the phase of the 20 contrasts. We demodulated the fitted phase before projecting the complex images onto the real-valued axis. Data points with less than 5\% of the maximum intensity of each pixel were ignored to avoid uncertainties of the signal's sign that occurs at its zero-crossing. Finally, we fitted the resulting real-valued data with Eq.~\eqref{eq:Bloch_McConnell}, which comprises the generalized Bloch model and reduces to a bi-exponential inversion recovery of the free spin pool after the inversion pulse.

We used a CPMG spin-echo sequence to acquire reference values for $R_2^f$. Here, the magnetization was excited by a slice selective $\pi/2$-pulse, followed by a train of 32 non-selective $\pi$-refocusing pulses. The refocusing pulses were spaced 16ms apart and flanked by crusher gradients on either side. We obtained relaxation rates by fitting the signal decay with an extended phase graph-based model.\cite{Hennig1991,Weigel.2015}

\subsubsection{Comparison to the chemical ground truth}
The second validation aims to compare our method to the chemical ground truth.
To this end, we built a second phantom with seven BSA concentrations (5\%, 10\%, \ldots, 35\%) using the above-described recipe. To maximize the number of BSA concentrations, we did not dope this phantom with MnCl\textsubscript{2}. Further, the tubes of this phantom were not embedded in a water bath and we scanned this phantom only with the proposed hybrid-state pulse sequence.

For each tube, we selected all voxels within the central 30 slices, eroded the outmost layer, and calculated the median value of $m_0^s$, $R_1$, and $R_2^f$.
We performed linear regression of each parameter as a function of the BSA concentration.

\subsection{In vivo experiments}
We tested our hybrid-state-based qMT approach with scans in two participants with clinically-established remitting-relapsing multiple sclerosis (MS) (female, ages 36 \& 25) and two controls (male, ages 36 \& 34). With informed consent obtained in agreement with our institutional review board, we performed scans with the MP-RAGE, FLAIR, and the proposed hybrid-state sequences.

\begin{table*}[tbp]
	\centering
	\newcommand{\g}{\cellcolor[gray]{0.8}}
	\begin{tabular}{l|c|c|c|c|c|c}
		                                                                        & \multicolumn{4}{c|}{hybrid state} & SIR\cite{Cronin2020} & sat.\cite{Yarnykh.2016}                                               \\
		\hline
		inversion pulse                                                         & \g yes                            & yes                  & yes                     & no             &                &           \\
		variable $\alpha$                                                       & \g yes                            & yes                  & no                      & yes            &                &           \\
		variable $T_\text{RF}$                                                  & \g yes                            & no                   & no                      & yes            &                &           \\
		\hline
		$\text{CRB}(m_0^s) \cdot \frac{M_0^2 T}{(m_0^s\sigma)^2}$~[s]           & \g $125$                          & $203$                & $4561$                  & $844$          & $170$          & $146$     \\
		$\text{CRB}(R_1  ) \cdot \frac{M_0^2 T}{(R_1\sigma)^2}$~[s]             & \g $75$                           & $89$                 & $9490$                  & $200$          & $26$           & $51$      \\
		$\text{CRB}(R_2^f) \cdot \frac{M_0^2 T}{(R_2^f\sigma)^2}$~[s]           & \g $46$                           & $41$                 & $57620$                 & $83$           & $+\infty$      & $*$       \\
		\hline
		$\text{CRB}(M_0       ) \frac{T}{\sigma^2}$~[s]                         & \g $49$                           & $57$                 & $9594$                  & $75$           & $22$           & $21$      \\
		$\text{CRB}(R_\text{x}) \cdot \frac{M_0^2 T}{(R_\text{x}\sigma)^2}$~[s] & \g $11940$                        & $15973$              & $47433$                 & $12379$        & $1048$         & $*$       \\
		$\text{CRB}(T_2^s     ) \cdot \frac{M_0^2 T}{(T_2^s\sigma)^2}$~[s]      & \g $8437$                         & $6251$               & $*$                     & $11355$        & $*$            & $*$       \\
		$\text{CRB}(B_0  )$                                                     & \g $+\infty^{\P}$                 & $+\infty^{\P}$       & $*$                     & $+\infty^{\P}$ & $+\infty$      & $\dagger$ \\
		$\text{CRB}(B_1^+       ) \cdot \frac{M_0^2 T}{(B_1\sigma)^2}$~[s]      & \g $29$                           & $17$                 & $*$                     & $26$           & $+\infty^{\P}$ & $\dagger$ \\
	\end{tabular}
	\caption{Comparison of the Cram\'er-Rao bound (CRB; Eq.~\eqref{eq:CRB}) values between pulse sequences.
		The left section analyzes optimizations for a minimal CRB of $m_0^s$, $R_1$, and $R_2^f$ with our framework, while the rightmost two columns provide CRB values for literature qMT methods for reference, namely selective inversion recovery (SIR)\cite{Cronin2020} and saturation-based qMT.\cite{Yarnykh.2016}
		In general, we assumed all 8 model parameters as unknown, and $*$ highlights parameters assumed to be heuristically fixed to literature values. $\dagger$ indicates parameters fixed to estimates from external calibration scans, which require an additional 5:34min of scan time,\cite{Yarnykh.2016} and $\P$ identifies CRB parameters that are large at $B_0=0$ and $B_1^+=1$, respectively, but decrease substantially when deviating from these values. The gray background highlights the proposed approach.
		The displayed CRB values are normalized by the squared value of the parameter, the squared magnetization $M_0$, and the noise variance of the time series in a voxel $\sigma^2$, as well as the scan time $T$, i.e., they reflect the inverse squared signal-to-noise ratio per unit time and for a unit signal noise variance.
	}
	\label{tab:CRB_seq}
\end{table*}

\section{Results}
\subsection{Cram\'er-Rao bound analysis} \label{sec:Results_CRB}
We performed numerical optimizations for minimal noise variance per unit time in the estimates of the target parameters $m_0^s$, $R_1$, and $R_2^f$.
First, we optimized the pulse sequence that utilizes a selective inversion-recovery combined with a bSSFP-like RF pattern with a variable flip angle and pulse duration.
The resulting CRB values are depicted in the first column of Tab.~\ref{tab:CRB_seq} and highlighted by the gray background.
This optimization has overall the lowest CRB values in the target parameters.
The second column shows the CRB values that result from an optimization with a fixed $T_\text{RF}$, which entails a slight degradation of the CRB for $m_0^s$ and $R_1$.
When also fixing the flip angle, i.e., when using the established inversion-recovery (IR) bSSFP sequence,\cite{Schmitt2004} the CRB values increase substantially (third column).
In the unconstrained case, they increased to the point of numerical instability, which is why we here heuristically fixed $T_2^s$, $B_0$, and $B_1^+$. Still, the resulting CRB values are substantially larger compared to the proposed method with a varying flip angle and $T_\text{RF}$.
The fourth column analyzes a pulse sequence with a variable flip angle and $T_\text{RF}$ but without an inversion pulse. Its performance is superior to the IR-bSSFP but again substantially inferior compared to the proposed sequence with an inversion pulse (first column).

The fifth column provides the CRB of the SIR\cite{Gochberg.2003} sequence for reference.
SIR and IR-bSSFP differ in one critical aspect: the former allows for an undisturbed inversion recovery, while the latter applies a train of RF pulses during the inversion recovery.
With SIR, we observe, in large, similar CRB values for $m_0^s$ and $R_1$ compared to the proposed approach (first column). However, there are two key differences between the CRB calculations. While we assume that all model parameters are fitted for the proposed approach, SIR relies on a heuristic constraint of $T_2^s$ to a fixed value. Further, the SIR data processing pipeline accounts only for the effect of $B_1^+$ inhomogeneities on the free spin pool but not for their effect on the semi-solid spin pool. For sufficiently miscalibrated $B_1^+$ values, such as $B_1^+=0.9$ or $B_1^+=1.1$, $B_1^+$ is well-encoded by the free pool's $B_1^+$-dependency and the $B_1^+$-dependence of the semi-solid spin pool has only a moderate effect on the CRB values.
When approaching $B_1^+=1$, however, the partial derivative of the free pool's spin dynamics wrt. $B_1^+$ vanishes,\cite{Dortch.2011} while the partial derivative of the semi-solid-pool's dynamics does not.
Consequently, the CRB diverges at $B_1^+=1$ when accounting for the $B_1^+$ dependence of the semi-solid spin pool's saturation.
We note that the CRB entails a linear approximation and, when considering higher order effects, the noise propagation to the estimated parameters is finite due to the finite derivative at $B_1^+\neq1$.
The noise properties in this non-linear system depend on the noise level itself; a more detailed discussion is beyond the scope of this paper.

The rightmost column of Tab.~\ref{tab:CRB_seq} analyzes a qMT pulse sequence that utilizes a saturation pulse to increase the MT effect.\cite{Henkelman1993, Graham1997, Yarnykh.2012, Yarnykh.2016}
It also has similar CRB values for $m_0^s$ and $R_1$ compared to the proposed approach. However, in contrast to the proposed approach, it relies on fixing $T_2^f/T_1$, $R_\text{x}$, and $T_2^s$ to heuristic values and requires external calibration scans for $B_0$ and $B_1^+$. Considering the scan time required for the latter (using the timings in Ref. \citen{Yarnykh.2016}) would increase the normalized CRB by roughly 30\%.

\begin{figure}[tbp]
	\centering
	\ifOL
		\includegraphics[]{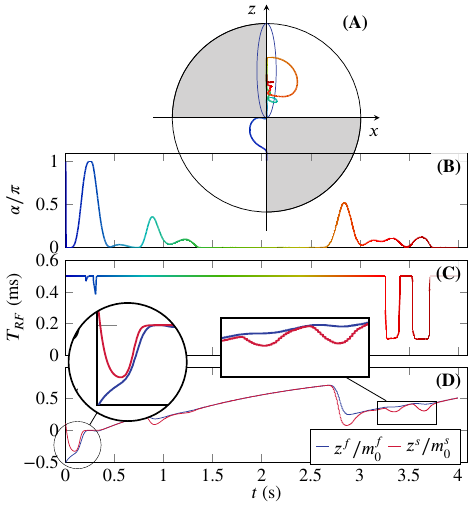}
	\else
		\begin{tikzpicture}[scale = .88]
    \def\TOne{1.6};
    \def\TTwo{0.067};

    \begin{axis}[
            width=\linewidth*0.8,
            axis equal image,
            axis lines=center,
            xmin = -1.2, xmax = 1.2,
            ymin = -1.2, ymax = 1.2,
            xtick=\empty, ytick=\empty, ztick=\empty,
            xlabel={$x$}, ylabel={$z$},
            every axis x label/.style={at={(ticklabel cs:0.975)}, anchor=north},
            every axis y label/.style={at={(ticklabel cs:0.975)}, anchor=east},
            colormap name = mybluered,
            point meta min=0,
            point meta max= 4,
            name=Bloch,
            clip=false,
            axis on top=true,
        ]

        \filldraw[fill=lightgray] (axis cs: 0,0) -- (axis cs: 0, 1) arc [start angle=  90, end angle= 180, radius={(1)}];
        \filldraw[fill=lightgray] (axis cs: 0,0) -- (axis cs: 0,-1) arc [start angle=270, end angle=360, radius={(1)}];

        \addplot[no marks, domain=0:360,samples=60, forget plot] ({sin(x)}, {cos(x)});

        \addplot[domain=0:1,samples=1000, UKLblue, forget plot] ({-sqrt(\TTwo * (1/(4*\TOne) - (x - .5)^2/\TOne)}, {x});
        \addplot[domain=0:1,samples=1000, UKLblue, forget plot] ({  sqrt(\TTwo * (1/(4*\TOne) - (x - .5)^2/\TOne)}, {x});

        \addplot[mesh, point meta=explicit, thick, forget plot]table[x=xf, y=zf, meta=t_s]{Figures/Spin_dynamics_Inv_varyFA_varyTRF.txt};

        \node[right, inner sep=0mm,minimum size=5mm] at (axis cs:  .8,1) {\textbf{(A)}};
    \end{axis}

    \begin{axis}[
            width=\linewidth*0.9,
            height=\textheight*0.075,
            scale only axis,
            xmin=0,
            xmax=4.1,
            ymin=0,
            xticklabel=\empty,
            ylabel={$\alpha/\pi$},
            colormap name = mybluered,
            point meta min=0,
            point meta max= 4,
            name=theta,
            at=(Bloch.south),
            anchor=north,
            yshift = 1.5cm,
        ]
        \addplot [mesh, thick, point meta=explicit, solid]table[x=t_s, y=alpha_pi, meta=t_s]{Figures/Spin_dynamics_Inv_varyFA_varyTRF.txt};

        \node[anchor=north east, fill=white, minimum size=4mm, opacity=0.85, text opacity=1] at (rel axis cs: 1,1) {\textbf{(B)}};
    \end{axis}

    \begin{axis}[
            width=\linewidth*0.9,
            height=\textheight*0.075,
            scale only axis,
            xmin=0,
            xmax=4.1,
            ymin=0,
            ymax=0.6,
            xticklabel=\empty,
            ylabel={$T_{RF}~(\text{ms})$},
            colormap name = mybluered,
            point meta min=0,
            point meta max= 4,
            name=TRF,
            at=(theta.below south east),
            anchor= north east,
        ]
        \addplot [mesh, thick, point meta=explicit, solid]table[x=t_s, y=TRF_ms, meta=t_s]{Figures/Spin_dynamics_Inv_varyFA_varyTRF.txt};

        \node[anchor=north east, fill=white, minimum size=4mm, opacity=0.85, text opacity=1] at (rel axis cs: 1,1) {\textbf{(C)}};
    \end{axis}

    \begin{scope}[spy using outlines={magnification=2.5, connect spies}]
        \begin{axis}[
                width=\linewidth*0.9,
                height=\textheight*0.075,
                scale only axis,
                xmin=0,
                xmax=4.1,
                ymin=-.5,
                ymax=0.99,
                xlabel={$t~\text{(s)}$},
                xlabel style={yshift=0.15cm},
                name=rz,
                at=(TRF.below south east),
                anchor= north east,
                legend entries = {$z^f/m_0^f$, $z^s/m_0^s$},
                legend pos = south east,
                legend columns=2,
                legend style={
                        xshift = 0.1cm,
                        yshift = -0.1cm,
                        legend image code/.code={\draw[##1,line width=0.6pt] plot coordinates {(0cm,0cm) (0.25cm,0cm)};}
                    },
            ]
            \addplot [color=UKLblue,            solid]table[x=t_s, y=zf_m0f]{Figures/Spin_dynamics_Inv_varyFA_varyTRF.txt};
            \addplot [color=UKLred,              solid]table[x=t_s, y=zs_m0s]{Figures/Spin_dynamics_Inv_varyFA_varyTRF.txt};

            \coordinate (spypoint_bE) at (axis cs:0.05,-0.35);
            \coordinate (spyviewer_bE) at (axis cs:0.5,0.8);
            \spy[circle, black,size=2.0cm] on (spypoint_bE) in node [fill=white] at (spyviewer_bE);

            \coordinate (spypoint_sat) at (axis cs:3.0,0.1);
            \coordinate (spyviewer_sat) at (axis cs:2.0,1);
            \spy[rectangle,black,height=1.0cm,width=2.5cm] on (spypoint_sat) in node [fill=white] at (spyviewer_sat);

            \node[anchor=north east, opacity=0.85, text opacity=1] at (rel axis cs: 1,1) {\textbf{(D)}};
        \end{axis}
    \end{scope}
\end{tikzpicture}
	\fi
	\caption{Spin trajectory and the corresponding control for a pulse sequence with a varying flip angle and $T_\text{RF}$, and an inversion pulse. \textbf{(A)} The dynamics of the free pool on the Bloch sphere, color-coded to provide a reference to the time axis in (B,C), and the steady-state ellipse in blue; \textbf{(B)} the flip angle $\alpha$ and \textbf{(C)} the pulse duration $T_\text{RF}$ controlling the spin dynamics. \textbf{(D)} The normalized $z$-magnetization of the two pools. The spherical and rectangular magnifications highlight segments that utilize a bi-exponential inversion-recovery\cite{Gochberg.2003} and saturation,\cite{Gloor2008} respectively, to encode the MT effect.}
	\label{fig:MT_Spin_Dynamics}
\end{figure}
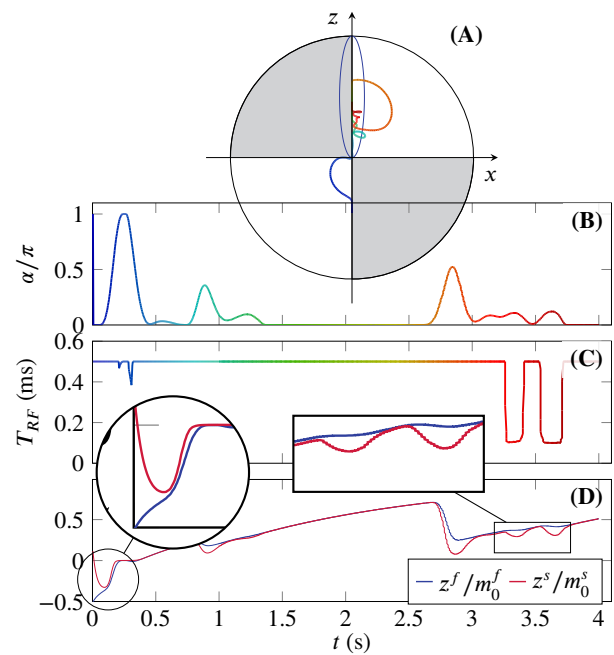

\subsection{Optimized RF pattern} \label{sec:Results_Optimization}
Fig.~\ref{fig:MT_Spin_Dynamics} depicts the optimized pulse sequence that incorporates an inversion pulse and a varying flip angle and $T_\text{RF}$ (1\textsuperscript{st} and highlighted column in Tab.~\ref{tab:CRB_seq}), along with the evoked spin dynamics.
At the beginning of the RF pulse train, $\alpha$ is small, which results in a bi-exponential recovery after the inversion pulse, similar to SIR (spherical magnification in Fig.~\ref{fig:MT_Spin_Dynamics}(D)). Once the exchange approaches an equilibrium ($z^f/m_0^f \approx z^s/m_0^s$, i.e., the red and blue curves approach each other), the optimized $\alpha$ increases to convert the longitudinal to detectable transversal magnetization, along with the exchange-related information encoded therein. The saturation-based encoding of the MT effect can be observed around 3.5 seconds, where blips in $\alpha$ at short $T_\text{RF}$ saturate the semi-solid spin pool's magnetization (rectangular magnification).
These saturation segments exhibit similar behavior as described in Ref. \citen{Gochberg.1999}: the optimized pulse sequence strives for a repeated maximal separation of the two pools' magnetization.
During several parts of the pattern, the flip angles approach zero in this unbounded optimization.

\begin{figure}[tb]
	\centering
	\ifOL
		\includegraphics[]{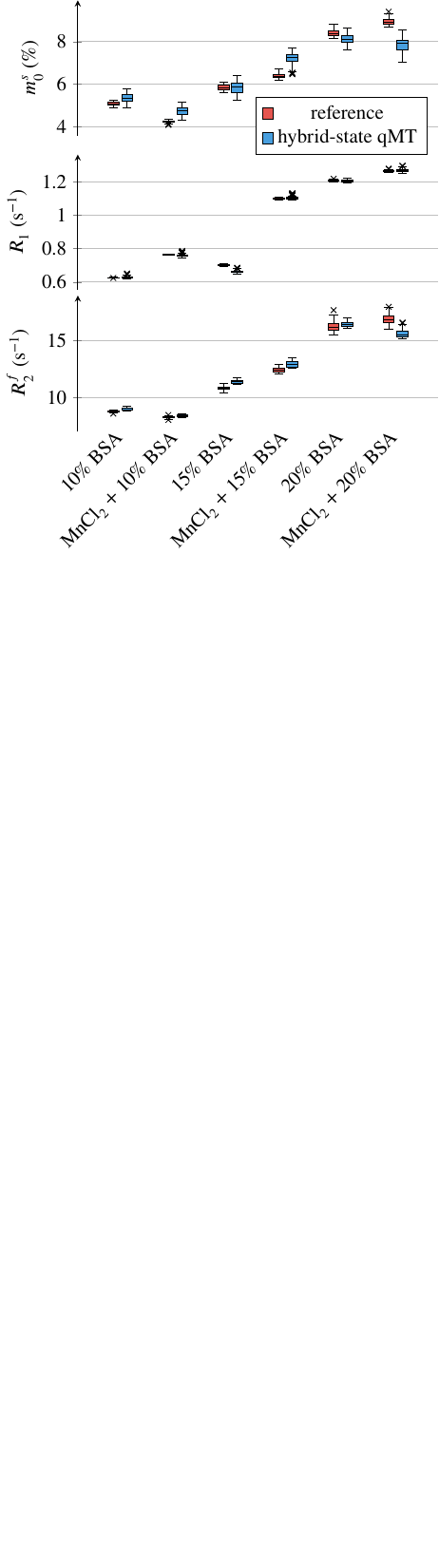}
	\else
		\input{Figures/Phantom_validation.tex}
	\fi
	\caption{Phantom validation. We compare $m_0^s$, $R_1$, and $R_2^f$ as measured with the proposed hybrid-state qMT approach to measurements with more established but slow reference scans. The reference values for $m_0^s$ and $R_1$ were measured with a selective inversion-recovery gradient recalled echo (SIR-GRE) sequence\cite{Gochberg.2003} and the reference for $R_2^f$ with a CPMG sequence.\cite{Meiboom.1958}}
	\label{fig:Phantom}
\end{figure}

\subsection{Phantom experiments} \label{sec:ResultsPhantom}
\subsubsection{Comparision to reference methods}
Fig.~\ref{fig:Phantom} validates our approach by comparing the estimates of $m_0^s$, $R_1$, and $R_2^f$ from the hybrid-state sequence to estimates from a SIR gradient echo\cite{Gochberg.2003} and a multi-echo CPMG\cite{Meiboom.1958} sequence.
We find good agreement between the two measurements.
The mean value in each sample deviates between the two measurements by 7.6\% in $m_0^s$, 1.9\% in $R_1$, and 5.4\% in $R_2^f$ (averaged over the absolute value of the relative deviations).
The deviations tend to be larger for the doped samples, and we found the largest deviation (12.0\%) in $m_0^s$ of the 20\% BSA sample with MnCl\textsubscript{2} doping.
We note that this sample exhibits fast $R_1$-relaxation, which is outside of the range expected for biological tissues.

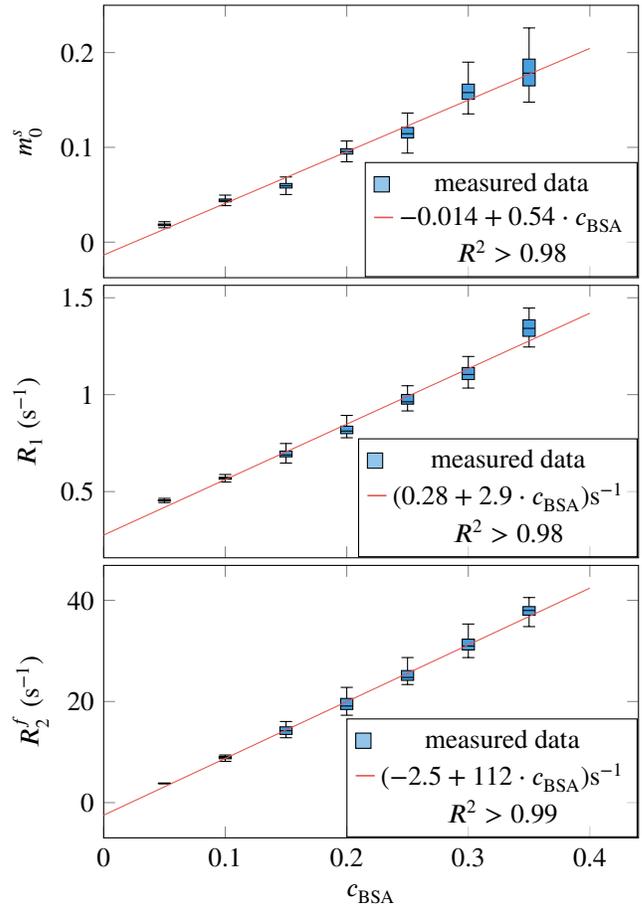
\begin{figure}
	\centering
	\ifOL
		\includegraphics[]{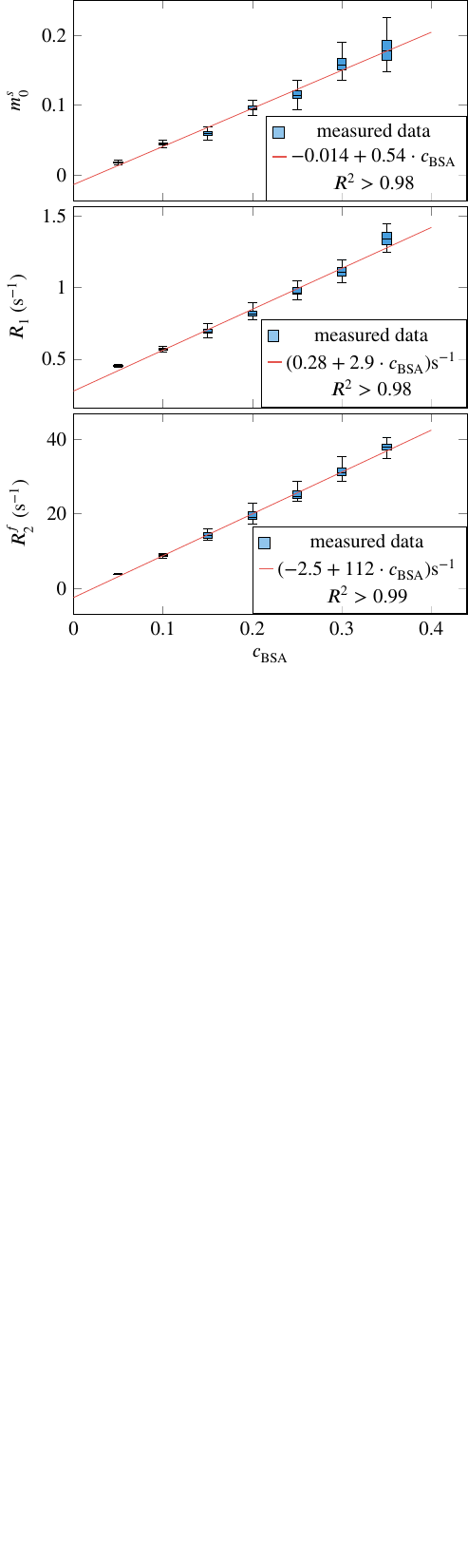}
	\else
		\begin{tikzpicture}
    \begin{axis}[%
        width=\columnwidth,
        height=\columnwidth*0.6,
        xmin=0,
        ylabel = {$m_0^s$},
        scaled y ticks=false,
        xticklabel = \empty,
        yticklabel style={/pgf/number format/fixed, /pgf/number format/precision=2},
        boxplot/draw direction = y,
        legend pos = south east,
        legend style={
            yshift=-0.1cm,
            xshift=0.2cm,
            fill opacity = 0.6,
            draw opacity = 1,
            text opacity = 1,
            },
        name=m0s,
        ]

        \addlegendimage{legend image code/.code={\draw[fill=TheLake, draw=black] (0cm,-0.1cm) rectangle (0.2cm,0.1cm);}}
        \addlegendentry{measured data}

        \addlegendimage{legend image code/.code={\draw[Pastrami,line width=0.6pt] plot coordinates {(0cm,0cm) (0.25cm,0cm)};}}
        \addlegendentry{$-0.014 + 0.54 \cdot c_\text{BSA}$}

        \addlegendimage{empty legend}
        \addlegendentry{$R^2 > 0.98$}

        \addplot+[boxplot={draw position=0.05, box extend=0.01, whisker extend=0.01}, fill, color=TheLake, draw=black, solid, mark=x] table[col sep=comma, y=5p_BSA] {Figures/Phantom_linearRegression_m0s_norm.txt};
        \addplot+[boxplot={draw position=0.10, box extend=0.01, whisker extend=0.01}, fill, color=TheLake, draw=black, solid, mark=x] table[col sep=comma,y=10p_BSA] {Figures/Phantom_linearRegression_m0s_norm.txt};
        \addplot+[boxplot={draw position=0.15, box extend=0.01, whisker extend=0.01}, fill, color=TheLake, draw=black, solid, mark=x] table[col sep=comma,y=15p_BSA] {Figures/Phantom_linearRegression_m0s_norm.txt};
        \addplot+[boxplot={draw position=0.20, box extend=0.01, whisker extend=0.01}, fill, color=TheLake, draw=black, solid, mark=x] table[col sep=comma,y=20p_BSA] {Figures/Phantom_linearRegression_m0s_norm.txt};
        \addplot+[boxplot={draw position=0.25, box extend=0.01, whisker extend=0.01}, fill, color=TheLake, draw=black, solid, mark=x] table[col sep=comma,y=25p_BSA] {Figures/Phantom_linearRegression_m0s_norm.txt};
        \addplot+[boxplot={draw position=0.30, box extend=0.01, whisker extend=0.01}, fill, color=TheLake, draw=black, solid, mark=x] table[col sep=comma,y=30p_BSA] {Figures/Phantom_linearRegression_m0s_norm.txt};
        \addplot+[boxplot={draw position=0.35, box extend=0.01, whisker extend=0.01}, fill, color=TheLake, draw=black, solid, mark=x] table[col sep=comma,y=35p_BSA] {Figures/Phantom_linearRegression_m0s_norm.txt};

        \addplot[domain=0:0.4,samples=2, color=Pastrami] ({x}, {-0.013584086937563735 + 0.5449545636240924 * x});
    \end{axis}

    \begin{axis}[%
        width=\columnwidth,
        height=\columnwidth*0.6,
        xmin=0,
        ylabel = {$R_1~(\text{s}^{-1})$},
        scaled y ticks=false,
        xticklabel = \empty,
        yticklabel style={/pgf/number format/fixed, /pgf/number format/precision=2},
        boxplot/draw direction = y,
        legend pos = south east,
        legend style={
            yshift=-0.1cm,
            xshift=0.2cm,
            fill opacity = 0.6,
            draw opacity = 1,
            text opacity = 1,
            },
        name=R1,
        at=(m0s.south),
        anchor=north,
        yshift = -0.1cm,
        ]

        \addlegendimage{legend image code/.code={\draw[fill=TheLake, draw=black] (0cm,-0.1cm) rectangle (0.2cm,0.1cm);}}
        \addlegendentry{measured data}

        \addlegendimage{legend image code/.code={\draw[Pastrami,line width=0.6pt] plot coordinates {(0cm,0cm) (0.25cm,0cm)};}}
        \addlegendentry{$(0.28 + 2.9 \cdot c_\text{BSA})\text{s}^{-1}$}

        \addlegendimage{empty legend}
        \addlegendentry{$R^2 > 0.98$}

        \addplot+[boxplot={draw position=0.05, box extend=0.01, whisker extend=0.01}, fill, color=TheLake, draw=black, solid, mark=x] table[col sep=comma, y=5p_BSA] {Figures/Phantom_linearRegression_R1a_s-1.txt};
        \addplot+[boxplot={draw position=0.10, box extend=0.01, whisker extend=0.01}, fill, color=TheLake, draw=black, solid, mark=x] table[col sep=comma,y=10p_BSA] {Figures/Phantom_linearRegression_R1a_s-1.txt};
        \addplot+[boxplot={draw position=0.15, box extend=0.01, whisker extend=0.01}, fill, color=TheLake, draw=black, solid, mark=x] table[col sep=comma,y=15p_BSA] {Figures/Phantom_linearRegression_R1a_s-1.txt};
        \addplot+[boxplot={draw position=0.20, box extend=0.01, whisker extend=0.01}, fill, color=TheLake, draw=black, solid, mark=x] table[col sep=comma,y=20p_BSA] {Figures/Phantom_linearRegression_R1a_s-1.txt};
        \addplot+[boxplot={draw position=0.25, box extend=0.01, whisker extend=0.01}, fill, color=TheLake, draw=black, solid, mark=x] table[col sep=comma,y=25p_BSA] {Figures/Phantom_linearRegression_R1a_s-1.txt};
        \addplot+[boxplot={draw position=0.30, box extend=0.01, whisker extend=0.01}, fill, color=TheLake, draw=black, solid, mark=x] table[col sep=comma,y=30p_BSA] {Figures/Phantom_linearRegression_R1a_s-1.txt};
        \addplot+[boxplot={draw position=0.35, box extend=0.01, whisker extend=0.01}, fill, color=TheLake, draw=black, solid, mark=x] table[col sep=comma,y=35p_BSA] {Figures/Phantom_linearRegression_R1a_s-1.txt};

        \addplot[domain=0:0.4,samples=2, color=Pastrami] ({x}, {0.275512056691306 + 2.8618527735982617 * x});
    \end{axis}

    \begin{axis}[%
        width=\columnwidth,
        height=\columnwidth*0.6,
        xmin=0,
        xlabel={$c_\text{BSA}$},
        ylabel = {$R_2^f~(\text{s}^{-1})$},
        scaled y ticks=false,
        xticklabel style={/pgf/number format/fixed, /pgf/number format/precision=2},
        yticklabel style={/pgf/number format/fixed, /pgf/number format/precision=2},
        boxplot/draw direction = y,
        legend pos = south east,
        legend style={
            yshift=-0.1cm,
            xshift=0.2cm,
            fill opacity = 0.6,
            draw opacity = 1,
            text opacity = 1,
            },
        name=R2f,
        at=(R1.south),
        anchor=north,
        yshift = -0.1cm,
        ]

        \addlegendimage{legend image code/.code={\draw[fill=TheLake, draw=black] (0cm,-0.1cm) rectangle (0.2cm,0.1cm);}}
        \addlegendentry{measured data}

        \addlegendimage{legend image code/.code={\draw[Pastrami,line width=0.6pt] plot coordinates {(0cm,0cm) (0.25cm,0cm)};}}
        \addlegendentry{$(-2.5 + 112 \cdot c_\text{BSA})\text{s}^{-1}$}

        \addlegendimage{empty legend}
        \addlegendentry{$R^2 > 0.99$}

        \addplot+[boxplot={draw position=0.05, box extend=0.01, whisker extend=0.01}, fill, color=TheLake, draw=black, solid, mark=x] table[col sep=comma, y=5p_BSA] {Figures/Phantom_linearRegression_R2f_s-1.txt};
        \addplot+[boxplot={draw position=0.10, box extend=0.01, whisker extend=0.01}, fill, color=TheLake, draw=black, solid, mark=x] table[col sep=comma,y=10p_BSA] {Figures/Phantom_linearRegression_R2f_s-1.txt};
        \addplot+[boxplot={draw position=0.15, box extend=0.01, whisker extend=0.01}, fill, color=TheLake, draw=black, solid, mark=x] table[col sep=comma,y=15p_BSA] {Figures/Phantom_linearRegression_R2f_s-1.txt};
        \addplot+[boxplot={draw position=0.20, box extend=0.01, whisker extend=0.01}, fill, color=TheLake, draw=black, solid, mark=x] table[col sep=comma,y=20p_BSA] {Figures/Phantom_linearRegression_R2f_s-1.txt};
        \addplot+[boxplot={draw position=0.25, box extend=0.01, whisker extend=0.01}, fill, color=TheLake, draw=black, solid, mark=x] table[col sep=comma,y=25p_BSA] {Figures/Phantom_linearRegression_R2f_s-1.txt};
        \addplot+[boxplot={draw position=0.30, box extend=0.01, whisker extend=0.01}, fill, color=TheLake, draw=black, solid, mark=x] table[col sep=comma,y=30p_BSA] {Figures/Phantom_linearRegression_R2f_s-1.txt};
        \addplot+[boxplot={draw position=0.35, box extend=0.01, whisker extend=0.01}, fill, color=TheLake, draw=black, solid, mark=x] table[col sep=comma,y=35p_BSA] {Figures/Phantom_linearRegression_R2f_s-1.txt};

        \addplot[domain=0:0.4,samples=2, color=Pastrami] ({x}, {-2.487049511500755 + 112.27983491761339 * x});
    \end{axis}
\end{tikzpicture}
	\fi
	\caption{Linear regression of $m_0^s$ $R_1$, and $R_2^f$, measured with the proposed hybrid-state qMT approach, vs. the bovine serum albumin (BSA) concentration. The data are depicted as box plots, comprising the median, 1\textsuperscript{st}, and 3\textsuperscript{rd} quartile. We performed linear regression on the median value, treating each tube as an independent sample.
	}
	\label{fig:Phantom_m0s_linear_fit}
\end{figure}

\subsubsection{Comparison to the chemical ground truth}
In line with our theoretical understanding of the MT effect, $m_0^s$ is well-described by a linear model of the BSA concentration (Fig.~\ref{fig:Phantom_m0s_linear_fit}). The 95\% confidence interval of the intercept with the y-axis is $[-0.029, 0.0014]$, which includes the presumed intercept with the origin.

The apparent longitudinal relaxation rate $R_1$ is also well-described by a linear model and the intercept with the y-axis of 0.28/s in the expected range of water relaxivity.
This finding is also in line with a model of spin relaxation induced by interactions with the BSA surface.\cite{Koenig1990}

Last, the transversal relaxation rate of the free spin pool $R_2^f$ also follows a linear dependency on the BSA concentration.
The y-intercept is, however, negative (95\% confidence interval of $[-4.28, -0.69]\text{s}^{-1}$), which is not physical. Instead, we would expect a y-intercept similar to the one of $R_1$.
We note, however, that the steep linear component of $122\text{s}^{-1} / c_\text{BSA}$ entails a sensitivity of the y-intercept to small systematic biases.
A further explanation could be a non-linearity of $R_2^f$ at low BSA concentrations.

\subsection{In vivo experiments} \label{sec:ResultsInVivo}
\subsubsection{$B_0$ and $B_1^+$ correction} \label{sec:ResultsB0B1}
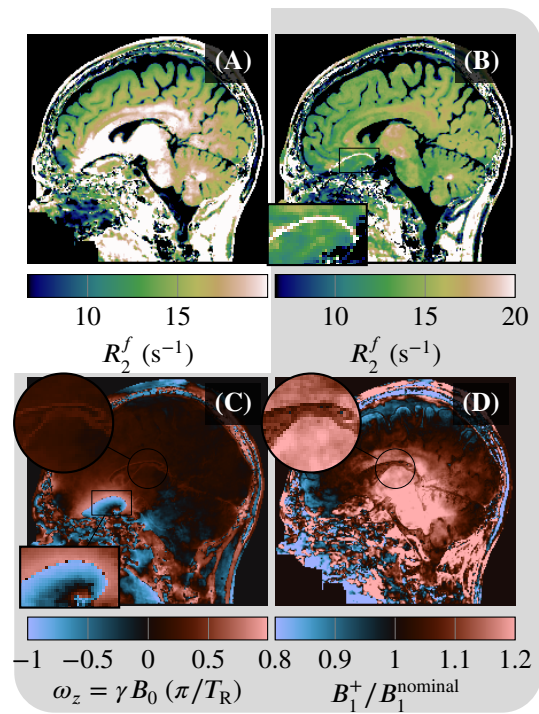
\begin{figure}[tbp]
	\centering
	\ifOL
		\includegraphics[]{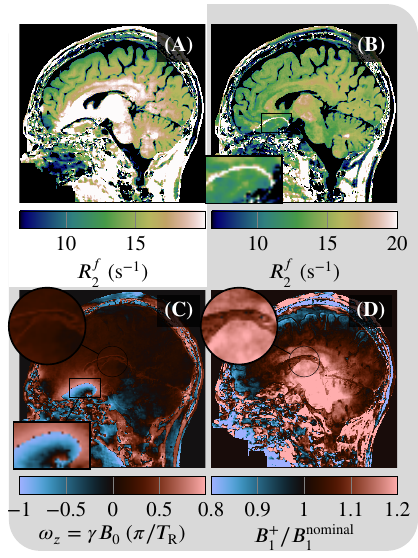}
	\else
		\begin{tikzpicture}[scale = 1]
    \begin{axis}[
            width={3cm*1.05},
            height={3cm},
            axis on top,
            scale only axis,
            xmin=0,
            xmax=1,
            ymin=0,
            ymax=1,
            hide axis,
            name=m0s_not_corrected,
            align=center,
            colormap name = gist_earth,
            colorbar horizontal,
            point meta min=6.67,
            point meta max=19.99,
            colorbar style={xlabel=$R_2^f~(\text{s}^{-1})$, height=0.3cm, yshift=0.15cm, xlabel style = {yshift = 0.1cm}, xticklabel style={/pgf/number format/fixed, /pgf/number format/precision=2}},
        ]
        \addplot graphics [xmin=0,xmax=1,ymin=0,ymax=1] {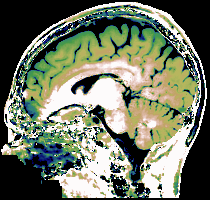};

        \node[fill=black,text=white, opacity=0.75, text opacity=1, anchor = north east] at (rel axis cs:  0.975,0.975) {\textbf{(A)}};
    \end{axis}

    \begin{scope}[spy using outlines={magnification=2.5, connect spies}]
        \begin{axis}[
                width={3cm*1.05},
                height={3cm},
                axis on top,
                scale only axis,
                xmin=0,
                xmax=1,
                ymin=0,
                ymax=1,
                hide axis,
                name=m0s_corrected,
                at=(m0s_not_corrected.right of north east),
                anchor=north west,
                xshift = 0.1cm,
                colormap name = gist_earth,
                colorbar horizontal,
                point meta min=6.67,
                point meta max=20,
                colorbar style={xlabel=$R_2^f~(\text{s}^{-1})$, height=0.3cm, yshift=0.15cm, xlabel style = {yshift = 0.1cm}, xticklabel style={/pgf/number format/fixed, /pgf/number format/precision=2}},
            ]
            \addplot graphics [xmin=0,xmax=1,ymin=0,ymax=1] {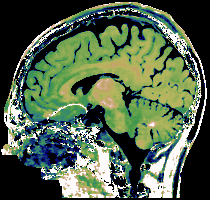};

            \coordinate (spypoint) at (axis cs:0.35,0.45);
            \coordinate (spyviewer) at (axis cs:0.175,0.125);
            \spy[rectangle,black,height=0.8cm,width=1.3cm] on (spypoint) in node at (spyviewer);

            \node[fill=black,text=white, opacity=0.75, text opacity=1, anchor = north east] at (rel axis cs:  0.975,0.975) {\textbf{(B)}};
        \end{axis}
    \end{scope}

    \begin{scope}[spy using outlines={magnification=2.5, connect spies}]
        \begin{axis}[
                width={3cm*1.05},
                height={3cm},
                axis on top,
                scale only axis,
                xmin=0,
                xmax=1,
                ymin=0,
                ymax=1,
                hide axis,
                name=B0_est,
                at=(m0s_not_corrected.below south west),
                anchor=north west,
                yshift = -1.5cm,
                align=center,
                colormap name = berlin,
                colorbar horizontal,
                point meta min=-1,
                point meta max=0.99,
                colorbar style={xlabel={$\omega_z = \gamma B_0~(\pi / T_{\text{R}})$}, height=0.3cm, yshift=0.15cm, xlabel style = {yshift = 0.1cm}, xticklabel style={/pgf/number format/fixed, /pgf/number format/precision=2}},
            ]
            \addplot graphics [xmin=0,xmax=1,ymin=0,ymax=1] {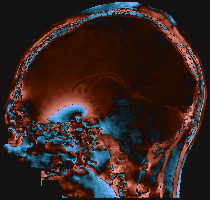};

            \coordinate (spypoint) at (axis cs:0.35,0.45);
            \coordinate (spyviewer) at (axis cs:0.175,0.125);
            \spy[rectangle,black,height=0.8cm,width=1.3cm] on (spypoint) in node at (spyviewer);

            \coordinate (spypoint2) at (axis cs:0.5,0.6);
            \coordinate (spyviewer2) at (axis cs:0.15,0.8);
            \spy[circle,black,height=0.8cm,width=1.3cm] on (spypoint2) in node at (spyviewer2);

            \node[fill=black,text=white, opacity=0.75, text opacity=1, anchor = north east] at (rel axis cs:  0.975,0.975) {\textbf{(C)}};
        \end{axis}
    \end{scope}

    \begin{scope}[spy using outlines={magnification=2.5, connect spies}]
    \begin{axis}[
        width={3cm*1.05},
        height={3cm},
        axis on top,
        scale only axis,
        xmin=0,
        xmax=1,
        ymin=0,
        ymax=1,
        hide axis,
        name=B1_est,
        at=(B0_est.right of north east),
        anchor=north west,
        xshift = 0.1cm,
        colormap name = berlin,
        colorbar horizontal,
        point meta min=0.8001,
        point meta max=1.2,
        colorbar style={xlabel=$B_1^+/B_1^{\text{nominal}}$, height=0.3cm, yshift=0.15cm, xlabel style = {yshift = 0.1cm}, xticklabel style={/pgf/number format/fixed, /pgf/number format/precision=2}},
        ]
        \addplot graphics [xmin=0,xmax=1,ymin=0,ymax=1] {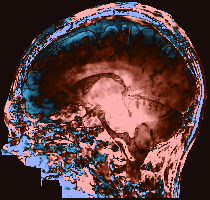};

        \coordinate (spypoint2) at (axis cs:0.5,0.6);
        \coordinate (spyviewer2) at (axis cs:0.15,0.8);
        \spy[circle,black,height=0.8cm,width=1.3cm] on (spypoint2) in node at (spyviewer2);

        \node[fill=black,text=white, opacity=0.75, text opacity=1, anchor = north east] at (rel axis cs:  0.975,0.975) {\textbf{(D)}};
    \end{axis}
\end{scope}

    \begin{scope}[
            on background layer,
        ]
        \pgfmathsetlengthmacro{\Shift}{10pt}
        \fill [black!15,rounded corners=15pt]
        ([shift={(-5pt,\Shift)}] m0s_not_corrected.north west)
        rectangle
        ([shift={(\Shift,-40pt)}] B1_est.below south east)
        ;
        \fill [white]
        ([shift={(-5pt,\Shift+2pt)}] m0s_not_corrected.north west)
        rectangle
        ([shift={(1pt,-41pt)}] m0s_not_corrected.south east)
        ;
    \end{scope}
\end{tikzpicture}
	\fi
	\caption{$B_0$- and $B_1^+$-correction. \textbf{(A)} The estimated $R_2^f=1/T_2^f$ is biased when assuming homogeneous magnetic fields. \textbf{(B)} Considering $B_0$ and $B_1^+$ as unknowns by varying them in the data used for training the neural network\cite{Zhang.2022} allows for the removal of the bias except for the few voxels in the center of the banding artifact (cf. magnifications). \mbox{\textbf{(C)}, \textbf{(D)}} $B_0$ and $B_1^+$ estimated from the hybrid-state data. The gray background groups the parameters that were jointly estimated (along with the other five parameters).
	}
	\label{fig:InVivo_B0}
\end{figure}

Fig.~\ref{fig:InVivo_B0} examines the feasibility of removing $B_0$- and $B_1^+$-induced biases by varying these parameters in the training data of the neural network:
The $R_2^f$-map (subplot A) was estimated with a neural network that was trained with $B_0 = 0$ and $B_1^+ = 1$ and exhibits a substantial over-estimation of $R_2^f$ in the brain's center, where $B_1^+$ is large, and the orbitofrontal lobe, where $\omega_z = \gamma B_0$ has substantial deviations from the resonance condition.
When treating the field variations as unknowns, however, this bias is substantially reduced (Fig.~\ref{fig:InVivo_B0}(B)).
The $B_0$ correction fails in the center of the bSSFP-typical banding artifact\cite{Carr1958} above the sinus frontalis, where $|\omega_z| = \pi / T_\text{R}$ (cf. magnifications Fig.~\ref{fig:InVivo_B0}(B) and (C)).

For verification, we also estimated $B_0$ and $B_1^+$ (Fig.~\ref{fig:InVivo_B0}(C--D)).
In the vicinity of the stopband (rectangular magnification), the $B_0$ estimates follow the expected pattern.
However, in the center of the brain, where $B_0$ is well-calibrated, we observe subtle variations at GM-CSF interfaces that are likely unphysical (spherical magnification).
The estimates of $B_1^+$ match the well-known pattern of low $B_1^+$ at the cortex and high $B_1^+$ in the brain's center, but exhibit spurious deviations in the CSF (circular magnification in Fig.~\ref{fig:InVivo_B0}(D)).

\begin{figure*}[tbp]
	\centering
	\ifOL
		\includegraphics[]{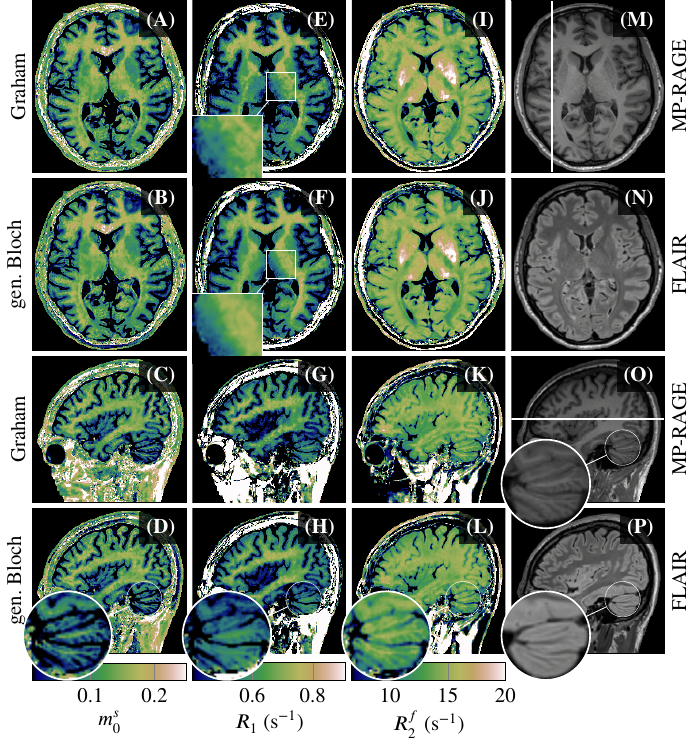}
	\else
		\input{Figures/In_Vivo_Control1.tex}
	\fi
	\vspace{-0.5cm}
	\caption{Size of the semi-solid spin pool $m_0^s$ (normalized by $m_0^s + m_0^f = 1$), the apparent longitudinal relaxation rate $R_1=1/T_1$, and the transversal relaxation rate of the free pool $R_2^f=1/T_2^f$. The depicted maps are part of a 3D whole-brain dataset of a healthy volunteer. While we fitted an 8-parameter model (Eq.~\eqref{eq:Bloch_McConnell}), we show here only the biophysical parameters for which the sequence was optimized.
		The rectangular magnifications in (E, F) highlight subtle differences between the estimates from Graham's spectral and the generalized Bloch model.
		The clinical contrasts in (M--P) are depicted for comparison and the spherical magnifications provide a visual resolution comparison to the clinical scans, which were measured with 1.0mm isotropic resolution.
		The white lines indicate the position of the sagittal and axial slices, respectively.
	}
	\label{fig:InVivo_Control1}
\end{figure*}

\begin{table*}[tbp]
	\centering
	\begin{tabular}{l|c|c|c|c|c|c}
		             & \multicolumn{3}{c|}{proposed hybrid-state approach} & \multicolumn{3}{c}{literature values}                                                                                                                               \\
		             & $m_0^s$ (\%)                                        & $T_1$ (s)                             & $T_2^f$ (ms)   & $m_0^s$ (\%)                        & $T_1$ (s)                            & $T_2^f$ (ms)                  \\ \hline
		entire WM    & $13.5 \pm 2.8$                                      & $1.52 \pm 0.16$                       & $70.1 \pm 5.5$ & $13.9  \pm 2.8$\cite{Stanisz.2005}  & $1.084 \pm 0.045$\cite{Stanisz.2005} & $69 \pm 3$\cite{Stanisz.2005} \\
		anterior CC  & $18.1 \pm 4.3$                                      & $1.47 \pm 0.16$                       & $66.1 \pm 6.3$ & $16.32 \pm 0.15$\cite{Yarnykh.2012} &                                      &                               \\
		posterior CC & $15.2 \pm 4.1$                                      & $1.52 \pm 0.26$                       & $72.1 \pm 7.6$ & $16.07 \pm 0.40$\cite{Yarnykh.2012} &                                      &                               \\
		cortical GM  & $ 6.6 \pm 3.0$                                      & $2.13 \pm 0.30$                       & $77 \pm 11$    & $ 5.0  \pm 0.5$\cite{Stanisz.2005}  & $1.820 \pm 0.11$\cite{Stanisz.2005}  & $99 \pm 7$\cite{Stanisz.2005} \\
		Caudate      & $ 7.9 \pm 2.6$                                      & $1.81 \pm 0.13$                       & $66.2 \pm 4.3$ & $ 6.89 \pm 0.13$\cite{Yarnykh.2012} &                                      &                               \\
		Putamen      & $ 9.1 \pm 2.0$                                      & $1.69 \pm 0.11$                       & $61.1 \pm 5.7$ & $ 7.22 \pm 0.01$\cite{Yarnykh.2012} &                                      &                               \\
		Pallidum     & $13.4 \pm 1.9$                                      & $1.407\pm 0.067$                      & $54.8 \pm 7.0$ &                                     &                                      &                               \\
		Thalamus     & $12.1 \pm 2.6$                                      & $1.73 \pm 0.18$                       & $64.7 \pm 5.8$ & $ 8.61 \pm 0.64$\cite{Yarnykh.2012} &                                      &                               \\
	\end{tabular}
	\caption{
		Region of interest (ROI) analysis in healthy controls. The ROIs were determined by segmenting MP-RAGE images with the \textit{FreeSurfer} software after co-registering it to the qMT maps. The values represent the mean and standard deviation of all voxels from 2 healthy subjects (Fig.~\ref{fig:InVivo_Control1} and Supporting Fig.~\ref{fig:InVivo_Control2}).
	}
	\label{tab:ROImean}
\end{table*}

\subsubsection{Healthy volunteers} \label{sec:ResultsHealthy}
Fig.~\ref{fig:InVivo_Control1} demonstrates the feasibility of whole-brain qMT imaging in a healthy volunteer, with 1.24mm effective isotropic resolution and a 12.6-minute scan time.
We observe overall good image quality in the target parameters $m_0^s$, $R_1$, and $R_2^f$. However, we do observe a slight increase of $m_0^s$ in the frontal lobe, which is likely a residual $B_1^+$ artifact.
We are omitting the depiction of $R_\text{x}$ and $T_2^s$ estimates, as their large CRB does not allow for meaningful estimates.

A comparison of the resolution to an MP-RAGE with 1.0mm isotropic resolution (circular magnifications) reveals subtle blurring in the qMT maps, which could, in part, stem from the reduced k-space coverage of the koosh-ball trajectory, as reflected by the effective resolution of 1.24mm. Additional blurring might stem from the undersampled reconstruction in combination with the regularized image reconstruction.
A more detailed analysis of the resolution and a comparison to other methods is, however, beyond the scope of this paper.

A region of interest (ROI) analysis (Tab.~\ref{tab:ROImean}) reveals good agreement of the $m_0^s$ estimates in white and cortical gray matter (WM, GM) between the proposed method and reference scans with the off-resonant saturation method.\cite{Stanisz.2005,Yarnykh.2012}
The $T_2^f$ estimates in WM also align well with literature values, though the difference slightly exceeds one standard deviation in cortical GM, possibly due to partial volume effects and imperfections of registration and segmentation.
By contrast, the apparent $T_1$ is overestimated with our method as compared to the here-shown reference measurement that utilizes an inversion-recovery experiment, paired with a mono-exponential fit.\cite{Stanisz.2005}
Beyond these two large ROIs, we also observe known patterns: e.g., $m_0^s$ is elevated in the genu of the corpus callosum (Fig.~\ref{fig:InVivo_Control1}(C), Tab.~\ref{tab:ROImean}) and $T_2^f$ shortened in the putamen, globus pallidus, and the thalamus, where iron depositions\cite{Yeung1994} facilitate fast transversal relaxation (Fig.~\ref{fig:InVivo_Control1}(J)).

A comparison between Graham's spectral and the generalized Bloch model in healthy brain tissue reveals overall good agreement, but we observe subtle differences in the $R_1$ maps: the fits with Graham's model exhibit noise-like patterns (magnification in (E)), which are less pronounced when fitting the generalized Bloch model (F). While this difference is subtle, it is present for all four volunteer subjects (cf. Sup. Figs.~\ref{fig:InVivo_Control2}--\ref{fig:InVivo_Patient2}).

\begin{figure}[tb]
	\centering
	\ifOL
		\includegraphics[]{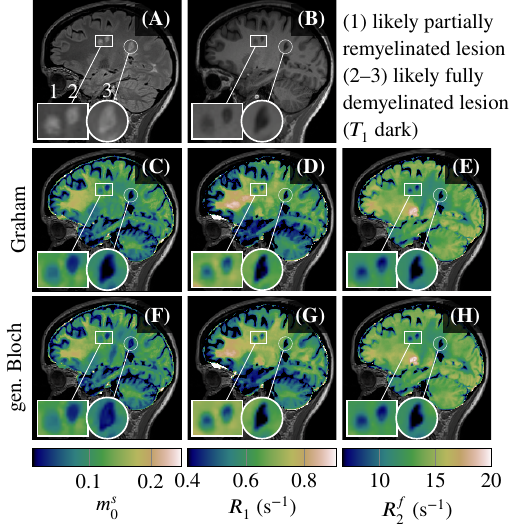}
	\else
		\input{Figures/In_Vivo_MS1.tex}
	\fi
	\vspace{-0.9cm}
	\caption{\textbf{(A)} A FLAIR, \textbf{(B)} an MP-RAGE, and \textbf{(C)}--\textbf{(H)} qMT maps of a multiple sclerosis subject.
		\textbf{(C)}--\textbf{(E)} were fitted with Graham's spectral model and \textbf{(F)}--\textbf{(H)} were fitted with the generalized Bloch model.
		The magnifications highlight three lesions with the same hyperintense appearance on FLAIR but show distinct characteristics on the quantitative MT maps.
	}
	\label{fig:InVivo_Patient1}
\end{figure}

\subsubsection{Participants with multiple sclerosis}
Fig.~\ref{fig:InVivo_Patient1} shows a sagittal slice of one of the MS subject scans, which includes three lesions.
In the lesions, $m_0^s$, $R_1$, and $R_2^f$ are reduced as expected.
Fig.~\ref{fig:InVivo_Patient1} further confirms several findings made in Fig.~\ref{fig:InVivo_Control1}: we observe overall good image quality in the qMT maps, but we also observe a slight and, presumed, $B_1^+$-related anterior-posterior gradient in the maps.
In terms of resolution, we observe a sharp delineation of lesion 3 and the NAWM in the qMT maps, similar to the clinical scans that have the same nominal resolution.
By contrast, lesions 1 and 2 show more heterogeneity in $R_1$ and $R_2^f$.
However, we do observe blurring in the qMT maps in comparison to the clinical scans, e.g., in the cerebellum, that might exceed the blurring of the data shown in Fig.~\ref{fig:InVivo_Control1}. The increased blurring is likely due to subject motion.

\section{Discussion}
\subsection{Scan efficiency} \label{sec:DiscussionScanEfficiency}
One aim of this paper is to identify an avenue toward a quantitative magnetization transfer imaging approach with few constraints on model parameters, yet, with high spatial resolution and short scan times. The basis for this scan efficiency is a \textit{hybrid state}\cite{Asslander2019a} of the free pool, which combines the encoding efficiency of the transient state\cite{Schmitt2004, Ehses2013, Ma2013, Asslander2020} with the steady state's robustness to magnetic field inhomogeneities, especially to inhomogeneous lineshape broadening.\cite{Carr1958,Scheffler2003}

The hybrid-state pulse sequence has a CRB comparable to or even slightly better than established approaches (Tab.~\ref{tab:CRB_seq}).
We note, though, that the SNR in the parameter maps depends on several sequence details---chiefly the readout bandwidth---which was assumed equal for all sequences.
The key benefit of the hybrid state is that it maintains desirable CRB values while performing an 8-parameter fit of the full Henkelman two-pool model.\cite{Henkelman1993}
By contrast, SIR,\cite{Cronin2020} single-point saturation-based qMT,\cite{Yarnykh.2016} and the recently proposed MT-MRF approach\cite{Hilbert.2019} only fit 3, 4, and 5 parameters, respectively, while constraining other parameters to literature values or values from external calibration scans.
In contrast to the first two reference methods, our approach additionally provides an $R_2^f$ map, which correlates to the iron content\cite{Vymazal.1999, Gossuin.2000, Gossuin.2002} and, thus, provides additional value for a comprehensive analysis of brain tissue.

For sampling efficiency, we implemented the hybrid-state sequence with a data acquisition scheme that is based on the concept of MR Fingerprinting:\cite{Ma2013} we sparsely sample different parts of k-space in each time frame and use temporal subspace modeling to reconstruct parameter maps from this data.\cite{McGivney.2014,Tamir2017,Asslander2018}
We further utilize parallel imaging\cite{Sodickson1997,Pruessmann.2001,Uecker.2014} and locally low-rank\cite{Trzasko2011} flavored compressed sensing\cite{Lustig2007} to reduce noise and undersampling artifacts.

Jointly, these advances allow us to extract three parameter maps of the whole brain with an effective resolution of 1.24mm isotropic from 12.6 minutes worth of data while considering 8 parameters as unknown (cf. Figs.~\ref{fig:InVivo_Control1} and \ref{fig:InVivo_Patient1}).
Note, there is a slight decrease of this effective resolution in comparison to the nominal 1.0mm as we sample only the inner sphere of the 1.0mm k-space cube, similar to the common \textit{elliptical scanning} approach.
A more detailed analysis of the effective spatial resolution would need to consider the undersampling and the regularized image reconstruction.
Since we use a non-linear image reconstruction, the effective resolution is also a function of the imaged object.
While this will be a subject of future research, we limit the resolution analysis here to a visual comparison to the clinical MPRAGE and FLAIR scans with the same nominal resolution (Fig.~\ref{fig:InVivo_Control1}).

\subsection{Model accuracy} \label{sec:Discussion_ModelAccuracy}
The second aim of this work was to improve model accuracy.
As discussed in Appendix \ref{sec:unmodeled_effects}, modeling errors can bias the parameter estimates. This bias depends on the pulse sequence and protocol settings, and we highlight three ways the current work improves model accuracy compared to prior methods.

First, our approach does not constrain any parameters of the two-pool model besides the common $R_1^s=R_1^f$ assumption.
By contrast, SIR heuristically sets $T_2^s$ to literature values and neglects the $B_1^+$-dependence of the semi-solid pool's saturation.
Yarnykh's single-point-saturation qMT approach heuristically sets the ratio $T_1/T_2^f$, $T_2^s$, and $R_\text{x}$ to literature values and relies on external estimates of $B_0$ and $B_1^+$.
Both approaches were extensively validated, and Yarnykh carefully optimized the pulse sequence of the single-point-saturation qMT approach for a minimal bias despite several model constraints and successfully reduced the bias to 8-8.5\%, averaged over the entire brain.\cite{Yarnykh.2012}
Nevertheless, pathological tissue changes might entail a disproportionate bias.
For example, iron accumulation shortens $T_2^f$ more than it shortens $T_1$,\cite{Vymazal.1999, Gossuin.2000, Gossuin.2002} as observed here for the globus pallidus in Fig.~\ref{fig:InVivo_Control1}(I, J).
Consequent alteration of the ratio $T_1$/$T_2^f$ entails the risk of a larger than average bias in the $m_0^s$ estimate.
By contrast, an unconstrained approach to the estimation of the two-pool model's parameters reduces the risk of unpredictable bias in the presence of pathology. Our approach captures variations of $T_1$ and $T_2^f$, which promises an improved disentanglement of myelin and iron.

Second, our approach does not rely on the accuracy of external calibration scans.
Such calibration scans can themselves be biased, and this error can propagate to the qMT estimates.
Many efforts were made to reduce such biases, e.g., by carefully analyzing and optimizing the RF-spoiling.\cite{Preibisch2009, Yarnykh.2010, Heule.2016}
However, residual sources of bias remain, such as MT itself.
For the popular actual flip angle\cite{Yarnykh.2007} and variable flip angle\cite{Fram.1987} methods, we estimated MT biases of approximately 3\% and 10\% in $B_1^+$ and $T_1$, respectively (cf. Appendix \ref{sec:data_availability}).
Our approach avoids this issue by forgoing calibration scans and fitting the full qMT model, including $B_1^+$ and $T_1$, to the data, assuring that MT does not bias their estimates.

Last, we use the generalized Bloch model.
Most approaches in the literature use Graham's single frequency approximation (Eq.~(8) in Ref. \citen{Graham1997}), where the saturation of the semi-solid spin pool is proportional to the lineshape at the pulse's central frequency.
This approach is problematic on resonance where the super-Lorentzian diverges.\cite{Morrison1995a,Gloor2008}
We note that this divergence is not observed in experiments due to further (dipolar) interactions that are not described by the super-Lorentzian lineshape.\cite{Wennerstrom1973,Pampel.2015}
One can circumvent this problem with Graham's spectral model, in which the lineshape is multiplied by the pulse's power spectral density and then integrated (Eq.~(4) in Ref. \citen{Graham1997}).
In Ref. \citen{Asslander.2022u6f}, we showed that Graham's spectral model approximates the saturation of the semi-solid spin pool well for long RF pulses.
Pulse durations between 100 and 500$\upmu$s (as used here) are in the range where we start to observe a moderate amount of deviations from the generalized Bloch model (cf. rightmost column in Fig.~3 of Ref. \citen{Asslander.2022u6f}).
The present work is in line with these findings since we observe only subtle differences between in vivo estimates made with either model (rectangular magnifications in Figs.~\ref{fig:InVivo_Control1}).
The similarity between the parameter maps, for the given pulse sequence, confirms that the theoretically well-founded generalized Bloch model is in line with prior literature, and promises more accuracy when pushing the limit to shorter or stronger RF pulses.

\subsection{Estimating $B_0$ and $B_1^+$ vs. accounting for their variability} \label{sec:DiscussionB0B1}
In Ref. \citen{Nataraj.2018}, Nataraj et al. pointed out that neural networks estimate each parameter irrespective of the other parameters, which implies that precise estimates of $B_0$ and $B_1^+$ are not required for a precise $B_0$ and $B_1^+$ correction.
Mathematically, we can describe fixing, estimating, and correcting for, e.g., $B_0$ by the related requirements for the signal's derivative wrt. $B_0$.
Fixing the parameter to a nominal value without biasing the other parameters requires a vanishing derivative of the signal wrt. $B_0$ or its orthogonality to the derivatives wrt. all other model parameters.
A precise estimation of $B_0$ requires a low CRB of $B_0$ or, in other words, a large component of the signal's derivative wrt. $B_0$ that is orthogonal to the derivatives wrt. to all other model parameters.\cite{Scharf.1993}
By contrast, a $B_0$-corrected estimate of, e.g., $R_2^f$ requires merely a large component of the signal's derivative wrt. $R_2^f$ that is orthogonal to the derivative wrt. $B_0$.
This comparably loose requirement can be fulfilled in different ways, as is illustrated by the case of $B_0$: the hybrid-state signal is, by design, rather insensitive to $B_0$ within the bSSFP-typical pass-band ($|\omega_z| \lesssim 0.8 \pi / T_\text{R}$),\cite{Carr1958, Scheffler2003, Asslander2017, Asslander2019a} i.e., it has a small derivative wrt. $B_0$.
Under this condition, we can estimate $R_2^f$ without $B_0$ bias, but $B_0$ itself is hard to estimate, and small modeling errors---e.g., partial volume effects---can cause tissue-dependent errors of the $B_0$ estimates (spherical magnification in Fig.~\ref{fig:InVivo_B0}(C)).
In contrast, the hybrid-state signal is sensitive to $B_0$ in the vicinity of the stop-band (rectangular magnification) and in this frequency range, the pulse sequence optimization disentangled the derivatives wrt. $R_2^f$ and $B_0$ in order to maximize the orthogonal component of the former.
Here, the derivative wrt. $B_0$ is substantial so that $B_0$ is accurately and precisely estimated in this frequency range (rectangular magnification).
The same concept applies to $B_1^+$. We do not necessarily expect a good estimate of $B_1^+$ as we did not optimize the pulse sequence for estimating this parameter, and even less so in CSF as we did not optimize the pulse sequence for the estimation of any parameter at such long relaxation times. For this reason, the observed variations in the $B_1^+$ estimates (spherical magnification in Fig.~\ref{fig:InVivo_B0}(D)) are neither surprising nor do they indicate poor $B_1^+$ correction.

\subsection{Open questions and future directions} \label{sec:DiscussionOpenQuestions}
Based on the current data, we see two remaining challenges:
First, the qMT maps exhibit an anterior-posterior gradient which is likely a residual $B_1^+$ artifact.
In future work, we will address this issue with RF patterns optimized for $B_1^+$ uncertainty.
In addition, we plan to incorporate phase cycling\cite{Bangerter.2004, Elliott.2007, Benkert2015a, Shcherbakova2018, Kobzar2019} to remove the banding artifact above the anterior skull base (cf. magnifications Fig.~\ref{fig:InVivo_B0}(B) and (C) and Supporting Fig.~\ref{fig:InVivo_Control1_sinus_gBloch}).

Second, our phantom validation (Figs. \ref{fig:Phantom}, \ref{fig:Phantom_m0s_linear_fit}) revealed a slight mismatch between the parameter estimates based on the hybrid-state data and the SIR reference scans, particularly for doped samples (Fig.~\ref{fig:Phantom}).
Further, the $m_0^s$ estimates also vary with the MnCl\textsubscript{2} doping (Fig.~\ref{fig:Phantom}), which is likely spurious.
In WM and GM tissue, we observe a mismatch between the apparent $R_1$ or $T_1$ estimates with the proposed approach and a reference NMR measurement of tissue samples, measured with an inversion-recovery experiment and fitted with a mono-exponential function.\cite{Stanisz.2005}
These mismatches substantially decrease when dropping the assumption $R_1^f = R_1^s$.\cite{Flassbeck.2022,Asslander.2023} While this assumption is consistent with most literature,\cite{Henkelman1993, Morrison1995a, Stanisz1999, Gochberg.2003, Gochberg2007, Dortch.2011,Yarnykh.2012} more recent studies have suggested that $R_1^s$ is substantially faster.\cite{Helms2009, Gelderen2016, Manning2021, Samsonov2021}
The lack of clarity for $R_1^s$ mainly stems from the notorious difficulty of separating this parameter from other qMT parameters. Refs. \citen{Helms2009, Gelderen2016, Manning2021, Samsonov2021} performed elaborate and carefully executed ex vivo NMR experiments or in vivo studies that entailed a separate fit of $R_1^f$ and $R_1^s$ to large ROIs, averaged over multiple subjects.
Our ongoing work builds on the encoding power of the here-proposed approach to remove this last constraint. We seek to investigate whether the hybrid state provides an avenue for a voxel-by-voxel fit of separate $R_1^f$ and $R_1^s$ and if this separation solves the described mismatches.

Additional research directions include more efficient k-space trajectories and joint consideration of the flip angle optimization and spatial encoding. The present framework resulted in segments with zero flip angle, which appears optimal from a CRB perspective, but might be suboptimal in terms of k-space encoding efficiency.
Future work will also include a careful analysis of the stability of neural-network-based parameter fitting and the adaption of neural network approaches with built-in uncertainty quantification.\cite{Abdar.2020,Edupuganti.2021,Lambert.2022}
Further, we will build on our analysis of the network's bias\cite{Zhang.2022} and modify the cost function to explicitly target and reduce residual biases.

\section{Conclusion}
Over the last few decades, several efficient quantitative magnetization transfer approaches have been proposed and thoroughly studied.\cite{Yarnykh.2012, Yarnykh.2016, Gochberg.2003, Gochberg2007, Dortch.2011, Cronin2020, Gloor2008, Hilbert.2019}
Our work builds on these advances, and we have demonstrated that the hybrid state, paired with the generalized Bloch model, provides a promising avenue toward rapid qMT imaging without or with fewer constraints on the qMT model.

\appendix
\section*{Acknowledgements}
We thank Daniel Gochberg and Richard Dortch for fruitful discussions and support with the selective inversion-recovery simulation.

\section{Data availability statement} \label{sec:data_availability}
In order to promote reproducibility, we uploaded the code repository \url{https://github.com/JakobAsslaender/HSFP_qMT_SupportingNotebooks}, where we provide Jupyter notebooks that reproduce
\begin{itemize}
	\setlength{\parskip}{0pt}
	\setlength{\itemsep}{0pt plus 0pt}
	\item the pulse sequence optimizations (Fig.~\ref{fig:MT_Spin_Dynamics} and Supporting Figs.~\ref{fig:Spin_dynamics_Inv_varyFA_constTRF}-\ref{fig:Spin_dynamics_noInv_varyFA_varyTRF})
	\item the CRB analysis in Tab.~\ref{tab:CRB_seq}
	\item the bias estimation of the actual flip angle $B_1^+$-mapping and the variable flip angle $T_1$-mapping method (cf. Section \ref{sec:Discussion_ModelAccuracy})
\end{itemize}

The pulse sequence optimization utilizes our toolbox MRIgeneralizedBloch.jl, whose source code is published on \url{https://github.com/JakobAsslaender/MRIgeneralizedBloch.jl}.
For the here presented optimizations, we used v0.8.0 (DOI:10.5281/zenodo.7433494).
All supplied code is written in the open source language \href{https://julialang.org}{Julia}, and we registered the package ``MRIgeneralizedBloch.jl" with Julia's package manager. The package documentation and tutorials can be found on
\url{https://jakobasslaender.github.io/MRIgeneralizedBloch.jl/v0.8.0/}. The tutorials render the code in HTML format with interactive figures and provide links to Jupyter notebooks that can be launched in \textit{binder}, enabling interactive learning in a browser without any local installations.

\section{Effects not captured by the proposed model} \label{sec:unmodeled_effects}
Given the complexity of the spin dynamics in biological tissue, the proposed model is naturally an approximation that neglects numerous processes.
Unmodeled physical processes can bias the estimated biophysical processes, and this bias often depends on the pulse sequence and sequence settings.
Therefore, we discuss in the following those unmodeled processes that have, to our understanding, the biggest effect on the signal, and we provide a rationale on why we can neglect them for experiments with our pulse sequence.

\subsection{Diffusion}
In order to assess signal variations caused by the diffusion of water molecules in spatially varying magnetic fields, it is helpful to distinguish between different sources of field variations.

First, we consider externally applied imaging gradients.
In spoiled pulse sequences, gradient pulses that are spread over many $T_\text{R}$s act in concert towards the formation of echoes.\cite{Hennig1991,Weigel.2015}
From a diffusion perspective, this results in comparably large effective $b$-values.
Here, we mitigate this problem by balancing all gradient moments in each $T_\text{R}$, similar to balanced SSFP sequences.\cite{Carr1958}
This results in small effective $b$-values that foremost depend on the gradient moments and, thus, on the spatial resolution.
At clinically typical spatial resolutions in the order of 1mm, we can neglect diffusion, as demonstrated by B\"ar et al.\cite{Bar2015}

\newpage
Second, we consider field variations caused by the tissue micro\-structure, such as the vasculature.\cite{Kiselev1998, Kiselev2018} Miller and Jezzard\cite{Miller.2008} analyzed their effect on the signal in balanced SSFP sequences. They observed a reduction of the apparent $R_2^f$ that depends on $T_\text{R}$. For the here-used $T_\text{R} = 3.5 \text{ms}$, $R_2^f$ is reduced by approximately 1\%. Thus, we can neglect diffusion in the field inhomogeneities caused by the vasculature when using short repetition times.

\vspace{-4cm}
\subsection{Dipolar order}
In contrast to the free pool, the dipolar Hamiltonian of the semi-solid pool does not average to zero.\cite{Yeung1994,Morrison1995}
In a many-spin system, one can approximate the effect of a dipolar Hamiltonian with Provotorov's theory,\cite{Provotorov1962} which models the dipolar order as a separate spin pool.
In this model, RF-irradiation facilitates magnetization transfer between the \textit{Zeeman pool} and the \textit{dipolar pool}.
This exchange, however, vanishes if the RF pulses are on-resonant or have a symmetric spectral response around the resonance frequency.\cite{Morrison1995, Varma2015, Malik.2020}
Precisely this difference between symmetric and asymmetric RF-irradiation is utilized by \textit{inhomogeneous MT}\cite{Varma2015} or \textit{dipolar MT}\cite{Manning2017} imaging.

The RF pulses we use for this work are all on-resonant; thus, we neglect the dipolar order. However, we note that this, strictly speaking, cannot be neglected for off-resonant saturation MT,\cite{Wolff1989, Henkelman1993, Morrison1995, Yarnykh.2012, Yarnykh.2016} where the dipolar order can reduce the MT saturation.\cite{Soustelle.2023}
Hence, estimates of $m_0^s$ and $T_1$ with our approach might be slightly larger than the literature values estimated with off-resonant RF-saturation while neglecting the dipolar order.\cite{Soustelle.2023}

\vspace{-4cm}
\section{Graham's spectral model} \label{sec:AppendixGraham}
In order to provide a comparison to the generalized Bloch model (Eq.~\ref{eq:Bloch_McConnell}), we show the Bloch-McConnell equation that incorporates Graham's spectral model:
\begin{widetext}
	\begin{equation}
		\partial_t \begin{pmatrix} x^f \\ y^f \\ z^f \\ z^s \\ 1 \end{pmatrix} = \begin{pmatrix}
			-R_2^f    & -\omega_z & \omega_y                    & 0                                                                         & 0           \\
			\omega_z  & -R_2^f    & 0                           & 0                                                                         & 0           \\
			-\omega_y & 0         & -R_1^f - R_{\text{x}} m_0^s & R_{\text{x}} m_0^f                                                        & m_0^f R_1^f \\
			0         & 0         & R_{\text{x}} m_0^s          & -R_{\text{RF}}(R_2^s, \alpha, T_{\text{RF}}) - R_1^s - R_{\text{x}} m_0^f & m_0^s R_1^s \\
			0         & 0         & 0                           & 0                                                                         & 0
		\end{pmatrix} \begin{pmatrix} x^f \\ y^f \\ z^f \\ z^s \\ 1 \end{pmatrix}.
		\label{eq:Bloch_McConnell_Graham}
	\end{equation}
\end{widetext}
The MT-effect is here captured by $R_{\text{RF}}(R_2^s, \alpha, T_{\text{RF}})$ that is calculated with Eq.~(4) in Ref. \citen{Graham1997} and describes an exponential saturation of the semi-solid spin pool's $z^s$-magnetization.
For this reason, Graham's model does not require modeling the $x^s$ component.
More details on the definition of $R_{\text{RF}}$ and the differences to the generalized Bloch model can be found in Ref. \citen{Asslander.2022u6f}.

\bibliography{library}%
\makeatletter\@input{SupportingInformation_aux.tex}\makeatother
\end{document}


\title{\Papertitle}

\author[1,2]{Jakob Assl\"ander*}{\orcid{0000-0003-2288-038X}}
\author[3]{Cem Gultekin}{\orcid{0000-0002-2562-371X}}
\author[1,2,4]{Andrew Mao}{\orcid{0000-0002-1398-0699}}
\author[1,2]{Xiaoxia Zhang}{\orcid{0000-0003-3620-6994}}
\author[5]{Quentin Duchemin}{\orcid{0000-0003-3636-3770}}
\author[6]{Kangning Liu}{\orcid{0000-0002-0187-4602}}
\author[7]{Robert W Charlson}{}
\author[1]{Timothy Shepherd}{\orcid{0000-0003-0232-1636}}
\author[3,6]{Carlos Fernandez-Granda}{\orcid{0000-0001-7039-8606}}
\author[1,2]{Sebastian Flassbeck}{\orcid{0000-0003-0865-9021}}

\authormark{Jakob Assl\"ander \textsc{et al}}

\address[1]{\orgdiv{Center for Biomedical Imaging, Dept. of Radiology}, \orgname{NYU School of Medicine}, \orgaddress{\state{NY}, \country{USA}}}
\address[2]{\orgdiv{Center for Advanced Imaging Innovation and Research (CAI\textsuperscript{2}R), Dept. of Radiology}, \orgname{NYU School of Medicine}, \orgaddress{\state{NY}, \country{USA}}}
\address[3]{\orgdiv{Courant Institute of Mathematical Sciences}, \orgname{New York University}, \orgaddress{\state{NY}, \country{USA}}}
\address[4]{\orgdiv{Vilcek Institute of Graduate Biomedical Sciences}, \orgname{NYU School of Medicine}, \orgaddress{\state{NY}, \country{USA}}}
\address[5]{\orgdiv{Laboratoire d’analyse et de mathématiques appliquées}, \orgname{Université Gustave Eiffel}, \orgaddress{\country{France}}}
\address[6]{\orgdiv{Center for Data Science}, \orgname{New York University}, \orgaddress{\state{NY}, \country{USA}}}
\address[7]{\orgdiv{Department of Neurology}, \orgname{NYU School of Medicine}, \orgaddress{\state{NY}, \country{USA}}}



\abstract[]{}
\maketitle

\section{Hybrid state of the free pool}
Similar to Ref. \citen{Asslander2019a}, we can describe the spin dynamics of the free pool in spherical coordinates, which traps its entire dynamics in a single dimension (the radial dimension $r^f$). This reduces Eq.~\eqref{eq:Bloch_McConnell} to
\begin{widetext}
	\begin{equation}
		\partial_t \begin{pmatrix} r^f \\ x^s \\ z^s \\ 1 \end{pmatrix} = \begin{pmatrix}
			-R_1^f \cos^2 \vartheta^f -R_2^f \sin^2 \vartheta^f - m_0^s R_{\text{x}} \cos^2 \vartheta^f & 0                                        & m_0^f R_{\text{x}} \cos \vartheta^f & m_0^f R_1^f \cos \vartheta^f \\
			0                                                                                           & -R_2^{s,l}(R_2^s, \alpha, T_{\text{RF}}) & \omega_y                            & 0                            \\
			m_0^s R_{\text{x}} \cos \vartheta^f                                                         & -\omega_y                                & -R_1^s - m_0^f R_{\text{x}}         & m_0^s R_1^s                  \\
			0                                                                                           & 0                                        & 0                                   & 0
		\end{pmatrix} \begin{pmatrix} r^f \\ x^s \\ z^s \\ 1 \end{pmatrix},
		\label{eq:Hybrid_State_Model_gBloch}
	\end{equation}
\end{widetext}

where $\vartheta^f$ describes the polar angle or the angle between the $z$-axis and the magnetization of the free pool.
On resonance, the polar angle is half the flip angle ($\vartheta^f = \alpha/2$).
This notation exposes the control of the spin dynamics in such a coupled spin system:
As discussed in Ref. \citen{Asslander2019a}, the polar angle controls the relaxation processes.
As discussed in the following, it also controls the magnetization transfer between the two pools.
For a more detailed analysis of the magnetization transfer between the two pools, it is essential to note that the semi-solid pool does not establish a hybrid state due to its fast $T_2^s$ relaxation.
Instead, it aligns with the $z$-axis in between RF pulses.
Consequently, $\vartheta^f$ describes the angle between the two pools and plays a crucial role in their exchange of $z$-magnetization.
Assuming, as a Gedankenexperiment, $\omega_y=0$ and $R_1^s=0$ simplifies the third row of Eq.~\eqref{eq:Hybrid_State_Model_gBloch} to
$\partial_t z^s = m_0^s R_{\text{x}} \cos \vartheta^f r^f - m_0^f R_{\text{x}} z^s$.
The first summand indicates that the semi-solid pool can only gain magnetization by exchange if $\vartheta^f$ is small, i.e., if there is a substantial $z^f = r^f \cos \vartheta^f$ component to draw magnetization from.
In contrast, it always loses magnetization due to the second summand that does not depend on $\vartheta^f$.

More interestingly, if we repeat the same process for the first row, we obtain the differential equation $\partial_t r^f = - m_0^s R_{\text{x}} \cos^2 \vartheta^f r^f  + m_0^f R_{\text{x}} \cos \vartheta^f z^s$, which highlights that the free pool experiences no exchange with the semi-solid pool when approaching $\vartheta^f = \pi/2$.
An examination of the spin dynamics at $\vartheta^f = \pi/2$ can resolve this seeming paradox. In this case, we apply a $\pi$-pulse in each $T_\text{R}$ that flips any existing $z^f$-magnetization from the positive $z$-axis to the negative one and vice versa. In linear approximation, any $z^f$-magnetization that results from exchange in one $T_\text{R}$ is, thus, canceled out by the magnetization gained in the next. This mechanism breaks the conservation of magnetization that is otherwise intrinsic to transfers of $z$-magnetization between two pools.

\section{Legacy Pulse-Sequence Optimization}
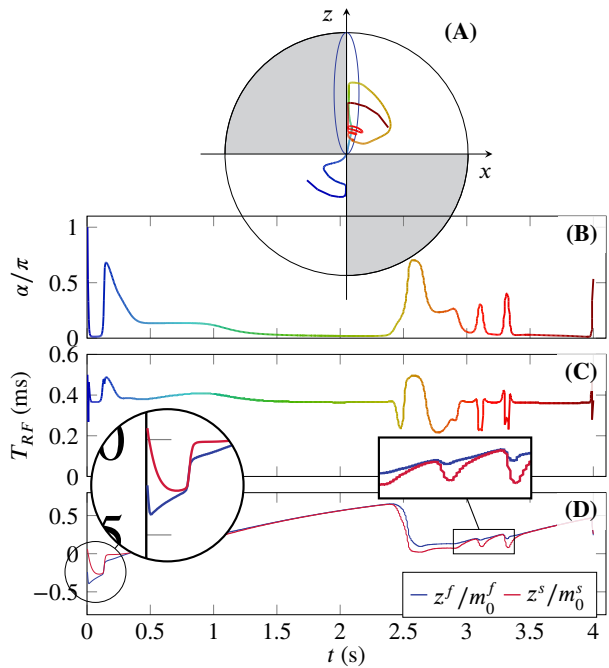
\begin{figure}[tbp]
	\centering
	\begin{tikzpicture}[scale = .88]
    \def\TOne{1.6};
    \def\TTwo{0.067};

    \begin{axis}[
            width=\linewidth*0.8,
            axis equal image,
            axis lines=center,
            xmin = -1.2, xmax = 1.2,
            ymin = -1.2, ymax = 1.2,
            xtick=\empty, ytick=\empty, ztick=\empty,
            xlabel={$x$}, ylabel={$z$},
            every axis x label/.style={at={(ticklabel cs:0.975)}, anchor=north},
            every axis y label/.style={at={(ticklabel cs:0.975)}, anchor=east},
            colormap name = mybluered,
            point meta min=0,
            point meta max= 4,
            name=Bloch,
            clip=false,
            axis on top=true,
        ]

        \filldraw[fill=lightgray] (axis cs: 0,0) -- (axis cs: 0, 1) arc [start angle=  90, end angle= 180, radius={(1)}];
        \filldraw[fill=lightgray] (axis cs: 0,0) -- (axis cs: 0,-1) arc [start angle=270, end angle=360, radius={(1)}];

        \addplot[no marks, domain=0:360,samples=60, forget plot] ({sin(x)}, {cos(x)});

        \addplot[domain=0:1,samples=1000, UKLblue, forget plot] ({-sqrt(\TTwo * (1/(4*\TOne) - (x - .5)^2/\TOne))}, {x});
        \addplot[domain=0:1,samples=1000, UKLblue, forget plot] ({  sqrt(\TTwo * (1/(4*\TOne) - (x - .5)^2/\TOne))}, {x});

        \addplot[mesh, point meta=explicit, thick, forget plot]table[x=xf, y=zf, meta=t_s]{Figures/Spin_dynamics_v3p2.txt};

        \node[right, inner sep=0mm,minimum size=5mm] at (axis cs:  .8,1) {\textbf{(A)}};
    \end{axis}

    \begin{axis}[
            width=\linewidth*0.9,
            height=\textheight*0.075,
            scale only axis,
            xmin=0,
            xmax=4.1,
            ymin=0,
            xticklabel=\empty,
            ylabel={$\alpha/\pi$},
            colormap name = mybluered,
            point meta min=0,
            point meta max= 4,
            name=theta,
            at=(Bloch.south),
            anchor=north,
            yshift = 1.25cm,
        ]
        \addplot [mesh, thick, point meta=explicit, solid]table[x=t_s, y=alpha_pi, meta=t_s]{Figures/Spin_dynamics_v3p2.txt};

        \node[anchor=north east, fill=white, minimum size=4mm, opacity=0.85, text opacity=1] at (rel axis cs: 1,1) {\textbf{(B)}};
    \end{axis}

    \begin{axis}[
            width=\linewidth*0.9,
            height=\textheight*0.075,
            scale only axis,
            xmin=0,
            xmax=4.1,
            ymin=0,
            ymax=0.6,
            xticklabel=\empty,
            ylabel={$T_{RF}~(\text{ms})$},
            colormap name = mybluered,
            point meta min=0,
            point meta max= 4,
            name=TRF,
            at=(theta.below south east),
            anchor= north east,
        ]
        \addplot [mesh, thick, point meta=explicit, solid]table[x=t_s, y=TRF_ms, meta=t_s]{Figures/Spin_dynamics_v3p2.txt};

        \node[anchor=north east, fill=white, minimum size=4mm, opacity=0.85, text opacity=1] at (rel axis cs: 1,1) {\textbf{(C)}};
    \end{axis}

    \begin{scope}[spy using outlines={magnification=2.5, connect spies}]
        \begin{axis}[
                width=\linewidth*0.9,
                height=\textheight*0.075,
                scale only axis,
                xmin=0,
                xmax=4.1,
                ymin=-.8,
                ymax=.8,
                xlabel={$t~\text{(s)}$},
                xlabel style={yshift=0.15cm},
                name=rz,
                at=(TRF.below south east),
                anchor= north east,
                legend entries = {$z^f/m_0^f$, $z^s/m_0^s$},
                legend pos = south east,
                legend columns=2,
                legend style={
                        xshift = 0.1cm,
                        yshift = -0.1cm,
                        legend image code/.code={\draw[##1,line width=0.6pt] plot coordinates {(0cm,0cm) (0.25cm,0cm)};}
                    },
            ]
            \addplot [color=UKLblue,            solid]table[x=t_s, y=zf_m0f]{Figures/Spin_dynamics_v3p2.txt};
            \addplot [color=UKLred,              solid]table[x=t_s, y=zs_m0s]{Figures/Spin_dynamics_v3p2.txt};

            \coordinate (spypoint_bE) at (axis cs:0,-0.4);
            \coordinate (spyviewer_bE) at (axis cs:0.5,0.6);
            \spy[circle, black,size=2.0cm] on (spypoint_bE) in node [fill=white] at (spyviewer_bE);

            \coordinate (spypoint_sat) at (axis cs:2.7,-0.05);
            \coordinate (spyviewer_sat) at (axis cs:2.5,0.8);
            \spy[rectangle,black,height=0.8cm,width=2.0cm] on (spypoint_sat) in node [fill=white] at (spyviewer_sat);

            \node[anchor=north east, fill=white, opacity=0.85, text opacity=1] at (rel axis cs: 1,1) {\textbf{(D)}};
        \end{axis}
    \end{scope}
\end{tikzpicture}
	\caption{Spin trajectory and the corresponding control that resulted from the legacy optimization are used for the experiments. \textbf{(A)} The dynamics of the free pool on the Bloch sphere with the steady-state ellipse in blue; \textbf{(B)} the flip angle $\alpha$ and \textbf{(C)} the pulse duration $T_\text{RF}$ control the spin dynamics. \textbf{(D)} The normalized magnetization of the two pools. The spherical and rectangular magnifications highlight segments that utilize a bi-exponential inversion- recovery\cite{Gochberg.2003} and saturation,\cite{Gloor2008} respectively, to encode the MT effect.}
	\label{fig:Spin_dynamics_v3p2}
\end{figure}

The experiments in this paper were performed with the pulse sequence shown in the Supporting Fig.~\ref{fig:Spin_dynamics_v3p2}, which resulted from a legacy optimization that did not account for $\omega_z$ and $B_1$ inhomogeneities and was based on the model

\begin{widetext}
	\begin{equation}
		\partial_t \begin{pmatrix} r^f \\ z^s \\ 1 \end{pmatrix} = \begin{pmatrix}
			-R_1 \cos^2 \vartheta^f -R_2^f \sin^2 \vartheta^f - m_0^s R_{\text{x}} \cos^2 \vartheta^f & m_0^f R_{\text{x}} \cos \vartheta^f                                     & m_0^f R_1 \cos \vartheta^f \\
			m_0^s R_{\text{x}} \cos \vartheta^f                                                       & -R_1 - R_{\text{RF}}(R_2^s, \alpha, T_{\text{RF}}) - m_0^f R_{\text{x}} & m_0^s R_1                  \\
			0                                                                                         & 0                                                                       & 0
		\end{pmatrix} \begin{pmatrix} r^f \\ z^s \\ 1 \end{pmatrix}.
		\label{eq:Hybrid_State_Model_Graham}
	\end{equation}
\end{widetext}

This model incorporates Graham's spectral model\cite{Graham1997} and was derived by transforming the free pool's magnetization in Eq.~\eqref{eq:Bloch_McConnell_Graham} to spherical coordinates (cf. Eq.~\eqref{eq:Hybrid_State_Model_gBloch}).

Supporting Tab.~\ref{tab:CRB_v3p2} analyzes the performance of the legacy pulse sequence (Supporting Fig.~\ref{fig:Spin_dynamics_v3p2}).
The leftmost column contains the CRB values that were used for the legacy optimization, i.e., using Eq.~\eqref{eq:Hybrid_State_Model_Graham} and assuming that $\omega_z=0$ and $B_1=1$ are fixed.
The rightmost column contains CRB values of the same pulse sequence but calculated with Eq.~\eqref{eq:Bloch_McConnell} and under the assumption that $\omega_z$ and $B_1$ are fitted to the data, which is the setup we used throughout this paper.
Comparing the two columns reveals slightly higher (worse) CRB values in the latter case (rightmost column).
When comparing these values to our latest optimization (leftmost column of Tab.~\ref{tab:CRB_seq} in the main manuscript), that were performed with Eq.~\eqref{eq:Bloch_McConnell} and under the assumption that $\omega_z$ and $B_1$ are fitted to the data, we find that the latest optimization promises slight improvements in image quality compared to the here-performed experiments.

\begin{table}[htbp]
	\centering
	\newcommand{\g}{\cellcolor[gray]{0.8}}
	\begin{tabular}{l|c|c|c}
		model                                                                   & Eq.~\eqref{eq:Hybrid_State_Model_Graham} & Eq.~\eqref{eq:Bloch_McConnell} & Eq.~\eqref{eq:Bloch_McConnell} \\
		\hline
		$\omega_z$ and $B_1$                                                    & known                                    & known                          & unknown                        \\
		\hline
		$\text{CRB}(m_0^s) \cdot \frac{M_0^2 T}{(m_0^s\sigma)^2}$~[s]           & \g$176$                                  & $368$                          & $368$                          \\
		$\text{CRB}(R_1  ) \cdot \frac{M_0^2 T}{(R_1\sigma)^2}$~[s]             & \g$72$                                   & $68$                           & $68$                           \\
		$\text{CRB}(R_2^f) \cdot \frac{M_0^2 T}{(R_2^f\sigma)^2}$~[s]           & \g$26$                                   & $44$                           & $136$                          \\
		\hline
		$\text{CRB}(M_0       ) \frac{T}{\sigma^2}$~[s]                         & $29.6$                                   & $60$                           & $88$                           \\
		$\text{CRB}(R_\text{x}) \cdot \frac{M_0^2 T}{(R_\text{x}\sigma)^2}$~[s] & $49132$                                  & $10756$                        & $11764$                        \\
		$\text{CRB}(T_2^s     ) \cdot \frac{M_0^2 T}{(T_2^s\sigma)^2}$~[s]      & $6168$                                   & $9092$                         & $12560$                        \\
		$\text{CRB}(\omega_z  )$                                                & $*$                                      & $*$                            & $+\infty$                      \\
		$\text{CRB}(B_1       ) \cdot \frac{M_0^2 T}{(B_1\sigma)^2}$~[s]        & $*$                                      & $*$                            & $40$                           \\
	\end{tabular}
	\caption{Cram\'er-Rao bound (CRB) values of the legacy pulse sequence (Supporting Fig.~\ref{fig:MT_Spin_Dynamics}) that was used for the experiments.
	The table compares the CRB calculated with the legacy model (Eq.~\eqref{eq:Hybrid_State_Model_Graham}), as used for the optimization, to the model that was used for analyzing the data (Eq.~\eqref{eq:Bloch_McConnell}).
	During the optimization, we minimized the CRB of $m_0^s$, $R_1$, and $R_2^f$ (highlighted in gray), assuming that all biophysical parameters will be fitted, but assuming that $\omega_z = 0$ and $B_1 = 1$ are known and fixed.
	All fits shown in this paper, in contrast, also fit $\omega_z$ and $B_1$, and the corresponding CRB values are shown in the rightmost column.
	The displayed CRB values are normalized by the squared value of the parameter, the squared magnetization $M_0$, and the noise variance of the time series in a voxel $\sigma^2$, as well as the scan time $T$, i.e. they reflect the inverse squared signal-to-noise ratio per unit time and for a unit signal noise variance.
	}
	\label{tab:CRB_v3p2}
\end{table}

\begin{figure}[tbp]
	\centering
	\begin{tikzpicture}[scale = .88]
    \def\TOne{1.6};
    \def\TTwo{0.067};

    \begin{axis}[
            width=\linewidth*0.8,
            axis equal image,
            axis lines=center,
            xmin = -1.2, xmax = 1.2,
            ymin = -1.2, ymax = 1.2,
            xtick=\empty, ytick=\empty, ztick=\empty,
            xlabel={$x$}, ylabel={$z$},
            every axis x label/.style={at={(ticklabel cs:0.975)}, anchor=north},
            every axis y label/.style={at={(ticklabel cs:0.975)}, anchor=east},
            colormap name = mybluered,
            point meta min=0,
            point meta max= 4,
            name=Bloch,
            clip=false,
            axis on top=true,
        ]

        \filldraw[fill=lightgray] (axis cs: 0,0) -- (axis cs: 0, 1) arc [start angle=  90, end angle= 180, radius={(1)}];
        \filldraw[fill=lightgray] (axis cs: 0,0) -- (axis cs: 0,-1) arc [start angle=270, end angle=360, radius={(1)}];

        \addplot[no marks, domain=0:360,samples=60, forget plot] ({sin(x)}, {cos(x)});

        \addplot[domain=0:1,samples=1000, UKLblue, forget plot] ({-sqrt(\TTwo * (1/(4*\TOne) - (x - .5)^2/\TOne)}, {x});
        \addplot[domain=0:1,samples=1000, UKLblue, forget plot] ({  sqrt(\TTwo * (1/(4*\TOne) - (x - .5)^2/\TOne)}, {x});

        \addplot[mesh, point meta=explicit, thick, forget plot]table[x=xf, y=zf, meta=t_s]{Figures/Spin_dynamics_Inv_constFA_constTRF.txt};

        \node[right, inner sep=0mm,minimum size=5mm] at (axis cs:  .8,1) {\textbf{(A)}};
    \end{axis}

    \begin{axis}[
            width=\linewidth*0.9,
            height=\textheight*0.075,
            scale only axis,
            xmin=0,
            xmax=4.1,
            ymin=0,
            xticklabel=\empty,
            ylabel={$\alpha/\pi$},
            colormap name = mybluered,
            point meta min=0,
            point meta max= 4,
            name=theta,
            at=(Bloch.south),
            anchor=north,
            yshift = 1.25cm,
        ]
        \addplot [mesh, thick, point meta=explicit, solid]table[x=t_s, y=alpha_pi, meta=t_s]{Figures/Spin_dynamics_Inv_constFA_constTRF.txt};

        \node[anchor=north east, fill=white, minimum size=4mm, opacity=0.85, text opacity=1] at (rel axis cs: 1,1) {\textbf{(B)}};
    \end{axis}

    \begin{axis}[
            width=\linewidth*0.9,
            height=\textheight*0.075,
            scale only axis,
            xmin=0,
            xmax=4.1,
            ymin=0,
            ymax=0.6,
            xticklabel=\empty,
            ylabel={$T_{RF}~(\text{ms})$},
            colormap name = mybluered,
            point meta min=0,
            point meta max= 4,
            name=TRF,
            at=(theta.below south east),
            anchor= north east,
        ]
        \addplot [mesh, thick, point meta=explicit, solid]table[x=t_s, y=TRF_ms, meta=t_s]{Figures/Spin_dynamics_Inv_constFA_constTRF.txt};

        \node[anchor=north east, fill=white, minimum size=4mm, opacity=0.85, text opacity=1] at (rel axis cs: 1,1) {\textbf{(C)}};
    \end{axis}

    \begin{scope}[spy using outlines={magnification=2.5, connect spies}]
        \begin{axis}[
                width=\linewidth*0.9,
                height=\textheight*0.075,
                scale only axis,
                xmin=0,
                xmax=4.1,
                ymin=-.8,
                ymax=.8,
                xlabel={$t~\text{(s)}$},
                xlabel style={yshift=0.15cm},
                name=rz,
                at=(TRF.below south east),
                anchor= north east,
                legend entries = {$z^f/m_0^f$, $z^s/m_0^s$},
                legend pos = south east,
                legend columns=2,
                legend style={
                        xshift = 0.1cm,
                        yshift = -0.1cm,
                        legend image code/.code={\draw[##1,line width=0.6pt] plot coordinates {(0cm,0cm) (0.25cm,0cm)};}
                    },
            ]
            \addplot [color=UKLblue,            solid]table[x=t_s, y=zf_m0f]{Figures/Spin_dynamics_Inv_constFA_constTRF.txt};
            \addplot [color=UKLred,              solid]table[x=t_s, y=zs_m0s]{Figures/Spin_dynamics_Inv_constFA_constTRF.txt};

            \coordinate (spypoint_bE) at (axis cs:0,-0.5);
            \coordinate (spyviewer_bE) at (axis cs:0.6,0.8);
            \spy[circle, black,size=1.5cm] on (spypoint_bE) in node [fill=white] at (spyviewer_bE);


            \node[anchor=north east, fill=white, opacity=0.85, text opacity=1] at (rel axis cs: 1,1) {\textbf{(D)}};
        \end{axis}
    \end{scope}
\end{tikzpicture}
	\caption{Spin trajectory and the corresponding control for a pulse sequence with an inversion pulse and with a constant flip angle and $T_\text{RF}$. \textbf{(A)} The dynamics of the free pool on the Bloch sphere with the steady-state ellipse in blue; \textbf{(B)} the flip angle $\alpha$ and \textbf{(C)} the pulse duration $T_\text{RF}$ control the spin dynamics. \textbf{(D)} The normalized magnetization of the two pools. The spherical magnification highlights a segment that utilizes a bi-exponential inversion-recovery\cite{Gochberg.2003} to encode the MT effect.}
	\label{fig:Spin_dynamics_Inv_constFA_constTRF}
\end{figure}
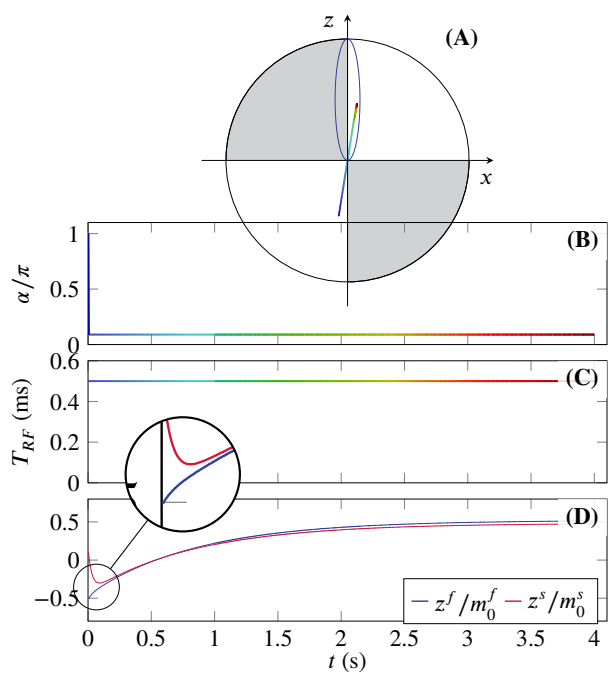

\begin{figure}[tbp]
	\centering
	\begin{tikzpicture}[scale = .88]
    \def\TOne{1.6};
    \def\TTwo{0.067};

    \begin{axis}[
            width=\linewidth*0.8,
            axis equal image,
            axis lines=center,
            xmin = -1.2, xmax = 1.2,
            ymin = -1.2, ymax = 1.2,
            xtick=\empty, ytick=\empty, ztick=\empty,
            xlabel={$x$}, ylabel={$z$},
            every axis x label/.style={at={(ticklabel cs:0.975)}, anchor=north},
            every axis y label/.style={at={(ticklabel cs:0.975)}, anchor=east},
            colormap name = mybluered,
            point meta min=0,
            point meta max= 4,
            name=Bloch,
            clip=false,
            axis on top=true,
        ]

        \filldraw[fill=lightgray] (axis cs: 0,0) -- (axis cs: 0, 1) arc [start angle=  90, end angle= 180, radius={(1)}];
        \filldraw[fill=lightgray] (axis cs: 0,0) -- (axis cs: 0,-1) arc [start angle=270, end angle=360, radius={(1)}];

        \addplot[no marks, domain=0:360,samples=60, forget plot] ({sin(x)}, {cos(x)});

        \addplot[domain=0:1,samples=1000, UKLblue, forget plot] ({-sqrt(\TTwo * (1/(4*\TOne) - (x - .5)^2/\TOne)}, {x});
        \addplot[domain=0:1,samples=1000, UKLblue, forget plot] ({  sqrt(\TTwo * (1/(4*\TOne) - (x - .5)^2/\TOne)}, {x});

        \addplot[mesh, point meta=explicit, thick, forget plot]table[x=xf, y=zf, meta=t_s]{Figures/Spin_dynamics_Inv_varyFA_constTRF.txt};

        \node[right, inner sep=0mm,minimum size=5mm] at (axis cs:  .8,1) {\textbf{(A)}};
    \end{axis}

    \begin{axis}[
            width=\linewidth*0.9,
            height=\textheight*0.075,
            scale only axis,
            xmin=0,
            xmax=4.1,
            ymin=0,
            xticklabel=\empty,
            ylabel={$\alpha/\pi$},
            colormap name = mybluered,
            point meta min=0,
            point meta max= 4,
            name=theta,
            at=(Bloch.south),
            anchor=north,
            yshift = 1.0cm,
        ]
        \addplot [mesh, thick, point meta=explicit, solid]table[x=t_s, y=alpha_pi, meta=t_s]{Figures/Spin_dynamics_Inv_varyFA_constTRF.txt};

        \node[anchor=north east, fill=white, minimum size=4mm, opacity=0.85, text opacity=1] at (rel axis cs: 1,1) {\textbf{(B)}};
    \end{axis}

    \begin{axis}[
            width=\linewidth*0.9,
            height=\textheight*0.075,
            scale only axis,
            xmin=0,
            xmax=4.1,
            ymin=0,
            ymax=0.6,
            xticklabel=\empty,
            ylabel={$T_{RF}~(\text{ms})$},
            colormap name = mybluered,
            point meta min=0,
            point meta max= 4,
            name=TRF,
            at=(theta.below south east),
            anchor= north east,
        ]
        \addplot [mesh, thick, point meta=explicit, solid]table[x=t_s, y=TRF_ms, meta=t_s]{Figures/Spin_dynamics_Inv_varyFA_constTRF.txt};

        \node[anchor=north east, fill=white, minimum size=4mm, opacity=0.85, text opacity=1] at (rel axis cs: 1,1) {\textbf{(C)}};
    \end{axis}

    \begin{scope}[spy using outlines={magnification=2.5, connect spies}]
        \begin{axis}[
                width=\linewidth*0.9,
                height=\textheight*0.075,
                scale only axis,
                xmin=0,
                xmax=4.1,
                ymin=-.8,
                ymax=.8,
                xlabel={$t~\text{(s)}$},
                xlabel style={yshift=0.15cm},
                name=rz,
                at=(TRF.below south east),
                anchor= north east,
                legend entries = {$z^f/m_0^f$, $z^s/m_0^s$},
                legend pos = south east,
                legend columns=2,
                legend style={
                        xshift = 0.1cm,
                        yshift = -0.1cm,
                        legend image code/.code={\draw[##1,line width=0.6pt] plot coordinates {(0cm,0cm) (0.25cm,0cm)};}
                    },
            ]
            \addplot [color=UKLblue,            solid]table[x=t_s, y=zf_m0f]{Figures/Spin_dynamics_Inv_varyFA_constTRF.txt};
            \addplot [color=UKLred,              solid]table[x=t_s, y=zs_m0s]{Figures/Spin_dynamics_Inv_varyFA_constTRF.txt};

            \coordinate (spypoint_bE) at (axis cs:0,-0.375);
            \coordinate (spyviewer_bE) at (axis cs:0.6,0.8);
            \spy[circle, black,size=2cm] on (spypoint_bE) in node [fill=white] at (spyviewer_bE);

            \coordinate (spypoint_sat) at (axis cs:2.25,0.2);
            \coordinate (spyviewer_sat) at (axis cs:2.5,1.2);
            \spy[rectangle,black,height=0.8cm,width=2.0cm] on (spypoint_sat) in node [fill=white] at (spyviewer_sat);

            \node[anchor=north east, fill=white, opacity=0.85, text opacity=1] at (rel axis cs: 1,1) {\textbf{(D)}};
        \end{axis}
    \end{scope}
\end{tikzpicture}
	\caption{Spin trajectory and the corresponding control for an optimized pulse sequence with an inversion pulse, a varying flip angle, and a constant $T_\text{RF}$. \textbf{(A)} The dynamics of the free pool on the Bloch sphere with the steady-state ellipse in blue; \textbf{(B)} the flip angle $\alpha$ and \textbf{(C)} the pulse duration $T_\text{RF}$ control the spin dynamics. \textbf{(D)} The normalized magnetization of the two pools. The spherical and rectangular magnifications highlight segments that utilize a bi-exponential inversion-recovery\cite{Gochberg.2003} and saturation,\cite{Gloor2008} respectively, to encode the MT effect.}
	\label{fig:Spin_dynamics_Inv_varyFA_constTRF}
\end{figure}

\begin{figure}[tbp]
	\centering
	\begin{tikzpicture}[scale = .88]
    \def\TOne{1.6};
    \def\TTwo{0.067};

    \begin{axis}[
            width=\linewidth*0.8,
            axis equal image,
            axis lines=center,
            xmin = -1.2, xmax = 1.2,
            ymin = -1.2, ymax = 1.2,
            xtick=\empty, ytick=\empty, ztick=\empty,
            xlabel={$x$}, ylabel={$z$},
            every axis x label/.style={at={(ticklabel cs:0.975)}, anchor=north},
            every axis y label/.style={at={(ticklabel cs:0.975)}, anchor=east},
            colormap name = mybluered,
            point meta min=0,
            point meta max= 4,
            name=Bloch,
            clip=false,
            axis on top=true,
        ]

        \filldraw[fill=lightgray] (axis cs: 0,0) -- (axis cs: 0, 1) arc [start angle=  90, end angle= 180, radius={(1)}];
        \filldraw[fill=lightgray] (axis cs: 0,0) -- (axis cs: 0,-1) arc [start angle=270, end angle=360, radius={(1)}];

        \addplot[no marks, domain=0:360,samples=60, forget plot] ({sin(x)}, {cos(x)});

        \addplot[domain=0:1,samples=1000, UKLblue, forget plot] ({-sqrt(\TTwo * (1/(4*\TOne) - (x - .5)^2/\TOne)}, {x});
        \addplot[domain=0:1,samples=1000, UKLblue, forget plot] ({  sqrt(\TTwo * (1/(4*\TOne) - (x - .5)^2/\TOne)}, {x});

        \addplot[mesh, point meta=explicit, thick, forget plot]table[x=xf, y=zf, meta=t_s]{Figures/Spin_dynamics_noInv_varyFA_varyTRF.txt};

        \node[right, inner sep=0mm,minimum size=5mm] at (axis cs:  .8,1) {\textbf{(A)}};
    \end{axis}

    \begin{axis}[
            width=\linewidth*0.9,
            height=\textheight*0.075,
            scale only axis,
            xmin=0,
            xmax=4.1,
            ymin=0,
            xticklabel=\empty,
            ylabel={$\alpha/\pi$},
            colormap name = mybluered,
            point meta min=0,
            point meta max= 4,
            name=theta,
            at=(Bloch.south),
            anchor=north,
            yshift = 1.0cm,
        ]
        \addplot [mesh, thick, point meta=explicit, solid]table[x=t_s, y=alpha_pi, meta=t_s]{Figures/Spin_dynamics_noInv_varyFA_varyTRF.txt};

        \node[anchor=north east, fill=white, minimum size=4mm, opacity=0.85, text opacity=1] at (rel axis cs: 1,1) {\textbf{(B)}};
    \end{axis}

    \begin{axis}[
            width=\linewidth*0.9,
            height=\textheight*0.075,
            scale only axis,
            xmin=0,
            xmax=4.1,
            ymin=0,
            ymax=0.6,
            xticklabel=\empty,
            ylabel={$T_{RF}~(\text{ms})$},
            colormap name = mybluered,
            point meta min=0,
            point meta max= 4,
            name=TRF,
            at=(theta.below south east),
            anchor= north east,
        ]
        \addplot [mesh, thick, point meta=explicit, solid]table[x=t_s, y=TRF_ms, meta=t_s]{Figures/Spin_dynamics_noInv_varyFA_varyTRF.txt};

        \node[anchor=north east, fill=white, minimum size=4mm, opacity=0.85, text opacity=1] at (rel axis cs: 1,1) {\textbf{(C)}};
    \end{axis}

    \begin{scope}[spy using outlines={magnification=2.5, connect spies}]
        \begin{axis}[
                width=\linewidth*0.9,
                height=\textheight*0.075,
                scale only axis,
                xmin=0,
                xmax=4.1,
                ymin=-.8,
                ymax=.8,
                xlabel={$t~\text{(s)}$},
                xlabel style={yshift=0.15cm},
                name=rz,
                at=(TRF.below south east),
                anchor= north east,
                legend entries = {$z^f/m_0^f$, $z^s/m_0^s$},
                legend pos = south east,
                legend columns=2,
                legend style={
                        xshift = 0.1cm,
                        yshift = -0.1cm,
                        legend image code/.code={\draw[##1,line width=0.6pt] plot coordinates {(0cm,0cm) (0.25cm,0cm)};}
                    },
            ]
            \addplot [color=UKLblue,            solid]table[x=t_s, y=zf_m0f]{Figures/Spin_dynamics_noInv_varyFA_varyTRF.txt};
            \addplot [color=UKLred,              solid]table[x=t_s, y=zs_m0s]{Figures/Spin_dynamics_noInv_varyFA_varyTRF.txt};


            \coordinate (spypoint_sat) at (axis cs:2.5,0.05);
            \coordinate (spyviewer_sat) at (axis cs:1.5,0.8);
            \spy[rectangle,black,height=0.8cm,width=2.0cm] on (spypoint_sat) in node [fill=white] at (spyviewer_sat);

            \node[anchor=north east, fill=white, opacity=0.85, text opacity=1] at (rel axis cs: 1,1) {\textbf{(D)}};
        \end{axis}
    \end{scope}
\end{tikzpicture}
	\caption{Spin trajectory and the corresponding control for an optimized pulse sequence without inversion pulse and with a varying flip angle and $T_\text{RF}$. \textbf{(A)} The dynamics of the free pool on the Bloch sphere with the steady-state ellipse in blue; \textbf{(B)} the flip angle $\alpha$ and \textbf{(C)} the pulse duration $T_\text{RF}$ control the spin dynamics. \textbf{(D)} The normalized magnetization of the two pools. The rectangular magnification highlights a segment that utilizes saturation\cite{Gloor2008} to encode the MT effect.}
	\label{fig:Spin_dynamics_noInv_varyFA_varyTRF}
\end{figure}

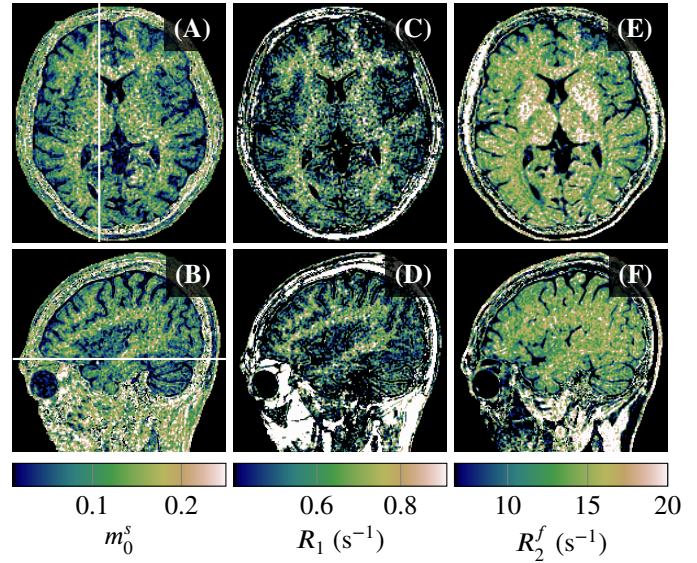
\begin{figure}[tbp]
	\centering
	\begin{tikzpicture}[scale = 1]
    \begin{axis}[%
            width={2.8cm},
            height={2.8cm*1.121},
            axis on top,
            scale only axis,
            xmin=0,
            xmax=1,
            ymin=0,
            ymax=1,
            hide axis,
            name=m0s,
            align=center,
        ]
        \addplot graphics [xmin=0,xmax=1,ymin=0,ymax=1] {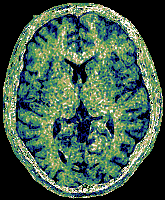};

        \draw [white, thick](axis cs: {(71-4)/165}, {0}) -- (axis cs: {(71-4)/165}, {1});
        \node[fill=black,text=white, opacity=0.75, text opacity=1, anchor = north east] at (rel axis cs:  0.975,0.975) {\textbf{(A)}};
    \end{axis}

    \begin{axis}[%
            width={2.8cm},
            height={2.8cm*0.949},
            axis on top,
            scale only axis,
            xmin=0,
            xmax=1,
            ymin=0,
            ymax=1,
            hide axis,
            name=m0s_sag,
            at=(m0s.south),
            anchor=north,
            yshift = -0.1cm,
            colormap name = gist_earth,
            colorbar horizontal,
            point meta min=0.01,
            point meta max=0.25,
            colorbar style={xlabel=$m_0^s$, height=0.3cm, yshift=0.15cm, xlabel style = {yshift = 0.1cm}},
        ]
        \addplot graphics [xmin=0,xmax=1,ymin=0,ymax=1] {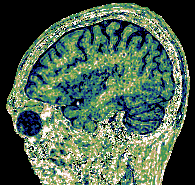};

        \draw [white, thick](axis cs: 0, {1-(110-10)/185}) -- (axis cs: 1, {1-(110-10)/185});
        \node[fill=black,text=white, opacity=0.75, text opacity=1, anchor = north east] at (rel axis cs:  0.975,0.975) {\textbf{(B)}};
    \end{axis}

    \begin{axis}[%
            width={2.8cm},
            height={2.8cm*1.121},
            axis on top,
            scale only axis,
            xmin=0,
            xmax=1,
            ymin=0,
            ymax=1,
            hide axis,
            name=T1,
            at=(m0s.east),
            anchor=west,
            xshift = 0.1cm,
        ]
        \addplot graphics [xmin=0,xmax=1,ymin=0,ymax=1] {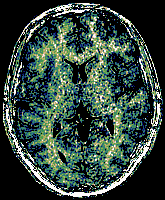};

        \node[fill=black,text=white, opacity=0.75, text opacity=1, anchor = north east] at (rel axis cs:  0.975,0.975) {\textbf{(C)}};
    \end{axis}

    \begin{axis}[%
            width={2.8cm},
            height={2.8cm*0.949},
            axis on top,
            scale only axis,
            xmin=0,
            xmax=1,
            ymin=0,
            ymax=1,
            hide axis,
            name=T1_sag,
            at=(T1.south),
            anchor=north,
            yshift = -0.1cm,
            colormap name = gist_earth,
            colorbar horizontal,
            point meta min=0.401,
            point meta max=0.91,
            colorbar style={xlabel=$R_1~(\text{s}^{-1})$, height=0.3cm, yshift=0.15cm, xlabel style = {yshift = 0.1cm}},
        ]
        \addplot graphics [xmin=0,xmax=1,ymin=0,ymax=1] {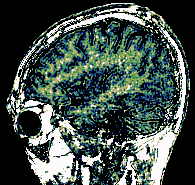};

        \node[fill=black,text=white, opacity=0.75, text opacity=1, anchor = north east] at (rel axis cs:  0.975,0.975) {\textbf{(D)}};
    \end{axis}

    \begin{axis}[%
            width={2.8cm},
            height={2.8cm*1.121},
            axis on top,
            scale only axis,
            xmin=0,
            xmax=1,
            ymin=0,
            ymax=1,
            hide axis,
            name=R2f,
            at=(T1.east),
            anchor=west,
            xshift = 0.1cm,
        ]
        \addplot graphics [xmin=0,xmax=1,ymin=0,ymax=1] {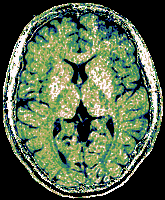};

        \node[fill=black,text=white, opacity=0.75, text opacity=1, anchor = north east] at (rel axis cs:  0.975,0.975) {\textbf{(E)}};
    \end{axis}

    \begin{axis}[%
            width={2.8cm},
            height={2.8cm*0.949},
            axis on top,
            scale only axis,
            xmin=0,
            xmax=1,
            ymin=0,
            ymax=1,
            hide axis,
            name=R2f_sag,
            at=(R2f.south),
            anchor=north,
            yshift = -0.1cm,
            colormap name = gist_earth,
            colorbar horizontal,
            point meta min=6.67,
            point meta max=20,
            colorbar style={xlabel=$R_2^f~(\text{s}^{-1})$, height=0.3cm, yshift=0.15cm, xlabel style = {yshift = 0.1cm}},
        ]
        \addplot graphics [xmin=0,xmax=1,ymin=0,ymax=1] {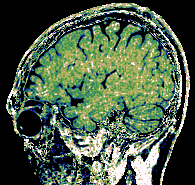};

        \node[fill=black,text=white, opacity=0.75, text opacity=1, anchor = north east] at (rel axis cs:  0.975,0.975) {\textbf{(F)}};
    \end{axis}

\end{tikzpicture}
	\vspace*{-0.75cm}
	\caption{Repetition of Fig.~\ref{fig:InVivo_Control1} reconstructed with an unregluarized conjugate gradient algorithm. Here, we show only the generalized Bloch model. The qMT maps contain, as expected, substantially more noise-like artifacts, but follow overall the same pattern as the maps in Fig.~\ref{fig:InVivo_Control1}, which were reconstructed with a locally low-rank regularization.
	The noise-like artifacts are a combination of thermal noise and undersampling artifacts. Since the latter is not Gaussian distributed, we do not expect the mean value of the qMT parameters to match those in Fig.~\ref{fig:InVivo_Control1}, where undersampling artifacts are suppressed.\cite{Lustig2007,Trzasko2011,Tamir2017}
	}
	\label{fig:InVivo_Control1_CG_gBloch}
\end{figure}

\begin{figure}[tbp]
	\centering
	\begin{tikzpicture}[scale = 1]
    \begin{axis}[%
            width={2.8cm},
            height={2.8cm*1.121},
            axis on top,
            scale only axis,
            xmin=0,
            xmax=1,
            ymin=0,
            ymax=1,
            hide axis,
            name=m0s,
            align=center,
        ]
        \addplot graphics [xmin=0,xmax=1,ymin=0,ymax=1] {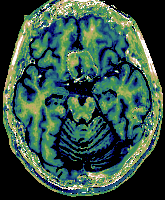};

        \draw [white, thick](axis cs: {(71-4)/165}, {0}) -- (axis cs: {(71-4)/165}, {1});
        \node[fill=black,text=white, opacity=0.75, text opacity=1, anchor = north east] at (rel axis cs:  0.975,0.975) {\textbf{(A)}};
    \end{axis}

    \begin{axis}[%
            width={2.8cm},
            height={2.8cm*0.949},
            axis on top,
            scale only axis,
            xmin=0,
            xmax=1,
            ymin=0,
            ymax=1,
            hide axis,
            name=m0s_sag,
            at=(m0s.south),
            anchor=north,
            yshift = -0.1cm,
            colormap name = gist_earth,
            colorbar horizontal,
            point meta min=0.01,
            point meta max=0.25,
            colorbar style={xlabel=$m_0^s$, height=0.3cm, yshift=0.15cm, xlabel style = {yshift = 0.1cm}},
        ]
        \addplot graphics [xmin=0,xmax=1,ymin=0,ymax=1] {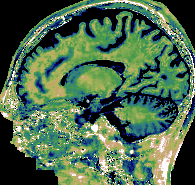};

        \draw [white, thick](axis cs: 0, {1-(110-10)/185}) -- (axis cs: 1, {1-(110-10)/185});
        \node[fill=black,text=white, opacity=0.75, text opacity=1, anchor = north east] at (rel axis cs:  0.975,0.975) {\textbf{(B)}};
    \end{axis}

    \begin{axis}[%
            width={2.8cm},
            height={2.8cm*1.121},
            axis on top,
            scale only axis,
            xmin=0,
            xmax=1,
            ymin=0,
            ymax=1,
            hide axis,
            name=T1,
            at=(m0s.east),
            anchor=west,
            xshift = 0.1cm,
        ]
        \addplot graphics [xmin=0,xmax=1,ymin=0,ymax=1] {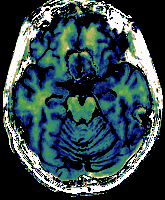};

        \node[fill=black,text=white, opacity=0.75, text opacity=1, anchor = north east] at (rel axis cs:  0.975,0.975) {\textbf{(C)}};
    \end{axis}

    \begin{axis}[%
            width={2.8cm},
            height={2.8cm*0.949},
            axis on top,
            scale only axis,
            xmin=0,
            xmax=1,
            ymin=0,
            ymax=1,
            hide axis,
            name=T1_sag,
            at=(T1.south),
            anchor=north,
            yshift = -0.1cm,
            colormap name = gist_earth,
            colorbar horizontal,
            point meta min=0.401,
            point meta max=0.91,
            colorbar style={xlabel=$R_1~(\text{s}^{-1})$, height=0.3cm, yshift=0.15cm, xlabel style = {yshift = 0.1cm}},
        ]
        \addplot graphics [xmin=0,xmax=1,ymin=0,ymax=1] {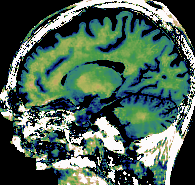};

        \node[fill=black,text=white, opacity=0.75, text opacity=1, anchor = north east] at (rel axis cs:  0.975,0.975) {\textbf{(D)}};
    \end{axis}

    \begin{axis}[%
            width={2.8cm},
            height={2.8cm*1.121},
            axis on top,
            scale only axis,
            xmin=0,
            xmax=1,
            ymin=0,
            ymax=1,
            hide axis,
            name=R2f,
            at=(T1.east),
            anchor=west,
            xshift = 0.1cm,
        ]
        \addplot graphics [xmin=0,xmax=1,ymin=0,ymax=1] {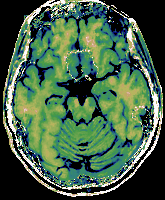};

        \node[fill=black,text=white, opacity=0.75, text opacity=1, anchor = north east] at (rel axis cs:  0.975,0.975) {\textbf{(E)}};
    \end{axis}

    \begin{axis}[%
            width={2.8cm},
            height={2.8cm*0.949},
            axis on top,
            scale only axis,
            xmin=0,
            xmax=1,
            ymin=0,
            ymax=1,
            hide axis,
            name=R2f_sag,
            at=(R2f.south),
            anchor=north,
            yshift = -0.1cm,
            colormap name = gist_earth,
            colorbar horizontal,
            point meta min=6.67,
            point meta max=20,
            colorbar style={xlabel=$R_2^f~(\text{s}^{-1})$, height=0.3cm, yshift=0.15cm, xlabel style = {yshift = 0.1cm}},
        ]
        \addplot graphics [xmin=0,xmax=1,ymin=0,ymax=1] {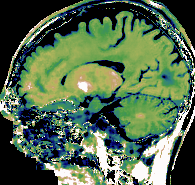};

        \node[fill=black,text=white, opacity=0.75, text opacity=1, anchor = north east] at (rel axis cs:  0.975,0.975) {\textbf{(F)}};
    \end{axis}

\end{tikzpicture}
	\vspace*{-0.75cm}
	\caption{Repetition of Fig.~\ref{fig:InVivo_Control1} with slices chosen to highlight the banding artifacts above the frontal sinuses, where the qMT parameter estimation fails. Here, we show only the generalized Bloch model.
	}
	\label{fig:InVivo_Control1_sinus_gBloch}
\end{figure}
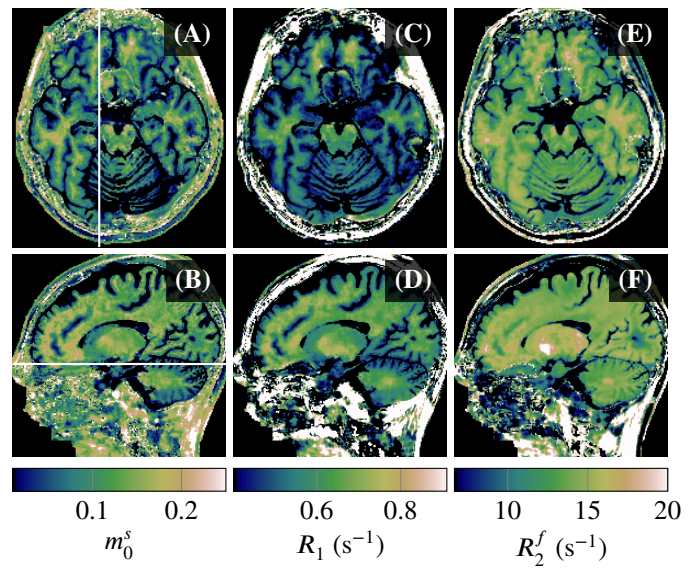

\begin{figure}[htbp]
	\centering
	\begin{tikzpicture}[scale = 1]
    \begin{axis}[%
        width={2.8cm},
        height={2.8cm*1.121},
        axis on top,
        scale only axis,
        xmin=0,
        xmax=1,
        ymin=0,
        ymax=1,
        xtick = \empty,
        ytick = \empty,
        ylabel={Graham},
        name=m0s_Graham,
        align=center,
    ]
    \addplot graphics [xmin=0,xmax=1,ymin=0,ymax=1] {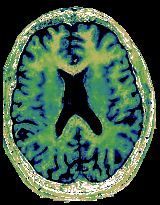};

    \node[fill=black,text=white, opacity=0.75, text opacity=1, anchor = north east] at (rel axis cs:  0.975,0.975) {\textbf{(A)}};
    \end{axis}

    \begin{axis}[%
            width={2.8cm},
            height={2.8cm*1.121},
            axis on top,
            scale only axis,
            xmin=0,
            xmax=1,
            ymin=0,
            ymax=1,
            xtick = \empty,
            ytick = \empty,
            ylabel={gen. Bloch},
            name=m0s_gBloch,
            at=(m0s_Graham.south),
            anchor=north,
            yshift = -0.1cm,
            align=center,
            colormap name = gist_earth,
            colorbar horizontal,
            point meta min=0.01,
            point meta max=0.25,
            colorbar style={xlabel=$m_0^s$, height=0.3cm, yshift=0.15cm, xlabel style = {yshift = 0.1cm}},
        ]
        \addplot graphics [xmin=0,xmax=1,ymin=0,ymax=1] {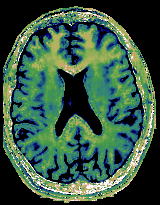};

        \node[fill=black,text=white, opacity=0.75, text opacity=1, anchor = north east] at (rel axis cs:  0.975,0.975) {\textbf{(B)}};
    \end{axis}

    \begin{axis}[%
            width={2.8cm},
            height={2.8cm*1.121},
            axis on top,
            scale only axis,
            xmin=0,
            xmax=1,
            ymin=0,
            ymax=1,
            hide axis,
            name=R1_Graham,
            at=(m0s_Graham.east),
            anchor=west,
            xshift = 0.1cm,
        ]
        \addplot graphics [xmin=0,xmax=1,ymin=0,ymax=1] {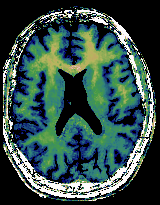};

        \node[fill=black,text=white, opacity=0.75, text opacity=1, anchor = north east] at (rel axis cs:  0.975,0.975) {\textbf{(C)}};
    \end{axis}

    \begin{axis}[%
            width={2.8cm},
            height={2.8cm*1.121},
            axis on top,
            scale only axis,
            xmin=0,
            xmax=1,
            ymin=0,
            ymax=1,
            hide axis,
            name=R1_gBloch,
            at=(m0s_gBloch.east),
            anchor=west,
            xshift = 0.1cm,
            colormap name = gist_earth,
            colorbar horizontal,
            point meta min=0.401,
            point meta max=0.91,
            colorbar style={xlabel=$R_1~(\text{s}^{-1})$, height=0.3cm, yshift=0.15cm, xlabel style = {yshift = 0.1cm}},
        ]
        \addplot graphics [xmin=0,xmax=1,ymin=0,ymax=1] {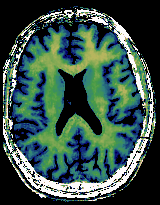};

        \node[fill=black,text=white, opacity=0.75, text opacity=1, anchor = north east] at (rel axis cs:  0.975,0.975) {\textbf{(D)}};
    \end{axis}

    \begin{axis}[%
            width={2.8cm},
            height={2.8cm*1.121},
            axis on top,
            scale only axis,
            xmin=0,
            xmax=1,
            ymin=0,
            ymax=1,
            hide axis,
            name=R2f_Graham,
            at=(R1_Graham.east),
            anchor=west,
            xshift = 0.1cm,
        ]
        \addplot graphics [xmin=0,xmax=1,ymin=0,ymax=1] {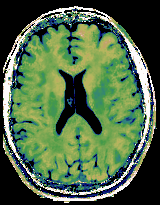};

        \node[fill=black,text=white, opacity=0.75, text opacity=1, anchor = north east] at (rel axis cs:  0.975,0.975) {\textbf{(E)}};
    \end{axis}

    \begin{axis}[%
            width={2.8cm},
            height={2.8cm*1.121},
            axis on top,
            scale only axis,
            xmin=0,
            xmax=1,
            ymin=0,
            ymax=1,
            hide axis,
            name=R2f_gBloch,
            at=(R1_gBloch.east),
            anchor=west,
            xshift = 0.1cm,
            colormap name = gist_earth,
            colorbar horizontal,
            point meta min=6.67,
            point meta max=20,
            colorbar style={xlabel=$R_2^f~(\text{s}^{-1})$, height=0.3cm, yshift=0.15cm, xlabel style = {yshift = 0.1cm}},
        ]
        \addplot graphics [xmin=0,xmax=1,ymin=0,ymax=1] {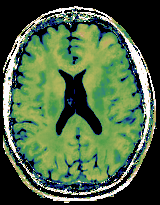};

        \node[fill=black,text=white, opacity=0.75, text opacity=1, anchor = north east] at (rel axis cs:  0.975,0.975) {\textbf{(F)}};
    \end{axis}

\end{tikzpicture}
	\vspace*{-0.75cm}
	\caption{Repetition of Fig.~\ref{fig:InVivo_Control1} for a second healthy volunteer. Here, we show only a transversal slice of the 3D volume.
	}
	\label{fig:InVivo_Control2}
\end{figure}
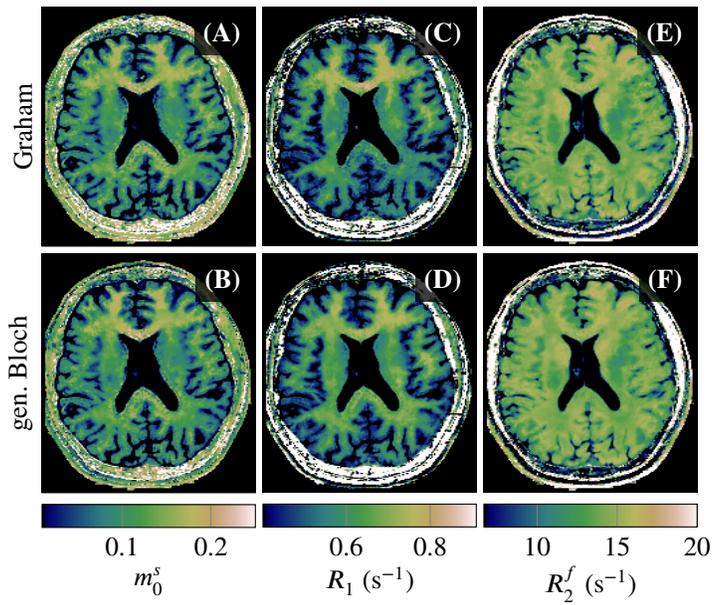

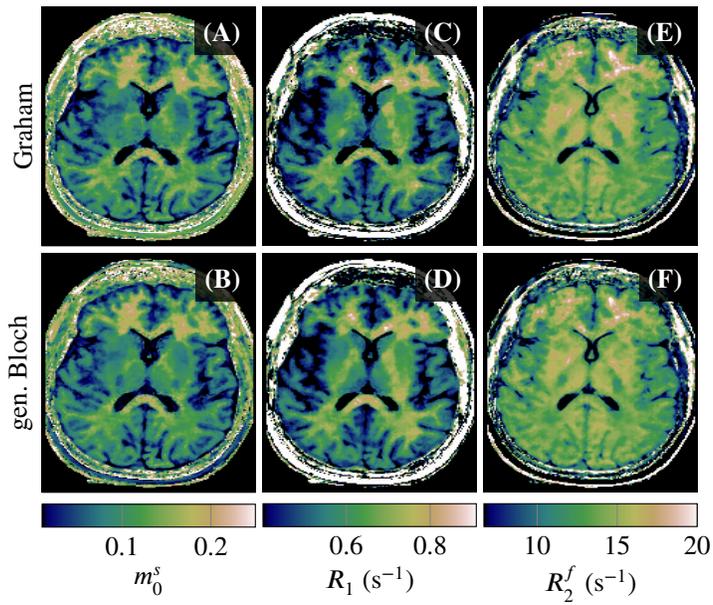
\begin{figure}[htbp]
	\centering
	\begin{tikzpicture}[scale = 1]
    \begin{axis}[%
        width={2.8cm},
        height={2.8cm*1.121},
        axis on top,
        scale only axis,
        xmin=0,
        xmax=1,
        ymin=0,
        ymax=1,
        xtick = \empty,
        ytick = \empty,
        ylabel={Graham},
        name=m0s_Graham,
        align=center,
    ]
    \addplot graphics [xmin=0,xmax=1,ymin=0,ymax=1] {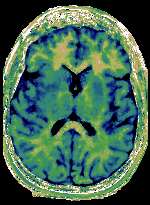};

    \node[fill=black,text=white, opacity=0.75, text opacity=1, anchor = north east] at (rel axis cs:  0.975,0.975) {\textbf{(A)}};
    \end{axis}

    \begin{axis}[%
            width={2.8cm},
            height={2.8cm*1.121},
            axis on top,
            scale only axis,
            xmin=0,
            xmax=1,
            ymin=0,
            ymax=1,
            xtick = \empty,
            ytick = \empty,
            ylabel={gen. Bloch},
            name=m0s_gBloch,
            at=(m0s_Graham.south),
            anchor=north,
            yshift = -0.1cm,
            align=center,
            colormap name = gist_earth,
            colorbar horizontal,
            point meta min=0.01,
            point meta max=0.25,
            colorbar style={xlabel=$m_0^s$, height=0.3cm, yshift=0.15cm, xlabel style = {yshift = 0.1cm}},
        ]
        \addplot graphics [xmin=0,xmax=1,ymin=0,ymax=1] {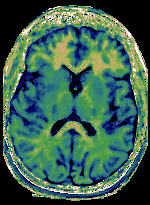};

        \node[fill=black,text=white, opacity=0.75, text opacity=1, anchor = north east] at (rel axis cs:  0.975,0.975) {\textbf{(B)}};
    \end{axis}

    \begin{axis}[%
            width={2.8cm},
            height={2.8cm*1.121},
            axis on top,
            scale only axis,
            xmin=0,
            xmax=1,
            ymin=0,
            ymax=1,
            hide axis,
            name=R1_Graham,
            at=(m0s_Graham.east),
            anchor=west,
            xshift = 0.1cm,
        ]
        \addplot graphics [xmin=0,xmax=1,ymin=0,ymax=1] {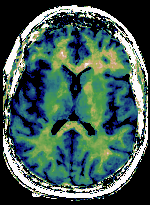};

        \node[fill=black,text=white, opacity=0.75, text opacity=1, anchor = north east] at (rel axis cs:  0.975,0.975) {\textbf{(C)}};
    \end{axis}

    \begin{axis}[%
            width={2.8cm},
            height={2.8cm*1.121},
            axis on top,
            scale only axis,
            xmin=0,
            xmax=1,
            ymin=0,
            ymax=1,
            hide axis,
            name=R1_gBloch,
            at=(m0s_gBloch.east),
            anchor=west,
            xshift = 0.1cm,
            colormap name = gist_earth,
            colorbar horizontal,
            point meta min=0.401,
            point meta max=0.91,
            colorbar style={xlabel=$R_1~(\text{s}^{-1})$, height=0.3cm, yshift=0.15cm, xlabel style = {yshift = 0.1cm}},
        ]
        \addplot graphics [xmin=0,xmax=1,ymin=0,ymax=1] {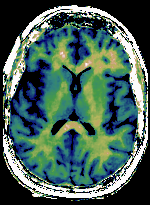};

        \node[fill=black,text=white, opacity=0.75, text opacity=1, anchor = north east] at (rel axis cs:  0.975,0.975) {\textbf{(D)}};
    \end{axis}

    \begin{axis}[%
            width={2.8cm},
            height={2.8cm*1.121},
            axis on top,
            scale only axis,
            xmin=0,
            xmax=1,
            ymin=0,
            ymax=1,
            hide axis,
            name=R2f_Graham,
            at=(R1_Graham.east),
            anchor=west,
            xshift = 0.1cm,
        ]
        \addplot graphics [xmin=0,xmax=1,ymin=0,ymax=1] {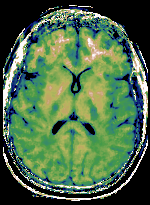};

        \node[fill=black,text=white, opacity=0.75, text opacity=1, anchor = north east] at (rel axis cs:  0.975,0.975) {\textbf{(E)}};
    \end{axis}

    \begin{axis}[%
            width={2.8cm},
            height={2.8cm*1.121},
            axis on top,
            scale only axis,
            xmin=0,
            xmax=1,
            ymin=0,
            ymax=1,
            hide axis,
            name=R2f_gBloch,
            at=(R1_gBloch.east),
            anchor=west,
            xshift = 0.1cm,
            colormap name = gist_earth,
            colorbar horizontal,
            point meta min=6.67,
            point meta max=20,
            colorbar style={xlabel=$R_2^f~(\text{s}^{-1})$, height=0.3cm, yshift=0.15cm, xlabel style = {yshift = 0.1cm}},
        ]
        \addplot graphics [xmin=0,xmax=1,ymin=0,ymax=1] {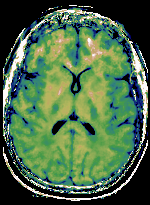};

        \node[fill=black,text=white, opacity=0.75, text opacity=1, anchor = north east] at (rel axis cs:  0.975,0.975) {\textbf{(F)}};
    \end{axis}

\end{tikzpicture}
	\vspace*{-0.75cm}
	\caption{Repetition of Fig.~\ref{fig:InVivo_Patient1} for a second participant with multiple sclerosis. Here, we show only a transversal slice of the 3D volume.
	}
	\label{fig:InVivo_Patient2}
\end{figure}

\bibliography{library}%
\makeatletter\@input{MT_HSFP_Paper_aux.tex}\makeatother